\documentclass[12pt,usegraphicx]{emulateapj}

\usepackage{natbib}
\usepackage{epsfig}
\usepackage{latexsym,amssymb}
\usepackage{amsmath}
\usepackage{textcomp}
\usepackage{dcolumn}
\usepackage{booktabs}
\usepackage{verbatim}
\usepackage{footmisc}

\newcommand{\galfit}{\textsc{galfit}}
\newcommand{\gfthree}{\textsc{galfit3}}
\newcommand{\sersic}{S\'{e}rsic}

\newcommand{\sex}{\textsc{SExtractor}}

\newcommand{\swarp}{\textsc{SWarp}}

\newcommand{\kms}{\mathrm{km\,s^{-1}}}

\newcommand{\mbh}{M_\mathrm{BH}} 

\newcommand{\eps}{\epsilon} 
\newcommand{\dd}{\mathrm{d}} 
\newcommand{\bas}{_\mathrm{bas}}
\newcommand{\bul}{_\mathrm{bul}}
\newcommand{\tot}{_\mathrm{tot}}

\newcommand{\lbul}{L\bul}
\newcommand{\ltot}{L\tot}
\newcommand{\lbas}{L\bas}
\newcommand{\mbul}{M\bul}
\newcommand{\mtot}{M\tot}
\newcommand{\mbas}{M\bas}

\newcommand{\magarcsec}{\,\mathrm{mag\,arcsec^{-2}}}
\newcommand{\mg}{\,\mathrm{mag}}
\newcommand{\msun}{\,M_\odot}

\newcommand{\dex}{\,\mathrm{dex}}

\newcommand{\pa}{\mathrm{PA}}
\newcommand{\pc}{\,\mathrm{pc}}
\newcommand{\kpc}{\,\mathrm{kpc}}
\newcommand{\Mpc}{\,\mathrm{Mpc}}

\newcommand{\UpsH}{\Upsilon_H}
\newcommand{\msigma}{$\mbh-\sigma$}

\newcommand{\hst}{\emph{HST}}
\newcolumntype{d}[1]{D{.}{.}{#1}}

\makeatletter
\def\@xfootnote[#1]{
  \protected@xdef\@thefnmark{#1}
  \@footnotemark\@footnotetext}
\makeatother

\lefthead{L\"asker et al.}
\righthead{Megamaser Host Galaxies}
\slugcomment{submitted to ApJ, Version: \today}

\begin{document}
 
\title{The Black Hole - Bulge Mass Relation in Megamaser Host Galaxies\footnotemark[*]}

\author{Ronald L\"{a}sker$^{1,7}$, Jenny E. Greene$^2$, Anil Seth$^3$, Glenn van de Ven$^1$, James A. Braatz$^4$, \\ Christian Henkel$^{5,6}$, and K. Y. Lo$^4$, }
\affil{${}^1$Max-Planck Institut f\"ur Astronomie, K\"onigstuhl 17, D-69117, Heidelberg, Germany; e-mail:laesker@mpia.de}
\affil{${}^2$Department of Astrophysical Sciences, 4 Ivy Lane, Peyton Hall, Princeton University, Princeton, NJ 08544, USA}
\affil{${}^3$Department of Physics \& Astronomy, 201 James Fletcher Bldg., 115 South 1400 East, University of Utah, Salt Lake City, UT 84112, USA}
\affil{${}^4$National Radio Astronomy Observatory, 520 Edgemont Rd., Charlottesville, VA 22903, USA}
\affil{${}^5$Max-Planck-Institut f{\"u}r Radioastronomie, Auf dem H{\"u}gel 69, D-31212 Bonn, Germany}
\affil{${}^6$Astron. Dept., King Abdulaziz University, P.O. Box 80203, Jeddah 21589, Saudi Arabia}
\affil{${}^7$Finnish Centre for Astronomy with ESO (FINCA), University of Turku, V\"ais\"al\"antie 20, 21500 Kaarina, Finland}

\footnotetext[*]{Based on observations made with the NASA/ESA Hubble Space Telescope, obtained at the Space Telescope Science Institute, which is operated by the Association of Universities for Research in Astronomy, Inc., under NASA contract NAS 5-26555. These observations are associated with program 12185.}

\begin{abstract}
We present \emph{HST} images for nine megamaser disk galaxies with the
primary goal of studying photometric BH-galaxy scaling relations. 
The megamaser disks provide the highest-precision extragalactic 
BH mass measurements, while our high-resolution \hst\ imaging affords 
us the opportunity to decompose the complex nuclei of their late-type 
hosts in detail.
Based on the morphologies and shapes of the galaxy nuclei, we argue that most
of these galaxies' central regions contain secularly evolving
components (pseudo-bulges), and in many cases we photometrically
identify co-existing ``classical'' bulge components as well. Using
these decompositions, we draw the following conclusions:  (1) The
megamaser BH masses span two orders of magnitude ($10^6$ --
$10^8\msun$) while the stellar mass of their spiral host galaxies are
all $\sim 10^{11}\msun$ within a factor of three; (2) the BH masses at
a given bulge mass or total stellar mass in the megamaser host spiral
galaxies tend to be lower than expected, when compared to an
extrapolation of the BH-bulge relation based on early-type galaxies;
(3) the observed large intrinsic scatter of BH masses in the megamaser
host galaxies raises the question of whether scaling relations exist
in spiral galaxies.
\end{abstract}


\section{Introduction}
\label{sec:intro}

\renewcommand{\thefootnote}{\arabic{footnote}}

Supermassive black holes (BHs) play a special role in galaxy
evolution.  They are a ubiquitous component of massive galaxies
\citep[e.g.,][]{Kormendy+Ho13}, and appear to approximately comprise a fixed fraction of the
mass of the spheroidal component of the galaxy
\citep[e.g.,][]{McConnell+Ma13}. Motivated by these scaling relations,
theory invokes energy injection from actively accreting BHs to
self-regulate BH growth \citep{Debuhr+10}, truncate star
formation in massive galaxies
\citep[e.g.,][]{Silk+Rees98,Springel+diMatteo+Hernquist05}, and keep
gas in clusters from cooling \citep[e.g.,][]{Fabian12}.

However, our understanding of BH demographics is far from
complete. In particular, while we now have dynamical BH mass
measurements for more than fifty galaxies, these galaxies convey a
biased view of the galaxy population.  They are skewed towards
dense elliptical galaxies \citep[][]{vdBosch+15}. They are also
biased towards massive systems with few exceptions
\citep[e.g.,][]{Seth+10,Seth+14}. Spiral galaxies are
particularly challenging; due to both the typically low BH masses and
the presence of dust, star formation, and non-axisymmetric components
(e.g., bars), stellar and gas dynamical modeling is far more
challenging. These limitations hinder our ability to diagnose the
underlying physical mechanisms driving the scaling relations.

There is one method that delivers high-precision, high-accuracy
dynamical BH masses in spiral galaxy nuclei: fitting the rotation
curves of megamaser disks \citep[e.g.,][]{Herrnstein+05,Kuo+11}. In
these special systems, we observe extremely luminous 22 GHz H$_2$O
masers in an edge-on accretion disk on $\sim 0.5$ pc scales from a
weakly accreting supermassive BH \citep[see review by][]{Lo+05}.  The
precision of the BH mass measurement is actually limited by our
knowledge of the galaxy distances. As first demonstrated with the
prototypical megamaser disk galaxy NGC4258 \citep[][]{Herrnstein+99},
it is also possible to measure a geometric distance using very-long-baseline observations (VLBI) in combination with single-dish monitoring of the acceleration of the systemic megamasers in the disk.  
These geometric distance measurements are the primary goal of the
Megamaser Cosmology Project \citep[MCP;][and associated follow-up
  publications]{Reid+09,Braatz+10}. Thus far, five galaxies have
direct distance measurements \citep[NGC4258, NGC5765b, NGC6264, NGC6323, 
  UGC3789 in][]{Herrnstein+99,Gao+15,Kuo+13,Kuo+15,Reid+13}. At least nine
galaxies have reliable megamaser-based BH mass measurements: NGC1194,
NGC2273, NGC2960, NGC4388, NGC6264 and NGC6323 \citep{Kuo+11}, as well
as NGC3993 \citep{Kondratko+08}, NGC4258
\citep[e.g.][]{Miyoshi+95,Herrnstein+05,Humphreys+13}, and UGC3789
\citep{Reid+09,Kuo+11}.

The megamaser disk galaxies allow us to peer through the gas, dust,
and star formation to directly measure BH masses in spiral galaxy
nuclei as well as get a handle on secular BH fueling mechanisms
\citep[][] {Greene+13a,Greene+14}. In previous work we have studied
the relationship between galaxy velocity dispersion $\sigma^*$ and BH
mass in megamaser disk galaxies \citep{Greene+10b}. Here we tackle the relationship between BH
mass and bulge mass in these objects. Although a number
of prior works have included bulge luminosities for many of these
galaxies \citep[e.g.,][]{Greene+10b,Kormendy+Ho13,Graham+Scott15}, in this paper we demonstrate that there is significant
substructure on sub-arcsecond scales, and that disentangling the
bulge components from other nuclear components requires the highest
possible spatial resolution provided by the Hubble Space Telescope.

This paper contains many technical sections.  In Section 2 we describe
the sample, our data sources, and data processing.  In Section 3 we
decompose the two-dimensional surface brightness profiles and in
Section 4 we investigate the nature of the bulge components of the
megamaser disk host galaxies. Those interested in the main results can
go directly to Section 5, where we discuss the megamaser disks in the
BH-galaxy mass plane, and then fit the scaling relations including the new
measurements presented here. In Section 6 we summarize and discuss our
results. The redshift distances to our targets are based on $H_0=70
\,\kms \Mpc^{-1}$, a value that is consistent with all the published
values of $H_0$ based on geometric distance determinations of
megamaser disk galaxies (see MCP references given above).

\section{Data}
\label{sec:data}

\subsection{Sample}
\label{subsec:data:sample}

Our sample of megamaser disk galaxies with BH masses is taken from
\cite{Greene+10b}, with most of the $\mbh$ measurements provided by
\citet[][Table \ref{tab:sample}]{Kuo+11}. We focus on these nine
galaxies because they have Keplerian megamaser rotation curves and
high-resolution imaging data, and thus well-determined $\mbh$. While the BH mass we adopt for IC2560 \citep[$10^{6.4\pm0.4}\msun$,][]{Kuo+11} is based on single-dish data, published and preliminary reductions of VLBI data yield a $\log\mbh/\msun$ of $6.54\pm0.06$ \citep{Yamauchi+12} and $6.64\pm0.03$ (Wagner et al. in prep). If we were to adopt those values, the results for the BH scaling relations (relation parameter and subsample offsets, see Section \ref{sec:bhscalerels}) barely change (within a few percent of the uncertainties). Several
of our targets are $>50\Mpc$ away, and most of them were previously
known or suspected to host small-scale structures (nuclear rings,
disks or bars) in addition to the bulge. High-resolution imaging is thus
essential for a robust analysis. We do not 
include Circinus, NGC1068, or NGC4945 in this work, since comparable 
high-resolution data are not readily available for them, but we will 
consider the first two as a secondary sample in \S 5. 

\begin{table*}
 \caption{The Megamaser Sample}
 \begin{tabular}{llld{3.1}llllll}
  \toprule
   Galaxy & RA & Dec & \multicolumn{1}{l}{$D$} & Hubble Type & $L_H$ & $R_e$ & $t_\mathrm{exp}$ & obs. date & wide-field imaging \\
   & & & \multicolumn{1}{l}{$[\Mpc]$} & & $[10^{10}\,L_{\odot,H}]$ & $[\kpc]$ & $[s]$ & & \\
   (1) & (2) & (3) & \multicolumn{1}{l}{(4)} & (5) & (6) & (7) & (8) & (9) & (10) \\
   \midrule
IC 2560     & $10:16:18.7$ & $-33:33:50$ &   41.8 & (R’)SB(r)b & 13.2 & 4.0 & $422$ & 2010-10-23 & \begin{tabular}{@{}l@{}} UVIS-F814W \\ LCO DuPont/Tek-I \end{tabular} \\
NGC1194  & $03:03:49.1$ & $-01:06:13$ &   52.0 & SA0+ & 5.75 & 2.8 & $422$ & 2011-11-23 & APO3.5m/NIC-K \\
NGC2273  & $06:50:08.6$ & $+60:50:45$ &   26.0 & SB(r)a: & 3.47 & 1.4 & $422$ & 2011-01-31 & APO3.5m/NIC-K \\
NGC2960  & $09:40:36.4$ & $+03:34:37$ &   71.0 & Sa? & 7.76 & 1.9 & $422$ & 2011-06-14 & APO3.5m/NIC-K \\
NGC3393  & $10:48:23.4$ & $-25:09:43$ &   53.6 & (R’)SB(rs) & 9.33 & 2.5 & $422$ & 2011-11-11 & UVIS-F814W \\
NGC4388  & $12:25:46.7$ & $-12:39:44$ &   19.0 & SA(s)b: & 3.09 & 1.4 & $422$ & 2011-06-08 & APO3.5m/NIC-K \\
NGC6264  & $16:57:16.1$ & $-27:50:59$ & 136.0 & S? & 8.71 & 4.0 & $422$ & 2011-07-30 & APO3.5m/NIC-K \\
NGC6323  & $17:13:18.1$ & $-43:46:57$ & 105.0 & Sab & 8.71 & 2.5 & $422$ & 2011-09-01 & WIYN/WHIRC-K \\
UGC03789 & $07:19:30.9$ & $-59:21:18$ &   50.0 & (R)SA(r)ab & 5.37 & 2.2 & $422$ & 2011-09-06 & APO3.5m/NIC-K \\
   \bottomrule
 \end{tabular}
 \label{tab:sample}
 \tablecomments{
Columns (1-5) give the galaxy name, coordinates, distance in Mpc, and
Hubble Type (RC3 catalog, \protect\citealt{RC3}). Column (6) gives the the galaxy (total) $H$-band
luminosity in units of $10^{10}\,L_{\odot,H}$, and column (7) the
galaxy effective radius in kpc. Both $L_H$ and $R_e$ values result from our
imaging data and multi-component fitting. The inner parts are imaged
with \emph{HST}/WFC3 in the F160W filter, with exposure times in seconds, and
dates given in columns (8) and (9). In order to get constraints on the
outer parts of the surface brightness that did not fit on the WFC3-IR
FOV of $70\farcs4 \times 62\farcs6$, we also include data that we
obtained on telescopes and instruments given in column (10).}
\end{table*}

\subsection{\emph{HST} imaging}
\label{subsec:data:hst}

The high-resolution \hst\ data (FWHM$=0\farcs15$ in the $H-$band 
corresponding to 50 pc at the median sample distance of 70 Mpc) were
taken between October 2010 and November 2011 (see Table
\ref{tab:sample}). We obtained \hst/WFC3-IR imaging with filters
F110W and F160W (similar to 2MASS $J-$ and $H-$band) for each
galaxy. Within the same program we also acquired WFC-UVIS exposures in
the F336W, F438W and F814W filters, (roughly $U$, $B$ and
$I-$band). We base our bulge luminosities and derived masses on the
$H-$band data, in order to reduce the uncertainties
associated with dust obscuration and variations in stellar population,
as compared to optical bands. Indeed, many of the late-type galaxies
in our sample are quite dusty and have complicated color
profiles. We use the color information contained in the UVIS bands for
a refined luminosity-mass conversion.

All of our galaxies were imaged in four exposures with F160W using sub-pixel
dithering and a total exposure time of $422$ sec. They were
co-added and cleaned of cosmic rays using the PyFits MultiDrizzle
pipeline\footnote{PyFits is a product of the Space Telescope Science
  Institute, which is operated by AURA for NASA.}. In order to fit our
five-band observations into two orbits, we utilize sub-arrays to avoid
buffer dumps.  The FOV of the subarray is $70\times62$ arcseconds,
which is filled by the target in all cases. MultiDrizzle provides
exposure time maps for the combined frame that we use to calculate the
noise map needed for model fitting with \galfit.

\subsection{Ground-based $K-$band imaging}
\label{subsec:data:k-band}

Since our \hst\ data do not cover the outskirts of our target
galaxies, we obtained additional wide-field imaging data. Most of these were ground-based $K$-band images,
acquired from 2010-2012 on the Apache Point Observatory (APO) and WIYN\footnote{The WIYN Observatory is a joint facility of the University of Wisconsin-Madison, Indiana University, the National Optical Astronomy Observatory and the University of Missouri.} {3.5-m} telescopes, using the NICFPS and WHIRC instruments respectively
\citep[http://nicfps.colorado.edu/,][]{WHIRC}. In two fields (IC2560 and NGC3993), sufficiently deep NIR images were not available; here we used \hst\ {WFC3 F814W} and Las Campanas Observatory (LCO) du Pont {100-inch} {SITe2K-CCD} $I$-band images. For details of instruments, exposure
times and observing dates, see Table \ref{tab:sample}. Apart from
providing valuable constraints on the derived galaxy parameters and
thus allowing robust fits to the complex morphologies we
encounter, their FOV is also large enough to allow reliable background
subtraction, unbiased by galaxy light. In ground-based $K$-band images, these advantages are tempered
by a bright ($\sim 14\magarcsec$) background that 
is variable both spatially and temporally. 
A carefully designed observing strategy and data reduction
is therefore required to achieve reliable background
subtraction.

\begin{figure*}
 \centering
 \includegraphics[width=8.5cm,trim=0 50 0 0,clip=true]{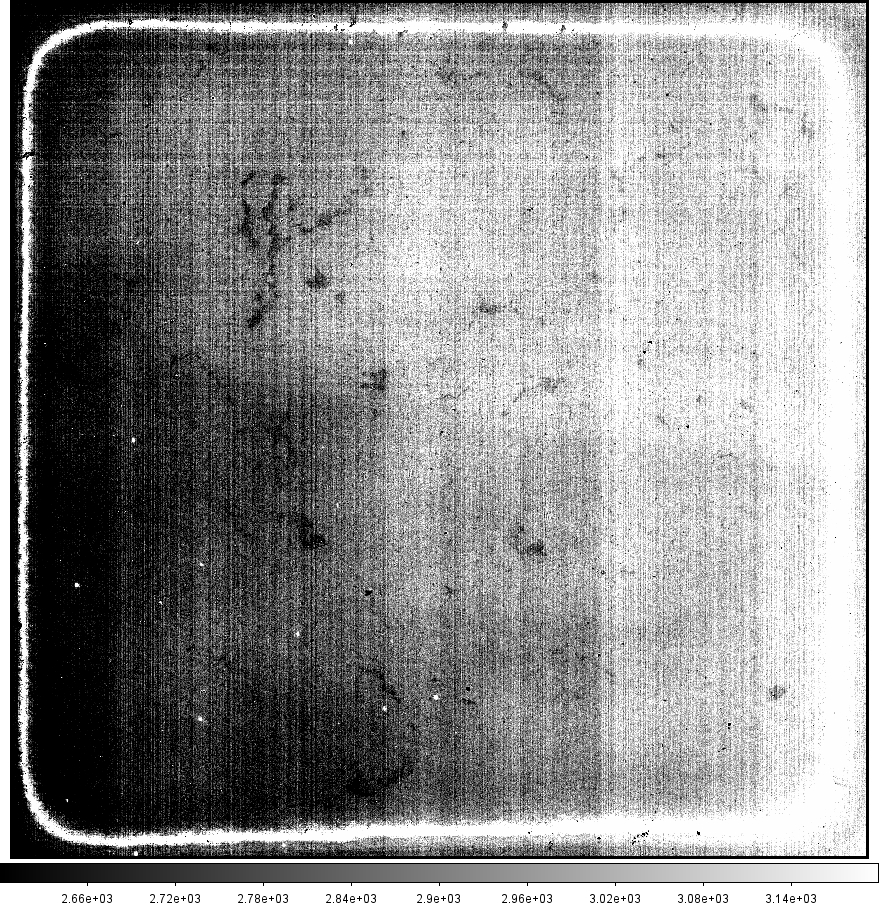}
 \includegraphics[width=8.5cm,trim=0 50 0 0,clip=true]{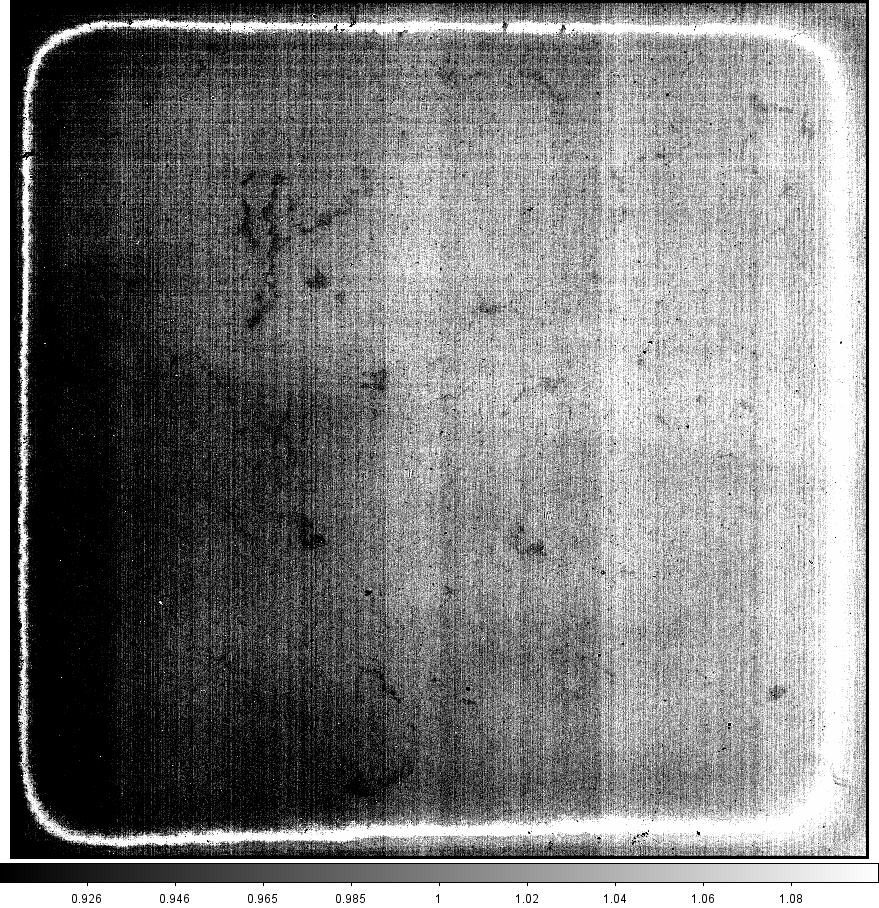}
 \includegraphics[width=8.5cm,trim=0 50 0 0,clip=true]{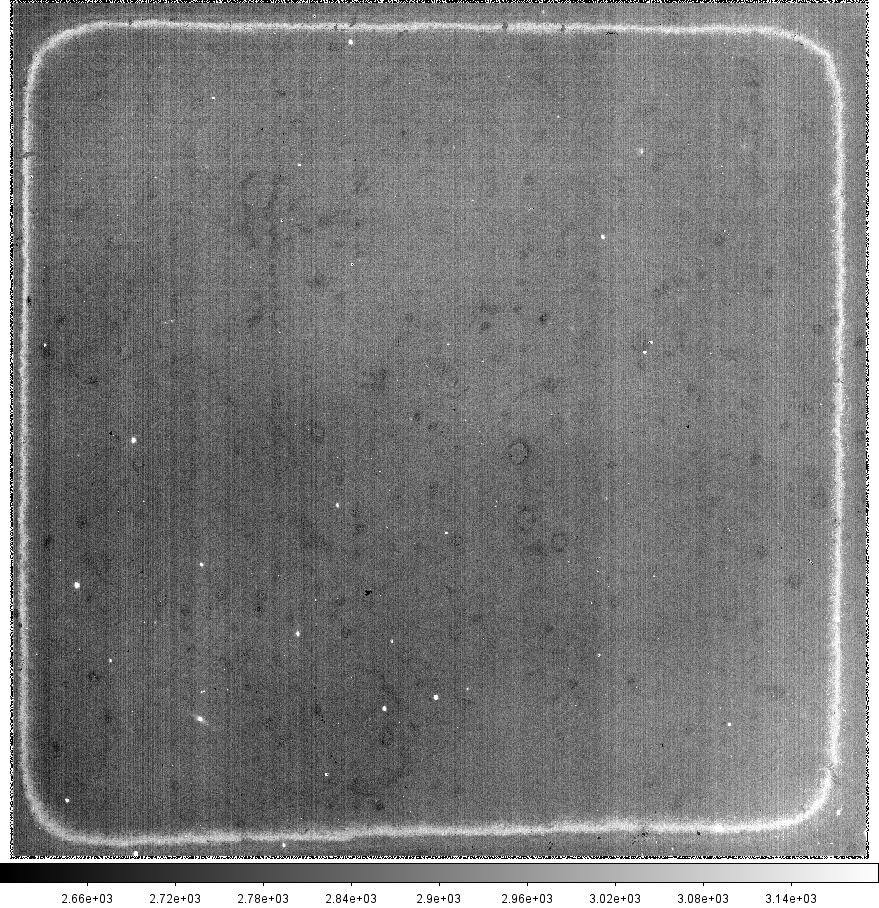}
 \includegraphics[width=8.5cm,trim=0 50 0 0,clip=true]{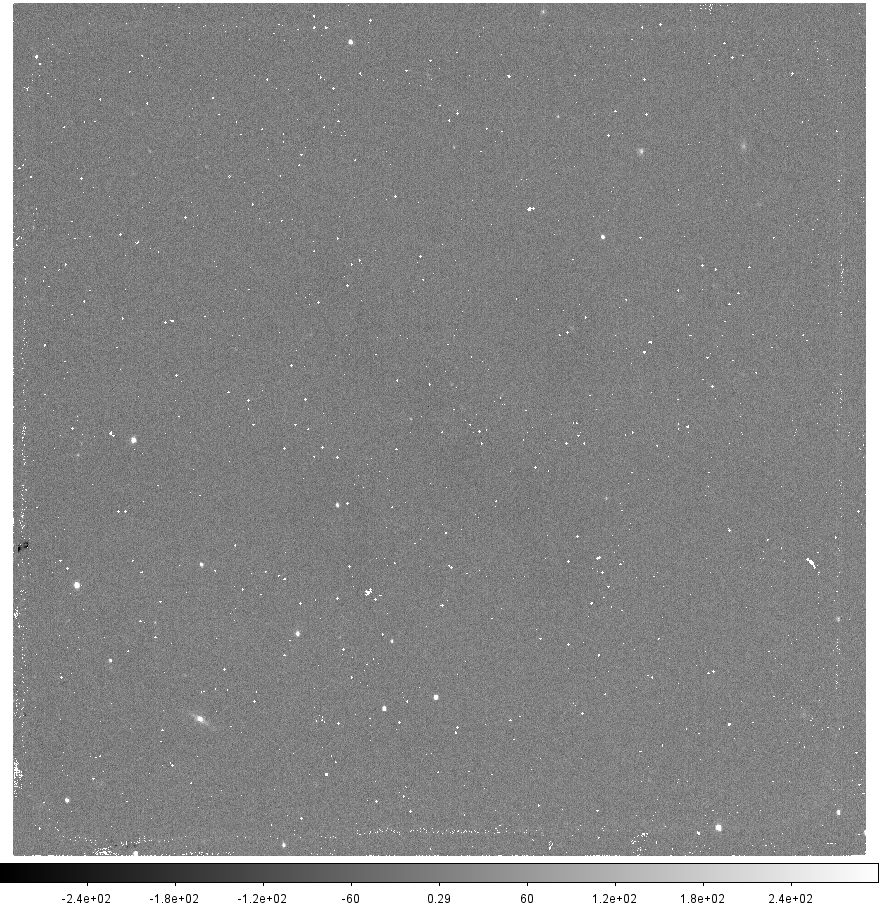}
 \caption{
Example of the need for subtracting the residual common sky
pattern after flatfielding, from APO data. Top left: raw exposure,
top right: twilight flatfield stack. The structure remaining after
dividing by the flatfield image is evident (bottom-left panel), as well
as the improvement after subtracting the mean sky pattern (the mean of
all frames of the OB, after source masking, bottom-right). Random temporal sky
pattern variation is negligible and not visible at this
contrast. The same relative greyscale was used in all panels
($\pm10\%$ of the mean).}
 \label{fig:skyflat}
\end{figure*}

In taking the data, we apply a large-scale dither pattern, with the
target first imaged near the detector center, and then moved towards
the corners of the FOV in a clockwise pattern. Before and after every
on-target exposure we perform an off-target exposure, and the whole
pattern is repeated several times. This results in $\sim25$
science frames and twice the number of sky frames. With this strategy,
we obtain a reliable reconstruction of the background, since every
part of the detector remains unoccupied by the relatively large target
galaxy at least in a sizeable fraction of the exposures, and we can
simultaneously monitor the evolution of the background level and 
the two-dimensional structure in the background.

Using the dark and twilight flat frames, we identify bad pixels, and
mask them in all subsequent exposures. After flatfielding the raw
frames, a background pattern (see bottom-left panel in Figure
\ref{fig:skyflat}) persists, as well as a much weaker, near-random
pattern that corresponds to the expected pattern from variable
non-uniform sky (atmospheric) emission. We ascribe the time-invariant
portion of the residual background pattern to a discrepancy between
the twilight flatfield image and the response of the detector to night
sky illumination. Therefore we need to subtract the mean background
pattern that persists after the frames have been flatfielded with the
twilight flat. The standard alternative, dome flatfield images,
provides a significantly poorer approximation to the nighttime sky
(and flatfield) pattern. We addressed
the residual time-invariant background pattern by flatfielding using
a ``superflat'', which is a flatfield image constructed from the
images taken over the course of the night. However, we found no improvement 
in the background subtraction using this method. To achieve the required S/N 
in the flat requires summing over most of the images over the course of the 
night, but there is sufficient change in the background pattern with time 
that residual patterns remain after flatfielding with the "superflat".

We construct an image of the mean residual background pattern as
follows.  We first normalize each frame by its mean background, and
then take a pixel-by-pixel median over all the normalized frames,
including all sky frames, and also on-target frames when the target
covers less then $40\%$ of the FOV. By averaging so many frames, we
can remove additional bad pixels and astronomical sources, while retaining a
high S/N in the averaged sky image. For masking of intervening objects
we use source detection by the \sex\ software, combined with by-hand
masks for bright stars and extended objects (galaxies).
The mask for measuring the mean must be a
combination of the masks of all frames, in order to avoid level offsets
caused by large masked objects covering different parts of the
detector on an uneven background. We subsequently subtract from each
frame this ``sky structure'' image, scaled by the frame's background
level.

This procedure is very effective at removing the residual
near-constant background pattern (see Figure \ref{fig:skyflat},
bottom-right panel). However, it requires that every part of the
detector is uncontaminated by celestial sources, which renders the
on-target frames unusable in some galaxies. For those fields, the sky
pattern can still be reconstructed using the interleaved sky
exposures. Naturally, for a successful background subtraction we also
need to measure the average sky level, apart from its spatial
structure. For this, we use on-target measurements rather than the
pure sky frames because the mean sky level fluctuates by $\sim1\%$
($\sim 16\text{-}17\magarcsec$) even on timescales of a minute or
less.

After subtracting sky structure and level, we coadd our frames. To
that end, we re-detect all astronomical sources using \sex, use the
resulting source catalogs to compute a 3rd-order polynomial
astrometric solution of the field distortion and true pointings
(different from the coordinates found in the headers by $\sim 1''$ on
average), and finally re-project and coadd the frames using \swarp.

After this first-pass sky subtraction and co-addition, we have
an image that is much deeper than any individual frame and
hence offers a much better opportunity to mask faint sources (stars
and small background galaxies), as well as low-level extended ``wings''
of bright stars and large galaxies. It also allows a better visual
identification of large galaxies in the field. In
fact, faint but previously unmasked source flux does sometimes
leave visible imprints in the sky-subtracted frames, which propagate
to the first-pass stack. After obtaining the improved masks from the
deep first-pass stack, we repeat the above procedure, this time using
the deep mask projected onto the individual frames. This improves the
sky structure model and removes remaining artifacts in the
background.

\subsection{Combining \emph{HST} and ground-based data}
\label{subsec:data:combined}

We combine \hst\ ($H-$band) with ground-based images, in order to
provide sufficient constraints during the fitting process. We
experimented and found that most of the real structures present in our
galaxies cannot be analyzed reliably with \galfit\ when either the
ground-based or \hst\ data are taken alone. For the medium-resolution
(median FWHM $\sim0\farcs8$) ground-based data the cause is clear:
small-scale structures are not resolved, and a basic bulge plus disk
model is the only feasible model in most cases.  However, the
\hst\ data alone are also insufficient: the lack of data at large
scales prevents a convergence of more complex models. In
surface-brightness profile modeling, the need for large-scale
information on both the galaxy light distribution and the background
has been discussed in \cite{GF3}.

\begin{figure*}
 \centering
 \includegraphics[width=17cm]{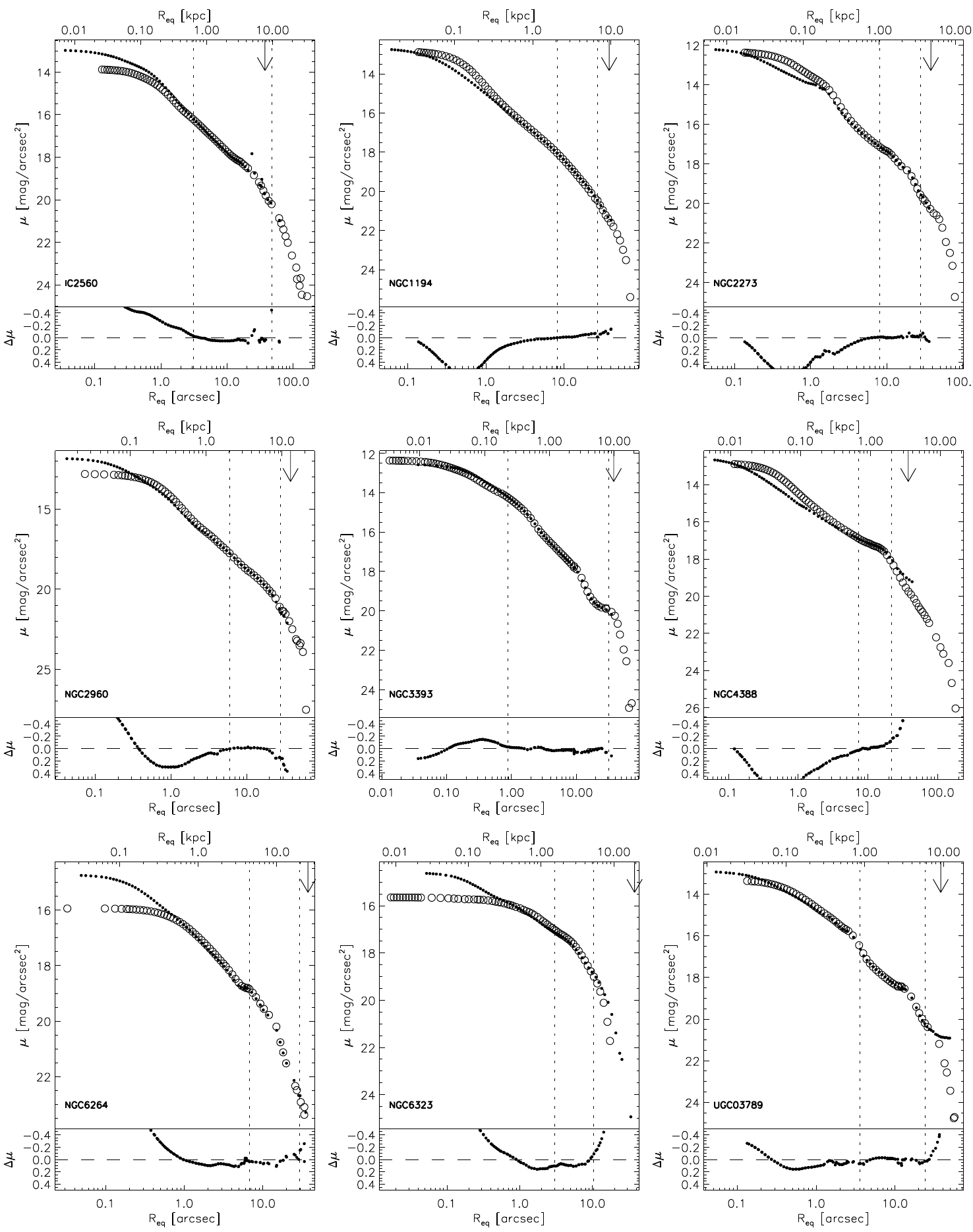}
 \caption{
Comparison of wide-field (open circles) and \hst/WFC3-F160W (filled dots) surface
brightness profiles $\mu$, plotted against equivalent circular radius, $R_{eq}=\sqrt{ab}$. The wide-field data are generally deep ground-based $K$-band images, except in NGC3393 (\hst/WFC3-UVIS F814W). In IC2560, the filled circles show F160W combined with \hst/WFC3-UVIS F814W data, and open circles ground-based LCO-100 $I$-band data. The profiles were matched by a linear fit (\hst\ background level and flux scaling) using a radial range chosen to minimize color gradients (dotted vertical lines). The arrow indicates the radius outside of which we replace the \hst\ with the wide-field data. Residuals of \hst\ minus wide-field data are plotted in the lower parts of the figures, exposing $H-K$ (or $H-I$) color variations and, in particular, how much better \hst\ resolves the nuclear region.
}
 \label{fig:rsc}
\end{figure*}

We combine each co-added \hst\ $J-$ or $H-$band image with ground-based
$K$-band data by first scaling the former to the latter: we measure the
radial surface brightness profiles in both, and pick by eye the part
of the profile in which both overlap and exhibit the same shape. The
matched profiles (Figure \ref{fig:rsc}) show that $J-$ and $H-$band
profiles are very similar to the $K-$band profiles, except in the very
center (active galactic nucleus or AGN 
light and dust) and around spiral arms (dust and young
stellar populations), and the similarity of the profiles justifies our
approach. After rescaling and background-subtracting the \hst\ images
using a linear fit to the profiles, 
$<HST> = <\rm{bkg}> + <\rm{scale}>*<\rm{ground}>$, we
resample the ground-based image stack onto the \hst\ WCS and replace
$K-$band by \hst\ data where they exist. Similarly, we re-project, scale,
and replace the noise and weight maps of $K-$band stacks with \hst\ data
where available.

\section{Image Fitting}
\label{sec:analysis}

\subsection{Decomposition philosophy}
\label{subsec:decomp_recap}

Although there is considerable interest in the correlations between BH
mass and bulge luminosity or mass, there are few works that
examine bulge masses in spiral galaxies with dynamical BH masses
\citep[they include ][]{Gadotti08,Hu2008,S11,Kormendy+Ho13}. Often,
spiral galaxies are excluded due to the difficulties in identifying
robust bulge parameters \citep[e.g.,][]{McConnell+Ma13}. For one
thing, the nuclei of spiral galaxies contain gas, dust, and ongoing 
or recent star formation \citep[e.g.,][]{Carollo+97b}. These alone 
complicate the task considerably, since it is difficult to determine the 
true bulge parameters or mass-to-light ratio. However, it is even 
more difficult to determine exactly what bulge means in the context 
of these galaxies.

All of our galaxies contain a large-scale disk, and all show a
steepening of the profile towards the center - a ``bulge'' by the most
general definition. These central light excesses, however, show a wide
range of physical properties, often more similar to disks than to
classical bulges and elliptical galaxies: they are flattened like
disks, tend to be fitted by an exponential \citep{AndredakisSanders94}
or \citet{Sersic63} with low index \citep[generally $n < 2$,
see][]{FisherDrory10}, they show bars, dust, or spiral arms
as disks do, and they have recent or ongoing star formation like disks
(\citealt{Carollo+97b}, and the review in \citealt{Kormendy+Kennicutt04}).

Together these attributes are interpreted as evidence that the central
light excess or bulge region is being built up slowly by secular,
gas-rich processes such as bar or spiral arm transport. To distinguish
these components from kinematically hot systems without gas or young
stars (e.g., elliptical galaxies) these systems have come to be known
as ``pseudobulges,'' while bulges that have the characteristics of
ellipticals are called ``classical bulges''. Because the
classification of a bulge depends on how it formed, it is certainly
possible that some late-type galaxies contain both a pseudo- and
classical bulge component, with the former having low mass-to-light
ratios and thus dominating the light \citep{Nowak+10,Erwin+15}.  Thus,
we first decompose these different components
\citep[e.g.][]{Weinzirl+09,Laurikainen+10,Lasker+14a}, and then ask
which are of interest in the context of BH-bulge scaling relations.

We take the following dual approach.  We first consider a ``basic''
model, comprising only a bulge and a large-scale disk, plus a nuclear
point source if it can be included\footnote{i.e., if the bulge+disk
model (without the point source) does not already overestimate the
central surface brightness}. In the basic fit, the small-scale
components are mostly fitted by the ``bulge'' component and only to
a small degree accounted for by the disk component. Therefore, the
magnitude of this "basic bulge" represents an approximate upper
limit to the true (classical) bulge magnitude.

With a basic model in hand, we construct a more complete model, including structures on small 
scales, which may help us isolate a bulge component. We
simultaneously model any central bars, rings, or disks, and look for
an additional light component that is rounder and more centrally
concentrated than the outer disk. If there is one, we call this the
``classical'' bulge. In the future, we will improve these
decompositions by combining spatially resolved kinematics
\citep[e.g.,][]{Greene+14} with our \hst\ imaging.

\subsection{Fitting}

We now describe our two-dimensional parametric modeling in more
detail to model the central components in each galaxy.

Working in two dimensions allows us to make full use of the
information contained in the data to reduce degeneracy, and is an
important advantage over one dimensional modeling for the complex
galaxy structures considered here. Each two-dimensional model
component is corrected for the effects of the point-spread function
(PSF), and then the parameters of the model are optimized using
$\chi^2$ minimization with the publicly available code
\gfthree\ \citep{GF3}.  At minimum (our ``basic" model), the models
include a bulge, a large-scale disk, and a central point source model
the AGN and/or an unresolved nuclear star cluster in those cases where
the bulge+disk model does not overestimate the cnetral light
already. In all of the megamaser host galaxies more components are
present. We identify these based on visual identification in the
science image, the residual image of the basic model, and the radial
profiles of surface brightness (SB), ellipticity ($e$) and position
angle ($\pa$).

An accurate PSF model is an important ingredient in the modeling.  We
construct a PSF model by combining a number of stars in our images,
which gives us a model of the PSF that is local both in space and in
time to the observations of the galaxy. In the Appendix we describe
various PSF comparisons to show that we are recovering the PSF
accurately (see Figure \ref{fig:psf_comparison} in the
appendix). Similarly, masking of intervening objects, background
subtraction, availability of a sufficiently large image area around
the target galaxy, and an appropriate noise image impact the
feasibility and accuracy of the fit results. We describe these
ancillary data in the appendix, Section
\ref{subsec:analysis:ancillary}.

We follow wide-spread convention in prescribing the radial
surface-brightness profile of \sersic\ form for the bulge, as well as
an exponential profile for the large-scale disk. The \sersic\ profile
is defined as

\begin{equation}
I(R) = I_e \exp \{ -b_n [(R/R_e)^{1/n} - 1] \}
\label{eqn:sersic}
\end{equation}

\noindent
in terms of three independent parameters: the effective radius $R_e$
along the semi-major axis (SMA), the surface brightness $I_e$ at
$R_e$, and the \sersic\ index $n$. The value of $b_n$ is numerically determined
such that the area inside of $R_e$ contains half of the total flux. 
The large-scale disk is modeled with an
exponential profile, which is equivalent to a
\sersic\ profile with $n \equiv 1$ and the scale radius ($R_s$) 
is related to $R_e = b_{n=1}R_s =
1.678 R_s$. The two-dimensional surface brightness profile
follows from
(\ref{eqn:sersic}) by additionally specifying the center $(x_0,y_0)$,
axis ratio ($q=1-e$), and $\pa$ of the elliptical
isophotes. In \galfit, $I_e$ is replaced by the
profile surface brightness magnitude, 
$\mu_\mathrm{ser}=\mu_0-2.5\log F_\mathrm{ser}$,
where $\mu_0$ is the photometric zero point, and $F_\mathrm{ser}$ the
total flux, which is calculated by
\begin{equation}
F_\mathrm{ser} = 2\pi R_e^2 q I_e n b_n^{-2n} e^{-b_n} \Gamma(2n)
\label{eqn:F_ser}
\end{equation}
with $\Gamma(2n)=\int_0^\infty x^{2n-1}e^{-x} \dd x$ the Gamma function.

The full image model is the sum of the fluxes from all included
components.  Apart from the two-dimensional PSF for the AGN, we
exclusively employ profiles of \sersic\ form, albeit with fixed
\sersic\ index if an exponential ($n\equiv1$) or Gaussian
($n\equiv0.5$) profile is desired. We sometimes use profile
modifications, including Fourier modes, truncations, coordinate
rotation, and bending modes.  For more details on these perturbations
see \citet{GF3}.

In Table \ref{tab:gf}, we list a selection of the resulting
\galfit\ parameters: component magnitudes, sizes (bulge $R_e$ and disk
$R_s$), and the bulge \sersic\ index. Total apparent magnitudes are
also listed (column 2).

\begin{table*}
 \caption{\galfit\ best-fit parameters of galaxy image models}
 \begin{tabular}{l*{15}c} 
  \toprule
  Galaxy & \multicolumn{2}{c}{Total} & \multicolumn{6}{c}{Bulge} & \multicolumn{4}{c}{Disk} & \multicolumn{2}{c}{Psf} & \multicolumn{1}{c}{Additional components} \\ 
  \cmidrule(lr){4-9}\cmidrule(lr){10-13}\cmidrule(lr){14-15}
   & \multicolumn{2}{c}{$m_t$} & \multicolumn{2}{c}{$m_b$} & \multicolumn{2}{c}{$R_e~[\,\arcsec\,]$} & \multicolumn{2}{c}{$n$} & \multicolumn{2}{c}{$m_d$} & \multicolumn{2}{c}{$R_s~[\,\arcsec\,]$} & \multicolumn{2}{c}{$m_p$} & \multicolumn{1}{c}{mag and type} \\ 
  & \multicolumn{2}{c}{(2)} & \multicolumn{2}{c}{(3)} & \multicolumn{2}{c}{(4)} & \multicolumn{2}{c}{(5)} & \multicolumn{2}{c}{(6)}  & \multicolumn{2}{c}{(7)} & \multicolumn{2}{c}{(8)} & \multicolumn{1}{c}{(9)} \\
    \cmidrule(lr){2-3}\cmidrule(lr){4-5}\cmidrule(lr){6-7}\cmidrule(lr){8-9}\cmidrule(lr){10-11}\cmidrule(lr){12-13}\cmidrule(lr){14-15}
  & bas & bul  & bas & bul  & bas & bul  & bas & bul  & bas & bul  & bas & bul & bas & bul & \\
  \midrule
  IC2560 & 8.83 & 8.62 & 10.71 & 11.50 & 0.99 & 0.59 & 2.9 & 1.5 & 9.04 & 9.65 & 6.2 & 5.6 & -- & 16.33 & \parbox[c][0.8cm][c]{3.5cm}{13.44, 11.77, 11.72, 9.57 \\ ND, XBul, SpRing, Env} \\
NGC1194 & 9.90 & 10.01 & 10.22 & 12.04 & 4.6 & 1.8 & 6.8 & 3.2 & 11.39 & 13.50 & 3.5 & 6.9 & -- & 17.30 & \parbox[c][0.8cm][c]{3.5cm}{12.75, 10.32 \\ Bar, Env} \\
NGC2273 & 9.13 & 9.04 & 10.96 & 10.81 & 0.25 & 0.33 & 1.0 & 2.0 & 9.36 & 9.75 & 1.4 & 2.2 & 14.84 & 15.24 & \parbox[c][0.8cm][c]{3.5cm}{12.85, 11.30, 11.31 \\ NR, Bar, Spir} \\
NGC2960 & 10.36 & 10.35 & 11.24 & 11.01 & 0.68 & 15.8 & 4.0 & 1.0 & 12.02 & 10.89 & 3.1 & 2.1 & -- & 14.84 & \parbox[c][0.8cm][c]{3.5cm}{12.44 \\ ND} \\
NGC3393 & 9.46 & 9.55 & 10.19 & 11.20 & 1.6 & 0.62 & 3.5 & 2.3 & 10.24 & 10.77 & 5.5 & 1.6 & 17.63 & 15.94 & \parbox[c][0.8cm][c]{3.5cm}{13.65, 13.01, 10.57 \\ Bar, SpRing, Env} \\
NGC4388 & 8.59 & 8.49 & 10.57 & 10.28 & 0.9 & 1.2 & 3.8 & 2.2 & 8.79 & 9.72 & 2.4 & 1.7 & 14.78 & 14.80 & \parbox[c][0.8cm][c]{3.5cm}{13.00, 10.43, 9.80 \\ ND, SpRing, Env} \\
NGC6264 & 11.64 & 11.64 & 12.97 & 14.44 & 3.2 & 0.65 & 3.1 & 1.3 & 12.02 & 12.67 & 5.6 & 5.1 & 18.48 & 18.04 & \parbox[c][0.8cm][c]{3.5cm}{13.59, 12.72 \\ Bar, SpRing} \\
NGC6323 & 11.08 & 11.08 & 14.05 & 14.14 & 0.75 & 0.66 & 1.2 & 1.1 & 11.16 & 11.82 & 4.2 & 3.4 & 17.45 & 17.40 & \parbox[c][0.8cm][c]{3.5cm}{13.05, 12.49 \\ Ring, Spir} \\
UGC03789 & 9.98 & 10.00 & 11.47 & 11.50 & 0.51 & 0.8 & 1.5 & 3.3 & 10.31 & 11.07 & 3.1 & 2.3 & 15.89 & 16.63 & \parbox[c][0.8cm][c]{3.5cm}{13.25, 13.66, 12.58, 11.76 \\ Bar, NR, Ring, Env} \\

  \bottomrule
 \end{tabular}
\tablecomments{In columns (2)-(8) we list $H-$band galaxy total magnitude, bulge magnitude,
effective radius and \sersic\ index, disk magnitude and scale radius,
and the magnitude of the central point source. Each of those columns
contains two values: the first one ("bas") derived from the basic
decomposition (bulge, disk, and the central point source if it could
be fitted), followed by the results of the full model ("bul") which
includes significant additional components and yields our best
estimate of the "classical" bulge parameters by virtue of separating
non-bulge and pseudobulge components. The additional components'
magnitudes of the full model are given in column (9), along with their
morphology in abbreviated form: ``NR'' for nuclear ring, ``ND'' for
nuclear disk, IDisk for inner disk, ``XBul'' for X-shaped/boxy bulge, 
``Bar'', ``Ring'' for medium-large scale ring, ``Spir'' for spiral
modified by rotation, ``SpRing'' for spiral arms that nearly form a
ring (no coordinate rotation fitted), and ``Env'' for envelope. See table \ref{tab:comptypes} for a short, and
Section \ref{subsec:imageana:decomposition} for a detailed presentation
of those morphologies and classification criteria.}
\label{tab:gf}
\end{table*} 

\begin{table*}
 \caption{Morphological types of non-standard components}
 \begin{tabular}{lll}
  \toprule
  Abbreviation & Type & description and criteria \\
  \midrule
  NR & nuclear ring & ring (see below) of $\lesssim 500\pc$ size, similar to nuclear disk with inner truncation \\
  ND & nuclear disk & $R_e\lesssim 500\pc$ \sersic\ profile with $n\lesssim1$, often not aligned with galaxy major axis \\
  IDisk & inner disk & flattened as the main disk but smaller ($\sim 1\kpc$); \\
        &            & only in NGC2960 where it exhibits spiral arm substructure \\
  XBul & X-shaped/boxy bulge & \sersic\ with $1\lesssim n \lesssim 2$, $q \approx q_\mathrm{disk}$ and $R_e\sim1\,\kpc$; identified in IC2560 \\
  Bar & bar & \sersic\ $n<1$, generally elongated and often misaligned from major axis, \\ & & in NGC1194 probably seen down its long axis \\
  Ring & ring & \sersic\ with inner truncation and $n=0.5$ fixed at intermediate and large $\gtrsim 1\kpc$ scales \\
  SpRing & spiral/ring & shows up in the profile as a ring but also spiral structure (tight winding) in the image, \\ & & $n$ fixed to $0.5$ or $1.0$ \\
  Spir & spiral & \sersic\ profile with coordinate rotation to emulate spiral arms, sometimes fixed $n=0.5$ or $n=1$ \\
  Env & envelope & extended disk and/or a halo depending on ellipticity $\eps$, \\ & & accounts for flux excess above the main disk at large radii ($\gtrsim 5\kpc$), \sersic\ $n$ varyingly fixed or free \\
  \bottomrule
 \end{tabular}
\tablecomments{We list here the abbreviations and morphologies of non-standard (beyond bulge, disk and PSF) components we used while fitting \galfit\ models to our galaxy images. Detailed descriptions of the morphological types, and the criteria we used to identify/classify them, are presented in Section \ref{subsec:imageana:decomposition}.}
\label{tab:comptypes}
\end{table*} 

\subsection{Detailed decompositions}
\label{subsec:imageana:decomposition}

\begin{figure*}
 \centering
 \includegraphics[width=17cm]{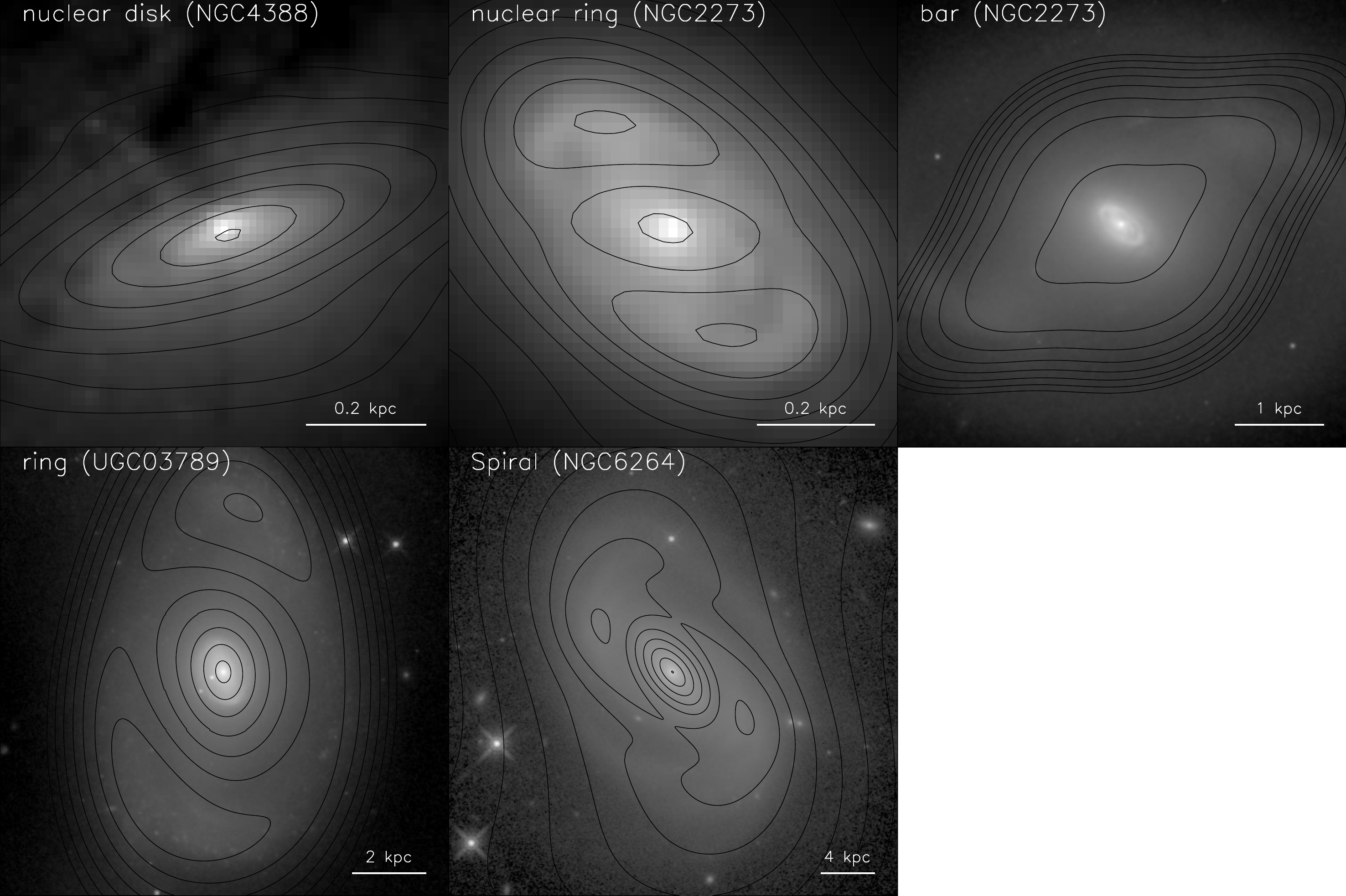}
 \caption{
Examples of observed and \galfit-modeled morphological components in
our sample. Shown are \emph{HST}/WFP3-F160W image cutouts, with contours of
their model counterparts overlaid. For detailed descriptions of the
component types and the fitting process, see Section
\ref{subsec:imageana:decomposition} and the appendix. The offset of
the model contours with respect to the brightest point in the image
seen in the small-scale nuclear disk and nuclear ring examples is a
result of constraining all model component center coordinates, except those of the central point source, to a
common value during the fit. 
}
 \label{fig:comp+cont}
\end{figure*}

It is not trivial to determine the number and type of components to
fit to these many-component profiles in the face of possible profile
mismatches and parameter degeneracies.  We generally begin with a very
simple two-component model, and then examine the fit residuals, the
ellipticity, and the $\pa$ profile to see whether additional
components are warranted.  We also refit, adding components in a
different order, to be sure that GALFIT robustly converges on the same
fit.  The most challenging task is to find the most probable and
physically realistic model while retaining acceptable alternatives for
the systematic uncertainty estimate. Compared to this systematic
(modeling) uncertainty, the formal parameter uncertainties from the
$\chi^2$ derivatives are very small.

In the following subsections, we describe the most common subcomponents 
and how we identify them. Each is demonstrated graphically in 
Figure \ref{fig:comp+cont}.

\subsubsection{Nuclear disks}
\label{subsec:imageana:nuclear_disks}

Nuclear disks have a scale of $\lesssim100\pc$. They are intrinsically
flattened, which is evident when they are observed nearly edge-on, and
may be misaligned with the large-scale disk (e.g. NGC4388). There is
kinematic evidence for a nuclear disk that we also recognize
photometrically in the case of both NGC2273
\citep[e.g.,][]{Barbosa+06} and NGC4388 \citep{Greene+14}.  Even when
seen nearly face on, nuclear disks are distinguished by their
relatively sharp boundaries and occasionally inset spiral arms (e.g.,
IC2560). Since they occur in the central regions of the galaxy and
have high surface brightness, they tend to bias larger scale
components to higher $n$ and smaller size if not modeled separately.
We model nuclear disks using a \sersic\ profile, typically with $n
\leq 1$.  Sometimes these disks are delineated by starforming regions
arranged in a ring (NGC2273, UGC3789) and we model them accordingly
(see \ref{subsec:imageana:rings}). We include a nuclear disk or ring
in five of the nine target galaxies (IC2560, Fig.\ \ref{fig:IC2560-1};
NGC2273, Fig.\ \ref{fig:NGC2273-1}; NGC2960,
Fig.\ \ref{fig:NGC2960-1}; NGC4388, Fig.\ \ref{fig:NGC4388-1};
UGC3789, Fig.\ \ref{fig:UGC3789-1}).

\subsubsection{Bars}
\label{subsec:imageana:bars}

A bar is present in the images of NGC2273 (Fig.\ 14), NGC3393
(Fig.\ 18) and UGC3789 (Fig.\ 26). The classic signatures of a bar are
a simultaneous increase in ellipticity and flat $\pa$ over the
radial range where the bar dominates the light
\citep[e.g.,][]{Delmestre+07}, followed by a sharp drop in ellipticity at
the outer radius where the bar ends. In certain cases, if the galaxy is
at high inclination or the bar is pointing towards us, then it is more
challenging to uncover, yielding an ambiguity in the final
decomposition (see for instance NGC1194, Fig.\ \ref{fig:NGC1194-1}).
Nuclear bars, similar to nuclear disks or rings, can particularly bias
the light profile of any underlying classical bulge component. Because
bars form only in disk structures, we exclude them from any classical
bulge estimate.

We model all bars as a \sersic\ profile with free $n$, and find them
all to have $n\lesssim1$ as expected. \galfit\ offers the (modified)
Ferrer profile, which becomes zero outside of the truncation radius
and is an alternative profile suited for bars due to its flat center
and steep outer profile \citep{GF3}. It does not provide an improvement of the fit over \sersic\ profiles in our data. We do allow a 4th-order Fourier mode to fit the boxiness of
the isophotes of the bar.

\subsubsection{Rings}
\label{subsec:imageana:rings}

A ring is prominent in NGC3393 (Fig.\ \ref{fig:NGC3393-1}) and NGC6323
(Fig.\ \ref{fig:NGC6323-1}). Three rings are present in NGC2273
\citep[see Fig.\ \ref{fig:NGC2273-1}, and][]{Erwin+Sparke03} and UGC3789
(Fig.\ \ref{fig:UGC3789-1}). The inner rings in each galaxy are associated
with a nuclear disk (see \ref{subsec:imageana:nuclear_disks}). In
NGC2273 we refrain from modeling the outermost ring, due to its low
surface brightness.  Nuclear rings generally bias the parameters for
other nuclear components (e.g., bars and bulges) to fainter and more
compact (lower $n$) profiles if not accounted for (UGC3789 is an
example).  At larger scales, omitting rings (or spiral arms) usually
biases the disk to higher flux or larger $R_s$.  We generally model
rings as an inner truncation multiplied by a \sersic\ profile with
fixed $n=0.5$ (a Gaussian profile). The compactness of this profile is
suitable to model the steep decline of the profile. We also test using
an exponential profile ($n=1$), but find that it does not improve the
fit. Finally, we never fit a nuclear disk and nuclear ring component
simultaneously, as they are degenerate.

\subsubsection{Spiral arms}
\label{subsec:imageana:spirals}

We detect and fit spiral arms in six out of our nine targets. Spiral
arms can be modeled in \galfit\ using coordinate rotation, which
changes the $\pa$ as a function of radius. Fitting spiral arms is not
merely a cosmetic measure, but also impacts the best-fit parameters of
the large-scale disk. Spiral arms often span a limited radial range in
the disk, and thus we use an inner truncation to limit their profile
towards small radii. In four cases (IC2560, Fig.\ \ref{fig:IC2560-1};
NGC2273, Fig.\ \ref{fig:NGC2273-1}; NGC3393,
Fig.\ \ref{fig:NGC3393-1}; and NGC4388, Fig.\ \ref{fig:NGC4388-1}),
the arms are tightly wound or partially dust-obscured. These prove
difficult to model with a free-parameter rotation function, so we
model them as a ring. This leaves only two galaxies (NGC6264,
Fig.\ \ref{fig:NGC6264-1}; NGC6323, Fig.\ \ref{fig:NGC6323-1}) where
we used a rotation function for the spiral component. The remaining three
galaxies have no significant spiral structure that could impact our
fits. The \sersic\ index of the arm model is fixed during the fit due
to degeneracy with the parameters of the required inner truncation. We
tested values of $n=0.5$, $1.0$ and $1.5$, and pick the value that
leads to the smallest residuals.

\subsubsection{Envelopes}
\label{subsec:imageana:env}

Some of our targets (IC2560, Fig.\ \ref{fig:IC2560-1}; NGC1194,
Fig.\ \ref{fig:NGC1194-1}; NGC4388, Fig.\ \ref{fig:NGC4388-1}; and
UGC3789, Fig.\ \ref{fig:UGC3789-1}) show a profile extension above the
large-scale disk that sets in at $> 5$ kpc scales (see Figure
1). These envelopes are identifiable visually on the image, and are
accompanied by a change in axis ratio $q=b/a$ coincident with a break
in the surface brightness profile, as seen in so-called type III
upbending or anti-truncated disks
\citep[e.g.,][]{Erwin+Beckman+Pohlen05,Pohlen+Trujillo06,
Erwin+Pohlen+Beckman08}.
They are sometimes rounder than the disk, like an outer halo, but are
often flattened and best fit by an exponential profile. Inclusion of
the envelope typically impacts the bulge parameters only weakly, due
to the large difference in scale. However, it may significantly
contribute to the total galaxy magnitude (IC2560, NGC4388).

\begin{table*}
 \centering
 \caption{Systematic uncertainties}
 \begin{tabular}{l*{12}c}
  \toprule
   Galaxy & \multicolumn{8}{c}{Bulge} & \multicolumn{2}{c}{Disk} & \multicolumn{2}{c}{PSF} \\
   \cmidrule(lr){2-9}
     & \multicolumn{4}{c}{$m_b$} & \multicolumn{2}{c}{$\log(R_e/\kpc)$} & \multicolumn{2}{c}{$\log n$} & \multicolumn{2}{c}{$m_d$} & \multicolumn{2}{c}{$m_p$} \\
     \cmidrule(lr){2-5}\cmidrule(lr){6-7}\cmidrule(lr){8-9}\cmidrule(lr){10-11}\cmidrule(lr){12-13}
     & psf & rsc & mod & tot & mod & tot & mod & tot & mod & tot & mod & tot \\
  IC2560 & 0.00 & 0.02 & 0.30 & 0.30 & 0.10 & 0.10 & 0.17 & 0.18 & 0.75 & 0.75 & 0.13 & 0.17 \\
 NGC1194 & 0.06 & 0.08 & 0.54 & 0.54 & 0.40 & 0.42 & 0.11 & 0.11 & 0.19 & 0.45 & 8.37 & 8.37 \\
 NGC2273 & 0.00 & 0.04 & 0.34 & 0.34 & 0.19 & 0.19 & 0.16 & 0.17 & 0.14 & 0.30 & 0.39 & 0.40 \\
 NGC2960 & 0.11 & 0.02 & 0.08 & 0.14 & 0.06 & 0.08 & 0.14 & 0.17 & 0.02 & 0.05 & 0.29 & 0.31 \\
 NGC3393 & 0.00 & 0.06 & 0.11 & 0.13 & 0.10 & 0.12 & 0.13 & 0.16 & 1.07 & 1.07 & 0.23 & 0.28 \\
 NGC4388 & 0.00 & 0.31 & 0.51 & 0.60 & 0.11 & 0.32 & 0.06 & 0.24 & 0.45 & 0.79 & 0.12 & 0.19 \\
 NGC6264 & 0.02 & 0.43 & 0.93 & 1.03 & 0.35 & 0.43 & 0.15 & 0.16 & 0.28 & 0.30 & 0.17 & 0.24 \\
 NGC6323 & 0.01 & 0.51 & 0.52 & 0.73 & 0.23 & 0.55 & 0.21 & 0.47 & 0.32 & 4.02 & 0.47 & 0.54 \\
UGC03789 & 0.03 & 0.12 & 0.41 & 0.43 & 0.18 & 0.23 & 0.06 & 0.08 & 0.21 & 0.22 & 0.14 & 0.20 \\

  \bottomrule
\end{tabular}
 \tablecomments{
Systematic uncertainties for five parameters of the
full decompositions: the bulge magnitude, logarithmic effective radius, and
\sersic\ index ($m_b$, $\log (R_e/\kpc)$, and $\log n$), as well as the
disk magnitude ($m_d$) and magnitude of the central point
source ($m_p$). Each total (``tot'') systematic error is the sum of
the errors resulting from PSF model (``psf''), background and flux
scaling (``rsc''), and model (``mod'') systematic uncertainties, added
in quadrature. In general, the modeling uncertainty (number and
profiles of the model components, see Section
\S\ref{subsubsec:moderr}) dominates the total systematic error, which
is why PSF- and background/scaling-related errors (Sections
\ref{subsubsec:psferr} and \ref{subsubsec:rscerr}) are shown only for
the most important parameter in our study, $m_b$.}
 \label{tab:syserr}
\end{table*} 

\subsection{Systematic uncertainties}
\label{subsec:syserr}

The uncertainty in our fitting is dominated not by measurement 
error but rather by systematic uncertainty. In general, degeneracy 
or deficiency in our modeling dominates the uncertainty, but the 
background level can also play a significant role in some cases 
(e.g., NGC4388).

\subsubsection{Modeling uncertainty}
\label{subsubsec:moderr}

In the previous subsection, and in the Appendix, we describe our
general approach to finding the most
suitable model for a given galaxy image. Even given the high
  spatial resolution and depth of our data, and the complexity of our
models, the true structure of the galaxy may not be fully represented.
Thus, our choice of number of components and boundary conditions
leaves a systematic uncertainty in the parameters we are interested
in. This systematic parameter uncertainty originating in the
choice of model is separate from the random uncertainties that
originate in pixel noise. The latter are directly provided by
\galfit\ and are completely subdominant to the systematic modeling
uncertainties\footnote{i.e by one to several orders of magnitude}.

We take a pragmatic approach and consider a handful (usually 2-4) of
alternative models to our chosen best-fit model that we consider
acceptable. This way, we obtain several alternative values of the
resulting bulge parameter of interest (chiefly the magnitude), and
take their standard deviation as a systematic uncertainty
estimate. These values are presented in Table
\ref{tab:syserr}. Typical systematic uncertainty estimates are a few
tenths of a magnitude. We use these uncertainties with the other
systematic uncertainties added in quadrature, to represent the
measurement error in the derived bulge mass ($\lbul$ and $\mbul$) when
investigating the $\mbh-M_\mathrm{bul}$ scaling relation in \S 5.

\subsubsection{PSF uncertainty}
\label{subsubsec:psferr}

We use a PSF image that is the weighted average of 14 individual star
cutouts, as we find them on the model-subtracted WFC3/IR F160W
images. Using this averaged image instead of any particular (fixed)
individual star cutout reduces random noise, finite sampling and
centering errors, as well as (correlated) background residuals, which
propagate to model parameter errors. We estimate the PSF-related
uncertainty of any parameter $p$ when fitted by a single PSF image
to be $\sigma^{(i)}_{p,\mathrm{PSF}}=\frac{1}{N-1}
\Sigma_{i=1,...,N}\,p_i$, where $p_i$ is the fit result when using the
$i$-th of $N$ available individual PSF images. Then, the uncertainty
in $p$ when fitting with the combined PSF image is approximately
$\sigma_{p,\mathrm{PSF}}=\frac{1}{\sqrt{N}}\sigma^{(i)}_{p,\mathrm{PSF}}$,
akin to the uncertainty of individual measurements propagating to the
uncertainty of the mean of measurements.

\subsubsection{Background uncertainty}
\label{subsubsec:rscerr}

The background uncertainties in our combined \emph{HST}+ground-based 
images largely stem from the matching of the \emph{HST} data to
the ground-based data. In the former, the sky is not well known
because the galaxy fills the field of view. In contrast, the
ground-based images have a wide FOV and a well-defined background.

Due to color gradients across the galaxy, the fitting of the sky
offset and relative flux scaling of the \emph{HST} data generally
depends on the radial range of the surface brightness profile chosen
to constrain the fit. We chose the fiducial range to minimize color
gradients, but even in the optimal range, fluctuations from
measurement noise or actual variations within the profile add
uncertainty in the fitted sky offset and relative flux scaling. In
order to estimate the resulting uncertainty to the sky offset of the
\emph{HST} data, we employ a ``Jackknife'' resampling method. In
each realization, one of the $N$ surface brightness measurements
within the fiducial range is omitted, resulting in $N$ sample
realizations of the fitted offset, $a_i$. Then, the uncertainty in
the \emph{HST} sky level $a$ is
$\sigma_a=\frac{n-1}{n}\sum_{i=1}{N}(a_i-\bar{a})$. These
background errors are listed in Table \ref{tab:syserr} in columns
labeled ``rsc''.

\subsection{Colors and conversion to stellar mass}
\label{subsec:colpluslups}

\begin{table*}
 \centering
 \caption{Bulge and total colors}
 \begin{tabular}{l*{17}{c}}
  \toprule
  Galaxy & $a_1 ['']$ & $a_2 ['']$  & $a_1 ['']$ & $a_2 ['']$ &\multicolumn{3}{c}{$F435-814W$} & \multicolumn{3}{c}{$F814-160W$} & \multicolumn{3}{c}{$g-i$} &\multicolumn{3}{c}{$i-H$} & $H-K$ \\
  \cmidrule(lr){2-3}\cmidrule(lr){4-5}\cmidrule(lr){6-8}\cmidrule(lr){9-11}\cmidrule(lr){12-14}\cmidrule(lr){15-17}
  & \multicolumn{2}{c}{bul} & \multicolumn{2}{c}{bas} & bul & bas & tot & bul & bas & tot & bul & bas & tot & bul & bas & tot \\
  \midrule
  IC2560 & 0.28 & 5.89 & 0.28 & 7.85 & 2.83 & 2.79 & 2.14 & 1.70 & 1.67 & 1.58 & 1.65 & 1.62 & 1.14 & 2.46 & 2.42 & 2.31 & 0.26 \\
 NGC1194 & 0.28 & 9.49 & 0.28 & 9.49 & 2.91 & 2.92 & 2.38 & 1.64 & 1.65 & 1.55 & 1.70 & 1.71 & 1.31 & 2.38 & 2.40 & 2.27 & 0.38 \\
 NGC2273 & 0.28 & 5.89 & 0.28 & 4.03 & 2.38 & 2.25 & 2.33 & 1.78 & 1.80 & 1.57 & 1.31 & 1.22 & 1.28 & 2.56 & 2.58 & 2.30 & 0.29 \\
 NGC2960 & 0.28 & 1.88 & 0.28 & 6.48 & 2.44 & 2.28 & 1.99 & 1.58 & 1.62 & 1.42 & 1.36 & 1.24 & 1.03 & 2.31 & 2.36 & 2.11 & 0.35 \\
 NGC3393 & 0.28 & 3.33 & 0.28 & 9.49 & 2.09 & 2.21 & 2.06 & 1.62 & 1.59 & 1.59 & 1.10 & 1.19 & 1.08 & 2.36 & 2.32 & 2.32 & 0.28 \\
 NGC4388 & 2.27 & 5.89 & 0.28 & 3.33 & 1.90 & 2.29 & 1.87 & 1.99 & 2.30 & 1.90 & 0.96 & 1.25 & 0.94 & 2.81 & 3.19 & 2.70 & 0.28 \\
 NGC6264 & 0.28 & 1.41 & 0.28 & 4.87 & 2.32 & 2.49 & 1.79 & 1.61 & 1.62 & 1.56 & 1.27 & 1.39 & 0.88 & 2.35 & 2.36 & 2.29 & 0.33 \\
 NGC6323 & 0.28 & 1.41 & 0.28 & 0.96 & 2.95 & 2.91 & 1.73 & 1.83 & 1.91 & 1.60 & 1.73 & 1.70 & 0.83 & 2.62 & 2.71 & 2.34 & 0.33 \\
UGC03789 & 0.27 & 4.67 & 0.28 & 4.43 & 2.03 & 2.01 & 1.94 & 1.69 & 1.67 & 1.47 & 1.06 & 1.04 & 0.99 & 2.45 & 2.42 & 2.18 & 0.27 \\

  \bottomrule
 \end{tabular}
\label{tab:colors}
\tablecomments{
Bulge colors were measured in elliptical annuli bounded by 
$a_1$ and $a_2$ where the respective
bulge component dominates, for the classical bulge ("bul") in our 
detailed decomposition and the \sersic\ component in the basic
\sersic +exponential ("bas") decomposition. Meanwhile, total colors 
("tot") were measured across the entire WFC3-IR
FOV. Colors in the \emph{HST}/WFC3 F435W, F814W and F160W (similar to Johnson
$B$, $I$ and 2MASS $H$) bands were converted to $g-i$ and $i-H$ as
described in Section \ref{subsec:colpluslups}. The $H-K$ color is
taken from the 2MASS catalog and measured within the
$20\,\mathrm{mag~arcsec^{-2}}$ isophote.  }
\end{table*} 

In order to facilitate direct comparison with early-type galaxies, we
attempt to mitigate the variability in mass-to-light ratio $\Upsilon
\equiv M_\star/L$ by using the galaxy color. Since the variation in
$\Upsilon$ with population properties and dust column density is
decreased at NIR wavelengths, we choose F160W (nearly $H-$band) to
measure the luminosity, and derive $\UpsH \equiv M_\star/L_H$ based on
optical colors. We use the color-$\Upsilon$ relation from
\cite{Bell+03c}, which determines conversions between a single galaxy
color and the galaxy $\Upsilon$ based on stellar population synthesis
models and an assumed dust model.  We also experimented with the
two-color relations from \cite{ZCR09}, but found that these
conversions yielded unrealistically low $\Upsilon$ values for some of
the most massive ellipticals in our comparison sample, for whom the
color is very red (\S 3.6).  We note that many systematic
uncertainties remain in stellar population synthesis modeling,
particularly regarding the contribution to the near-infrared light
from asymptotic giant branch stars
\citep[e.g.][]{Maraston05,Conroy+Gunn+White09,Kriek+10,Zibetti+13}. It
is possible that the \cite{Bell+03c} mass-to-light ratios are
overestimates at blue $g-i$ color because their models do not treat these
later stages of stellar evolution \citep[e.g.][]{Roediger+Courteau15}.

Since we have measured F435W (roughly $B-$band) and F814W (roughly
$I$-band) magnitudes, we are closest in color to the $g-i$ conversions
to $\Upsilon_{\rm H}$ from \cite{Bell+03c}. We derive a conversion 
between the \emph{HST}
colors and the Sloan Digital Sky Survey \citep[SDSS;][]{York+00} $g-i$
system using simple stellar population models from the Padova group
\citep{Marigo+08}, including extinction in each band ranging from an
$A_V$ of 0 to 3.  A tight linear relation is found (scatter of 0.05
mag) between the $g-i$ and $F438W-F814W$ color of
\begin{equation}
g-i = -0.443 + 0.738 (F438W-F814W)~. 
\end{equation}
Note that the $F438W,~F814W,~F160W$ and
$H-$band magnitudes here are all in the Vega system,
while the SDSS magnitudes are given in the AB system 
\citep{Oke+Gunn83,Fukugita+96a}.

For both the basic and classical bulges we measure the bulge color by
defining a radial range over which the given component dominates the
total flux. Using the same aperture on both bands we can derive a
color without performing full fits to each band. We ensure that PSF
and AGN corrections are negligible by imposing a lower limit of
$\gtrsim 2\,\mathrm{pix}$ ($0.26''$) on the semimajor axis of the
aperture, corresponding to $\sim 50$ pc at the typical distance of 50
Mpc. The magnitudes for the entire galaxy (total) are measured from
the total $H-$band magnitude within the WFC3-IR sky-subtracted
image. We note that a more precise method to determine component
colors would involve simultaneous fitting of images in multiple
bands, but this analysis is outside the scope of our study and left
to future work.

\begin{table}
 \centering
 \caption{Bulge and total mass-to-light ratios}
 \begin{tabular}{l*{3}{c}} 
  \toprule
  Galaxy & \multicolumn{3}{c}{$\log \UpsH(g-i)$}  \\
  \cmidrule(lr){2-4}
  & bul & bas & tot \\
  \midrule
  IC2560 &  0.11 &  0.10 &  0.02 \\
 NGC1194 &  0.12 &  0.05 &  0.05 \\
 NGC2273 &  0.05 &  0.03 &  0.04 \\
 NGC2960 &  0.06 &  0.04 & -0.00 \\
 NGC3393 &  0.01 &  0.03 &  0.01 \\
 NGC4388 & -0.01 &  0.04 & -0.02 \\
 NGC6264 &  0.04 &  0.06 & -0.03 \\
 NGC6323 &  0.12 &  0.12 & -0.04 \\
UGC03789 &  0.00 &  0.00 & -0.01 \\

  \bottomrule
 \end{tabular}
\label{tab:lups}
\tablecomments{Logarithmic $H$-band mass-to-light ratios ($\log \UpsH(g-i)$) based on the
\protect\cite{Bell+03c} prescription. The colors and
decompositions on which these are based are listed in Table
\ref{tab:colors}.}
\end{table} 

The inferred colors are listed in Table \ref{tab:colors}, along with
the inner and outer annuli in which bulge colors were measured, both
for the basic and classical bulges. Table \ref{tab:lups} shows the
$\UpsH(g-i)$ that are derived from the \cite{Bell+03c} models.  
The resulting $\log M_\star = \log L_H
+ \log\UpsH$ are given in Table \ref{tab:lummassrad} and used in
Section \ref{sec:bhscalerels} regarding their correlation with $\mbh$.

\begin{table*}
 \centering
 \caption{Luminosities, masses, and sizes}
 \begin{tabular}{l*{16}c}
  \toprule
  Galaxy & \multicolumn{2}{c}{$\log (\mbh/M_\odot)$} & \multicolumn{3}{c}{$\log (L_H/L_{\odot,H})$} & \multicolumn{2}{c}{$(B/T)_L$} & \multicolumn{4}{c}{$\log (M_\star/M_\odot)$} & \multicolumn{2}{c}{$(B/T)_{M \star}$} & \multicolumn{3}{c}{$\log(R_e/\kpc)$} \\
   \cmidrule(lr){2-3}\cmidrule(lr){4-6}\cmidrule(lr){7-8}\cmidrule(lr){9-12}\cmidrule(lr){13-14}\cmidrule(lr){15-17}
   & val & err & bas & bul & tot & bas & bul & bas & bul & err & tot & bas & bul & bas & bul & tot \\
  \midrule
  IC2560  &  6.40 &  0.40 & 10.29 &  9.97 & 11.12 &  0.18 &  0.07 & 10.40 & 10.08 &  0.14 & 11.14 &  0.22 &  0.09 & -0.14 & -0.33 &  0.60 \\
 NGC1194  &  7.82 &  0.05 & 10.67 &  9.94 & 10.76 &  0.74 &  0.15 & 10.72 & 10.39 &  0.20 & 10.81 &  0.74 &  0.38 &  0.58 & -0.34 &  0.44 \\
 NGC2273  &  6.88 &  0.05 &  9.77 &  9.83 & 10.54 &  0.19 &  0.19 &  9.82 &  9.88 &  0.12 & 10.58 &  0.19 &  0.20 & -0.70 & -0.56 &  0.15 \\
 NGC2960  &  7.05 &  0.05 & 10.53 & 10.63 & 10.89 &  0.44 &  0.55 & 10.59 & 10.21 &  0.22 & 10.89 &  0.50 &  0.21 & -0.28 &  0.73 &  0.29 \\
 NGC3393  &  7.49 &  0.12 & 10.71 & 10.30 & 10.97 &  0.51 &  0.21 & 10.72 & 10.31 &  0.05 & 10.98 &  0.51 &  0.21 &  0.14 & -0.21 &  0.40 \\
 NGC4388  &  6.93 &  0.05 &  9.66 &  9.78 & 10.49 &  0.16 &  0.19 &  9.65 &  9.77 &  0.22 & 10.47 &  0.17 &  0.20 & -0.11 & -0.00 &  0.15 \\
 NGC6264  &  7.45 &  0.05 & 10.41 &  9.82 & 10.94 &  0.30 &  0.08 & 10.45 &  9.86 &  0.41 & 10.91 &  0.35 &  0.09 &  0.33 & -0.26 &  0.60 \\
 NGC6323  &  6.96 &  0.05 &  9.75 &  9.72 & 10.94 &  0.06 &  0.06 &  9.87 &  9.84 &  0.27 & 10.90 &  0.09 &  0.09 & -0.20 & -0.24 &  0.40 \\
UGC03789  &  7.05 &  0.05 & 10.14 & 10.13 & 10.73 &  0.26 &  0.25 & 10.14 & 10.13 &  0.15 & 10.72 &  0.26 &  0.26 & -0.36 & -0.12 &  0.34 \\

  \bottomrule
 \end{tabular}
\label{tab:lummassrad}
\tablecomments{BH masses (log $\mbh/\msun$) and their errors, 
bulge and total $H$-band luminosities (log $L_H/L_{\odot,H}$), 
the ratio of bulge to total luminosity [$(B/T)_L$], 
stellar masses (log $M_*/\msun$), and effective $H$-band radii 
[$\log(R_e/\kpc)$] of
our sample based on our \galfit\ models. "Bas" indicates values 
based on basic bulge+disk(+point source) models while "bul" indicates those 
based on the full models of the classical bulge parameters. Bulge mass 
errors are based on the combined calibration, point-source and 
modeling uncertainties (see Table \ref{tab:syserr}). Luminosities 
were converted to masses via the color-based 
$\Upsilon_H(g-i)$ \protect\citep{Bell+03c}, 
with colors measured on our \emph{HST}/WFC3 F435W, F814W, and F160W 
images inside the appropriate apertures (bulge-dominated regions, 
see Section \ref{subsec:bulgeprop}). The low $B/T$
of the classical bulges conform to the late Hubble types of most of
our targets. Alternative values for $\log\mbh$ of IC2560 \protect\citep[$6.54\pm0.06$,][]{Yamauchi+12} and $6.643\pm0.025$, Wagner et al. in prep) based on VLBI data instead of the single-dish based $\mbh$ adopted here have no appreciable effect on the scaling relations (Section \ref{sec:bhscalerels}).
}
\end{table*} 

\subsection{Galaxies from the literature}
\label{subsec:literature_values}

The primary sample of megamaser disk galaxies that we have fitted
comprises nine disk galaxies of similar total stellar mass ($\sim
10^{11}\msun$). We wish to compare the results for our megamaser
sample with a larger sample of galaxies spanning a range of masses,
morphologies, and $\mbh$.  Many works have done two-dimensional
image decomposition of the hosts of galaxies with dynamical BH mass
measurements \citep[e.g.][]{MH03,S11,Vika+12,Beifiori+12}. Here we rely on the $\mbh-\lbul$ data of \citealt{Lasker+14b} (L14 hereafter), since of all available previous work,
their image analysis and quality are closest to our current
study. We will discuss the comparison with other literature data in
Section \ref{sec:disc+sum}. The \citealt{Lasker+14b} sample comprises 35
galaxies of all Hubble types (4 spiral, 11 lenticular, and 20
elliptical galaxies) selected based on the availability of a reliable
BH mass at the time the imaging data were taken. As in this paper, L14
provides total photometry, results based on basic bulge+disk models,
and detailed decompositions, all based on homogeneous deep
sub-arcsecond resolution $K-$band photometry. One of the galaxies in
L14, NGC4258, is the prototypical megamaser disk galaxy.  In the
following, we will include NGC4258 in our sample of maser galaxies.

The sample from L14 is not identical to that of 
\citet[][KH13 hereafter]{Kormendy+Ho13}. L14 includes the gas
emission-based $\mbh$ measurements of PGC49940 and Cygnus A, as well
as NGC2778, NGC4261, NGC6251 and NGC7052, which were omitted by KH13
due to doubts about their $\mbh$ reliability.  Conversely, several
$\mbh$ measurements are included in KH13 
that became available after the photometry for L14
was completed. 

We again use the \cite{Bell+03c} color conversions to calculate the
stellar masses of the L14 sample. In this case, we do not have uniform
\hst\ color imaging for the objects. Instead, we use the SDSS
photometry provided by the NASA-SDSS
Atlas\footnote{http://www.nsatlas.org/} \citep[NSA hereafter; see,
  e.g.,][]{Blanton+11}.  Eighteen of the L14 galaxies have SDSS
photometry. For these, we use the NSA single-component \sersic\ fits
for the galaxy magnitudes. We correct for galactic extinction
 \citep[][retrieved using the NED\footnote{http://ned.ipac.caltech.edu/}]{SF11}
and calculate a $g-i$ color. For the galaxies that are not in the NSA
catalog, we use the mean $g-i$ color corresponding to the appropriate
Hubble type: 1.17 mag (E), 1.15 mag (S0), and 1.0 mag (S).

The resulting $H-$band mass-to-light ratio $\Upsilon_H(g-i)$ ranges
from 0.99 to 1.06, and thus any assumptions made based on the color
likely have minor effects on the galaxy mass estimates.  We calculate
$L_H$ from the L14 $K-$band magnitudes assuming the
extinction-corrected $20\magarcsec$-isophotal $H-K$ color of the 2MASS
database. We find good agreement between the masses derived this way
and the empirical relation of \cite{Cappellari13}, which fitted a
relation between dynamical mass and $M_K$. The dynamical estimates are
$0.2\dex$ higher on average and related to our color-based masses by a
correlation with slope $0.9$ but with negligible scatter. The higher
dynamical masses are consistent with the presence of dark matter within 
$R_e$ increasing with increasing galaxy mass. 
Similarly, our color-based masses are consistent with the
values given in KH13, which are only $0.06\dex$ more massive on
average with a scatter of only $0.15\dex$. The least certain masses
are for the spirals, where again the stellar population modeling
harbors the most uncertainties. As one additional sanity check, we
consider the five megamaser galaxies in the NSA catalog; deriving a
stellar mass for them using the \cite{Bell+03c} color relations and
$L_i$, we find agreement at the $0.2\dex$ level, indicating that our stellar masses are all on the same relative scale.

In Section \ref{sec:bhscalerels} we also consider three additional
late-type galaxies: the Milky Way, Circinus, and NGC1068 (the
megamaser host NGC4258 is already included in the L14 sample). In the
first case the BH mass is known more precisely than any other
\citep[e.g.,][]{Ghez+08,Gillessen+09} but the bulge mass is quite
difficult to determine. In the other two systems, the BH masses are
based on megamaser disk modeling, but the fidelity of $\mbh$ has been
questioned.  In the case of NGC1068, the rotation curve presented by
\citet{Greenhill+96} is sub-Keplerian, perhaps because of a massive
self-gravitating disk \citep{Lodato+Bertin03}. Circinus
\citep{Greenhill+03} may have an uncertain inclination
\citep{Ferrarese+Ford05}.

The $M\bul$ of the MW is taken from \cite{Kormendy+Ho13}. Bulge
magnitudes and masses for Circinus and NGC1068 are based on the work
of \cite{S11}. We convert their apparent $[3.6\mu\mathrm{m}]$
magnitudes to $M_\star$ using their formula (5) in conjunction with
$\sigma_v$ (dynamical mass), or their equation (6) (empirical
$M_\star-L_{3.6}$). For NGC1068, both $M_\star$ values of the bulge
are nearly identical; for Circinus they differ by $0.5\dex$ so we use
their mean ($\log\mbul/\msun=10.00$).

\section{Results: Bulge Classification}
\label{sec:results_bulgeclass}

In this section, we investigate the general properties (luminosities,
sizes, stellar masses) of the megamaser disk host galaxies. First 
we discuss whether or not our galaxies contain pseudobulges and then 
we attempt to determine whether or not they contain classical 
bulges. The nature of these components was discussed above in \S 3.1.

\subsection{Presence of Pseudobulges}
\label{subsec:galfit_pseudobul}

In \S3, we described in detail the different physical components that
are required to fully model the central regions of our megamaser disk
galaxies. Because pseudobulges are defined by their
formation history, not by their appearance, we are forced to consider what
properties might indicate a pseudobulge. Here we mean primarily that
we have identified structural components (nuclear rings or disks,
as well as X-shaped/boxy bulges) that are associated with secular
evolution.  In the past it has been shown that \sersic\ indices 
$n < 2$ correlate with the presence of pseudobulges
\citep[e.g.,][]{Kormendy+Kennicutt04,FisherDrory10}, as well as low
effective surface brightnesses at a given size \citep{Gadotti09}, and so
we will also consider these specific criteria.

Below, we will go through the classification of each galaxy in
detail (\S 4.4). Here we simply summarize our main finding: all but
two of the galaxies show unambiguous signatures of secular
evolution. We find sub-kpc disks, bars, or spirals in all galaxies
with Hubble Type Sa or later (Table 1). We only identify one case (the
lenticular galaxy NGC1194) with no evidence for a pseudobulge. 
Thus, we are fairly confident that the 
large majority of these galaxies do contain a pseudobulge component.

As emphasized by \cite{Erwin+15} among others,
the presence of pseudobulge components does not preclude the presence
of a rounder, older, kinematically hot classical bulge. From the point
of view of BH scaling relations, it may well be that this classical
bulge component matters most \citep{Nowak+10}. Thus, we have
attempted to use our photometric fitting to identify such
components. In the next section, we characterize these
putative classical bulge components using color, shape, and structural
information.

\subsection{Bulge Luminosities, Sizes, and Colors}
\label{subsec:lums+sizes}

We derive galactic-extinction--corrected luminosities and linear sizes
from the apparent magnitudes and sizes presented in Table
\ref{tab:gf}, assuming $M_{\odot,H}$=3.32. As a reminder, the
``basic'' bulge is derived from a fit that includes just two or three
components: disk, bulge and central point source where it can be
fit. The point source may arise from the contribution of an AGN or be
stellar in nature \citep[e.g.,
  NGC3384][]{Graham+Driver07a,Ravindranath+01}.  Basic fits are
indicated with the subscript ``bas'' in tables and figures. The
``classical'' bulge component is the roundest and highest $n$
component in the multi-component fit. Classical bulge components are
indicated with subscript ``bul'' in figures and tables.  Finally, the
``total'' magnitude and size are calculated as the sum of all
components in our best fit multi-component model. We list the
resulting parameters in Table \ref{tab:lummassrad}, along with the
bulge-to-total ratios ($B/T$) for the classical and basic bulge
measurements.

In general, the basic bulges are both more extended (mean effective
radius $0.8\kpc$, ranging from $0.2$ to $3.8\kpc$) and more luminous
($0.5-5.1\times10^{10}\,L_{\odot,H}$) while the classical bulges, as
they comprise only a single component of the galaxy central region,
tend to be smaller (mean effective radius of $0.7\kpc$, ranging from
$0.3$ to $5.4\kpc$) and slightly fainter
($0.5-4.3\times10^{10}\,L_{\odot,H}$).  Likewise the median
bulge-to-total ratio ($B/T$) drops slightly from the basic bulge
(30\%) to the classical bulge (20\%). Generally these $B/T$ values are
completely consistent with our expectations for early-type spirals
\citep[e.g.,][]{Simien+deVaucouleurs86,Laurikainen+10}.

\begin{figure}
 \centering
 \includegraphics[width=7cm]{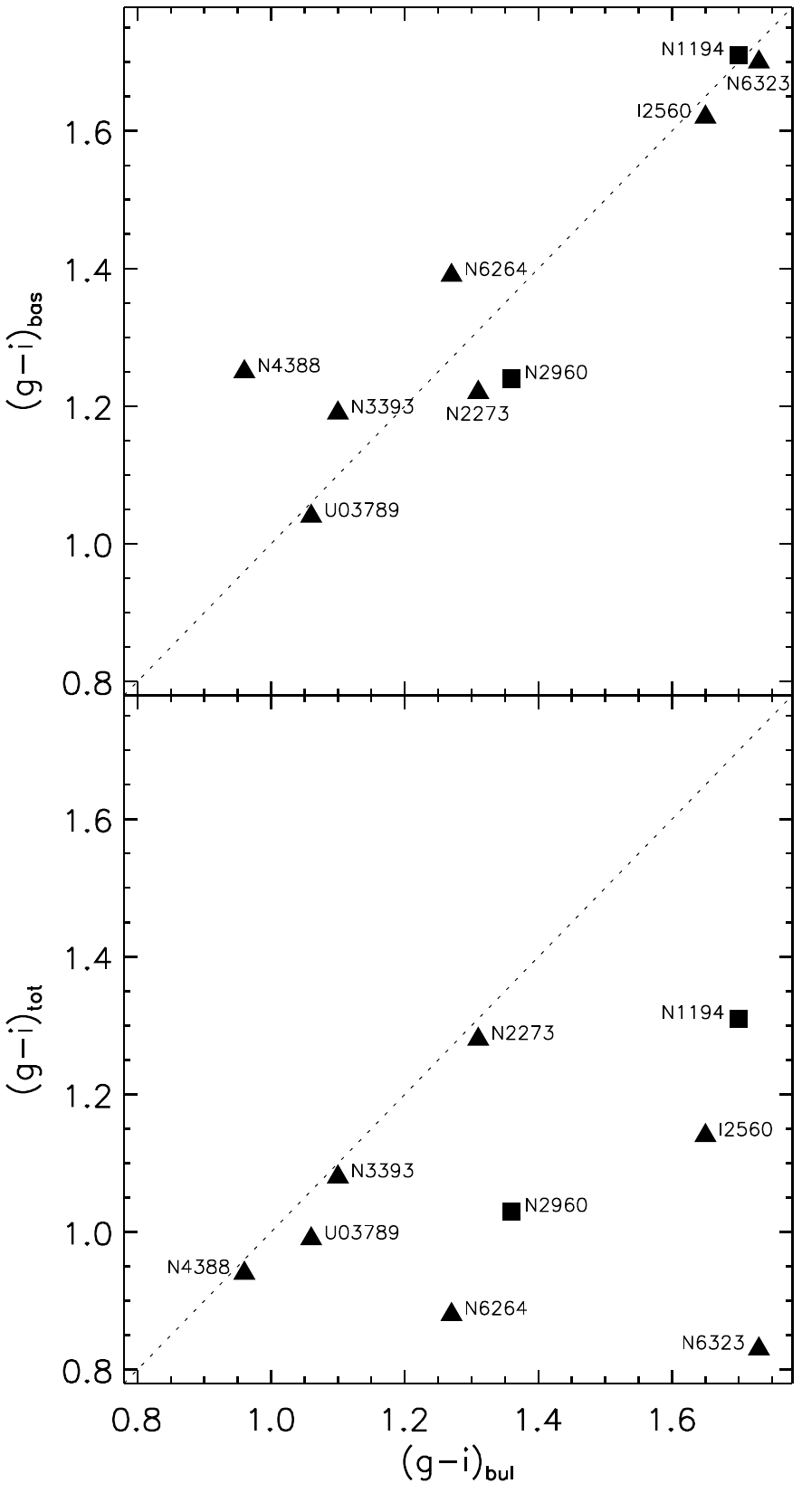}
 \caption{
Comparison of the $g-i$ color (mag) of the basic bulge (``bas''; top
panel) and total galaxy (``tot''; bottom panel) with the classical
bulge (``bul''; full decomposition) for our megamaser sample, as drawn
from Table \ref{tab:colors}. S0 galaxies are indicated with squares
while spiral galaxies are triangles.  All colors are measured in
apertures: the region where the respective bulge component dominates
in the \galfit\ model, while total color is measured within the
HST/WFC3-IR field-of-view. The colors of basic and classical bulge
differ only marginally, but all bulge colors are redder than the total
color, as expected.  }
 \label{fig:gicomp}
\end{figure}

The $g-i$ colors of the bulge and
total galaxy are compared in Figure \ref{fig:gicomp}. As expected,
the total color is always bluer than the bulge color, which
avoids light from the large-scale disk component.
We also check whether our detailed decompositions result in redder
bulges than the basic decompositions. The latter tend to include central
disks and rings, which are presumably younger and bluer. With our
aperture measurements, however, classical bulges are not appreciably
redder than the basic bulges. This similarity suggests that any
color differences are too subtle for our crude method to discern, or
that dust obscuration offsets real differences in the population.

\subsection{Bulge Shapes and Profiles}
\label{subsec:bulgeprop}

As discussed in \S3, we model a putative classical bulge component in
addition to the exponential kpc-scale disk and any other identifiable
morphological components, such as bars and nuclear disks.  We attempt
to understand the nature of the putative classical bulge components by
examining their intrinsic flattening, \sersic\ indices, and
photometric scaling relations. Recall that in Figures and Tables, the
classical bulge component is denoted as ``bul''.

\begin{figure*}
 \centering
 \includegraphics[width=16.5cm]{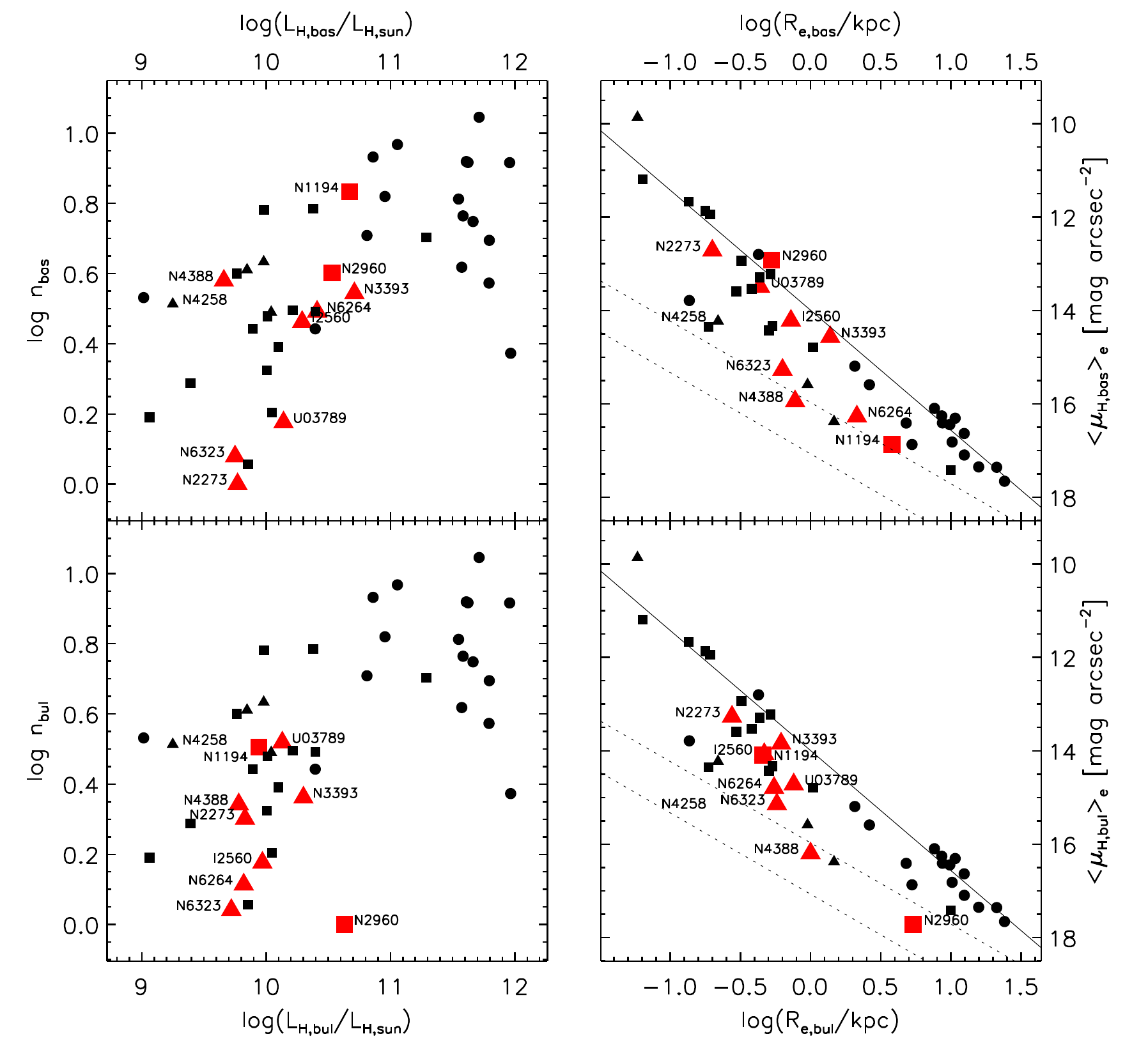}
 \caption{
Correlations of photometric bulge parameters. The large red symbols
represent the bulges of the present study, with ``basic'' bulges from
bulge+disk decompositions shown in the top panels (subscript ``bas'') 
and classical bulges
from full decompositions in the bottom panels (subscript ``bul''). 
The bulges of L14 are
overplotted as small black symbols, and the only megamaser-based 
$\mbh$ from L14 (NGC 4258) labeled along with the rest of the 
maser hosts. Circles indicate elliptical,
squares lenticular, and triangles spiral galaxies.
{\it Left panels:} relation between bulge \sersic\ index $n$ and
$H$-band bulge luminosity. Our measurements (bulges in mostly spiral
galaxies) broadly match the L14 relation (mostly ellipticals and
bulges in early-types), but the $n-L_H$ correlation is generally
weak.
{\it Right panels:} $H$-band surface brightness inside $R_e$ ($\mu_e$)
versus $R_e$ \protect\citep{Kormendy77}. Overplotted in solid is the fit for
early-type galaxies by \protect\cite{Khosroshahi+00}, transformed from
$K$ to $H$-band using an average $H-K=0.3$ mag and distances corrected
from their $H_0=50\,{\rm km~s^{-1} Mpc^{-1}}$ to our 
$H_0=70\,{\rm km~s^{-1} Mpc^{-1}}$. Dotted lines
mark the upper limit for pseudobulge brightness recommended by Gadotti
(2009), assuming $i-H=2.1$ and $i-H=3.2$ (the bluest and reddest bulge
colors, respectively, in our sample): the classical bulge measurements
for the megamaser hosts are only slightly below the early-type
relation and avoid the region of pseudobulges, suggesting we indeed
excluded the latter and recovered the classical bulge
component. However, this applies to the basic bulge measurements as
well, which have on average larger size and lower surface brightness
than the classical bulges from full decompositions.}
 \label{fig:n-lum+mu_re}
\end{figure*}

\subsubsection{Flattening}
\label{subsubsection:flattening}

As one determining factor, we consider the flattening of the classical
bulge component. A true classical bulge should be rounder than the disk. 
In contrast, if the
light is dominated by a disky component, we expect to see a wide
distribution of flattening $q=b/a$. The $q$ of the disk and 
pseudobulge component should be similar, since secular evolution is supposed
to bring material into the bulge from the disk.  In practice, all of
our classical bulge candidates have a higher
projected axis ratio than the main disk, with the difference in
bulge-disk $q$ ranging from $0.2-0.5$.

\subsubsection{\sersic\ $n$ and the $n-L$ relation}
\label{subsubsec:n-L}

The basic bulges are best-fit by \sersic\ indices ranging from 1.0 to
6.8 (Table 2).  The presence of compact disk-like central components
(e.g., rings, disks, or bars) tends to drive up the value of $n$ for
the basic bulge components. When we perform more detailed
decompositions, the effect of the nuclear components is mitigated,
leading in general to lower $n$. For example, in the galaxy NGC4388,
$n$ drops from 3.8 to 2.2 when the highly inclined nuclear disk is
modeled separately from the bulge.  These classical bulge components
have \sersic\ $n\in[1.0,4.4]$, with four of the nine classical bulge
candidates having $n<2$ (IC2560, NGC2960, NGC6264, NGC6323). The
classical bulges in \cite{Erwin+15} also tend to have $n < 2$,
suggesting that a high \sersic\ index is not a hard requirement for
low-mass classical bulges using our photometric definition of a
rounder, extra-light component (\S 3.1).

Most of the classical bulge \sersic\ indices in our late-type galaxies
follow the relation with $L_{\rm bulge}$ in early-type bulges (Figure
\ref{fig:n-lum+mu_re}, left panels). Yet, the $n<2$ bulges have lower
$n$ for their luminosity than classical bulges in general. NGC2960 is
an interesting outlier. As suggested by \cite{Kormendy+Ho13}, this
galaxy may be a merger remnant. We find a kpc-scale disk in NGC2960, plus a nuclear disk with a steep profile, neither of which is
easily interpreted as a classical bulge. If we instead consider its
basic bulge (bottom-left panel of Figure \ref{fig:n-lum+mu_re}), it
falls in line with the general trend.

\subsubsection{The $\mu_e-R_e$ relation}
\label{subsubsec:mu-Re}

We also investigate a projection of the Fundamental Plane
\citep{Dressler+87,Djorgovski+Davis87} to see whether we have isolated
components that scale like elliptical galaxies.  We use the
\cite{Kormendy77} relation between the surface brightness 
measured within the effective radius ($\mu_{\rm e}$) and
effective radius.  We have asserted based on
morphology that most of our galaxies harbor pseudobulges. Structural
studies find that pseudobulges tend to have lower effective
surface brightness at a given size than classical bulges and
elliptical galaxies \citep{Carollo99,Macarthur+03,Gadotti09,
  Laurikainen+10,FisherDrory10}. Thus, we turn to this projection to
test whether our galaxies have the structures of pseudobulges.

In Figure \ref{fig:n-lum+mu_re} (top right), we show the Kormendy
relation for our basic bulges. They reside marginally below the
relation for ellipticals and early-type bulges, but have relatively
high effective surface brightness compared with the pseudobulge
selection advocated by \cite{Gadotti09}. Interestingly,
\cite{Laurikainen+10} also find that $\mu_{\rm e}/R$ decreases 
in later-type spirals. Since our
objects are nearly all S0-Sab galaxies, perhaps we should not be
surprised that they do not fall far below the elliptical galaxy
relation.

Nevertheless, we find clear evidence in the bulge morphologies for
pseudobulge components. Why do these not translate into considerably
lower $\mu_{\rm e}$ at a given size as we might expect? We conclude
that while nuclear bars, disks, and rings are readily apparent in our
\hst\ images, they do not carry a substantial fraction of the central
mass. This is why we find only very small changes in mass and size
when going from the ``basic'' to the ``classical'' bulge component. We
can then investigate the nature of the dominant bulge components in
these galaxies. From photometry alone, it is difficult to characterize
these classical bulge components. In two cases, NGC2273 and NGC4388,
we have additional kinematic information.  Here we know that there is
a nuclear disk on $<200$pc scales, followed by $V/\sigma < 1$ further
out in the bulge, showing that the stars outside of the nuclear disk
are supported predominantly by random motions.  In the case of these
two galaxies, some of the stars are also part of the known kpc-scale
bar \citep{Veilleux+99, Falcon-Barroso+06,Greene+14}. For the rest of
our galaxies, we do not yet know the kinematics of the stars in our
putative classical bulges.

\subsection{Final Bulge Classification}
\label{subsec:bulgeprop}

For each galaxy, we will summarize the evidence that we have a
pseudobulge, a classical bulge, or both.  We established that in
nearly all cases megamaser disk host galaxies show unambiguous
pseudobulge signatures in the form of nuclear rings, nuclear or inner
($\lesssim1\kpc$) disks, central dust, and spiral structures.The
presence of these components is our criterion for classifying a galaxy
as containing a pseudobulge.

We have also attempted to isolate classical bulges.  As criteria for
classical bulge determination we use (1) shape: they should be rounder
than the large-scale disk; (2) surface brightness in relation to the
effective radius: they should be close to the relation between 
$\mu_e$ and $R_e$ defined by elliptical galaxies; and 
(3) \sersic\ index: they should have high 
($n > 2$) values.  In our view, 
the \sersic\ index is the least reliable discriminator of
classical bulges, due to the large scatter in the $n-L$ plane. We
determine that a galaxy contains a classical bulge if it satisfies two
of these three criteria.  For each galaxy, we assign a {\it P} to
indicate a likely pseudobulge and a {\it C} to indicate a likely
classical bulge.

{\it IC2560}: IC2560 is a relatively strongly inclined galaxy with a
readily apparent X-shaped bar/bulge, which provides clear evidence for
a pseudobulge component. We do identify a small classical bulge
candidate that is rounder than the disk ($q=0.6$ versus $q=0.4$). The position in the $\mu_e-R_e$ diagram, close to the relation of early-type galaxies, likewise points towards a classical bulge, while the \sersic\ index is somewhat low ($n=1.5$). However, despite the above classical bulge indicators, given the dust and complexity of IC2560 in the central regions we cannot confirm that this small component is indeed a classical bulge. (P)

{\it NGC1194}: This edge-on, bulge-dominated galaxy is difficult to
fit robustly due to significant dust contamination and a likely bar.
The basic and classical bulge fits are round and 
both have a high \sersic\ index. 
The effective surface brightness is not as high
as in massive ellipticals, but is higher than expected for a
 pseudobulge based on the work of \cite{Gadotti09}. 
Overall, we determine that NGC1194 contains a classical bulge. 
We do not find any pseudobulge component. (C)

{\it NGC2273}: 
While the $\mu_e-R_e$ position is consistent with expectations for
classical bulges, the axis ratio of the putative classical bulge
component matches that of the disk. The \sersic\ index is
intermediate with $n=2$. Superposition of the candidate bulge
component with the strong bar complicates significant detection of a
classical bulge, and given the generally complex inner structure, we
conclude that there is not sufficiently compelling evidence for a
classical bulge component. NGC2273 unambiguously contains
pseudobulge components. From photometry, it has long been known to
contain multiple rings and a nuclear disk \citep{Erwin+03}. More
recently, kinematic evidence has confirmed that the central $\sim
200$pc is dominated by a disk
\citep[][]{Barbosa+06,Falcon-Barroso+06}. There is also a kpc-scale
bar component.  (P)

{\it NGC2960}: Apart from a large-scale smooth and round component,
this galaxy contains two central disks, one nuclear and one $\sim
1\kpc$-sized with inset spiral structure. It is particularly
challenging in this case to determine what component comprises the
bulge. Treating the nuclear disk component as the classical bulge
yields a very low \sersic\ index, but a high central surface
brightness. The other option was suggested by KH13; the galaxy may be
an elliptical that just swallowed a spiral galaxy. In that case the
classical bulge may be the outer component and the inner ($1\kpc$)
disk a recently acquired addition. The outer component is round, and
the observed low $\mu_e$ is likely an artifact of the distortion to
the center caused by the merger.  Although technically this component
does not satisfy our classical bulge requirements, we treat it as a
classical bulge distorted by a merger. Since we detect an inner disk,
it has a clear pseudobulge component. (CP)

{\it UGC3789}: This galaxy is relatively face on, so using the shape
is challenging. However, given the high \sersic\ index $n=3.3$ and
relatively high $\mu_e$, we identify a classical bulge component.  We
see clear signs for pseudobulge components in the nuclear and
larger-scale rings. (CP)

{\it NGC3393}: Like UGC3789, this galaxy is close to face on.  The
classical bulge candidate is round and has a \sersic\ index
$n=2.3$. Because the galaxy is face on, we cannot infer much about the
intrinsic shape of the component.  The high effective surface
brightness, combined with $n>2$, argues for a classical bulge. NGC3393
contains multiple rings and a bar. The evidence for pseudobulge
components is clear. (CP)

{\it NGC4388}: The classical bulge fit to this galaxy is rounder than
the disk by a factor of two.  However, given the low $\mu_e$ of the
putative classical bulge, $n \approx 2$, and the significant dust
extinction towards the galaxy center, we do not find a convincing
classical bulge component.  This well-known galaxy has a kpc-scale bar
that has been studied kinematically \citep{Veilleux+99}. On the
smallest scales, we have identified a nuclear disk both from the
\hst\ photometry and our NIR spectroscopy \citep{Greene+14}, all
pointing to a pseudobulge component.  (P)

{\it NGC6264}: One of the most distant galaxies in the sample, this
galaxy clearly contains a kpc-scale bar, and thus a pseudobulge
component. The galaxy is close to face-on, so the axis ratio is 
not very constraining. The \sersic\ index is low ($n \approx 1$) 
but $\mu_e$ relatively high. Given the lack of consensus between 
the different indicators and the distance of the galaxy leading to 
extra degeneracy with the bar, we do not robustly
identify a classical bulge. (P)

{\it NGC6323}: We fit a very round central component as a classical
bulge candidate. The low \sersic\ index of $n\approx 1$ and the low
surface brightness combined suggest that we have not isolated a
classical bulge component. By contrast, the structural properties of
the bulge and the inner $\kpc$-scale disk/ring lead us to conclude that
NGC6323 harbors a pseudobulge (P).

Armed with these bulge classifications, luminosities, and
stellar masses for our different stellar components, we now turn to
the primary goal of this paper, the BH scaling relations.

\section{Results: BH scaling relations}
\label{sec:bhscalerels}

Our goal is to learn about the origin of the BH-galaxy scaling relations 
by studying their slope, zeropoint, and scatter over as wide a 
dynamic range as possible. Before this work, there were very few dynamical
BH masses in late-type galaxies. We add nine new systems with structural 
measurements.

\subsection{The Special Role of Megamaser Disk Galaxies}

In this subsection, we focus on the megamaser disk galaxies taken
alone (including NGC4258 from L14).  Megamasers play two important
roles. First and foremost, VLBI allows us to resolve much smaller spheres of influence than
optical or NIR stellar- or gas-dynamical methods, and the usually near-perfect Keplerian rotation curves traced by
the masers provide the most precise and accurate extragalactic black hole mass
measurements. Nuclear megamasers probe the full range of BH mass at a given 
galaxy property in a way that no other current technique can. 
Secondly, as a corollary, they allow us to probe BH mass in spiral 
galaxies, where gas, dust, and typically small spheres-of-influence 
all conspire to make stellar or gas-dynamical techniques especially 
challenging.

The megamaser disk galaxies span a stellar mass range of $\log M_{*}/\msun=
10.5$ -- $11.1$ (only a factor of four), and yet the BH masses span a
range of $\log (\mbh/\msun) \approx 6.4 - 7.8$, a factor of 25.  If we
take the scaling relations measured for predominantly early-type
galaxies from L14, then for the measured BH masses we expect a range
in bulge mass of $10^8$ to $10^{10}\msun$, corresponding to a range in
bulge-to-total light ($B/T$) of $10^{-3}$ (for IC2560) to 0.2 (NGC1194),
with a median value of 0.03.  In contrast, galaxies of these Hubble
types (S0-Sb) typically have $B/T \approx 0.1-0.2$
\citep{Simien+deVaucouleurs86, Laurikainen+10}.  {\it It is already
  clear before we perform any fitting that these BHs would need to
  have very small classical bulge components if they were to obey the
  scaling relations seen in early-type galaxies}. A similar conclusion
was reached for the $\mbh-\sigma_{\ast}$ relation by \citet[][see also
  KH13 for other refs]{Greene+10b}.

The maser galaxies occupy a range of bulge masses log $M_{*}/\msun=
9.5\,-\,11$.  The wide distribution in BH mass in a narrow range of
galaxy property\footnote{Within this range, two galaxies in the L14
  sample have $\mbh>10^8\msun$, increasing the total range of $\mbh$
  yet further (Figure \ref{fig:mbh_l+m_bas}).} shows us that there is
not a tight correlation between $\mbh$ and galaxy or bulge mass for
galaxies with $M_{*} \lesssim 10^{11}\msun$. Instead, the measured
correlation found for early-type galaxies defines an upper envelope
for the BH mass, with many of the megamaser galaxies scattering below
this relation.  As we will quantify in the following subsections, this
broad tail towards low $\mbh$ appears to hold even when we consider
bulge rather than total galaxy mass.

This tail to low BH mass, which is most clearly apparent in the
megamaser disk galaxies, raises a more basic question about the
scaling relations. We have to wonder whether the scaling relations
reflect the true distribution in nature, or whether they reflect our
inability to measure BHs at low mass in high-mass galaxies
\citep[][]{Batcheldor10}. \cite{Gultekin+11a} explore the possibility
that the $\mbh-\sigma$ relation is only an upper envelope, since
galaxies with unresolved gravitational spheres of influence may be
preferentially missing.  At high galaxy mass, 
$\sigma_{\ast}>250\,\mathrm{km\,s^{-1}}$ \citep[corresponding roughly
  to $\sim 10^{11}\,\msun$; e.g.,][]{HR04}, \cite{Gultekin+11a} 
show that the existing dynamical BH mass sample is
large enough to rule out a long tail to low $\mbh$\footnote{That is,
  barring dramatic problems with the BH masses, such as very steep IMF
  gradients \citep{McConnell+13,Martin-Navarro+15}}. However, at lower
stellar velocity dispersion, a large range in $\mbh$ at fixed galaxy
property is what we observe. We discuss the possibility of bias in the
low-mass galaxy sample, along with the possibility of bias in the
megamaser disk galaxies, in \S 6.1. First we quantitatively fit the
relationships between $\mbh$ and measured galaxy properties.

\subsection{Fitting method}
\label{subsec:fitmethod}

We use the IDL implementation of the Bayesian inference-based
LINMIX\_ERR \citep{Kelly07} for fitting our BH scaling relations
between $\mbh$ and galaxy properties. This routine naturally implements
intrinsic scatter in the $y$-coordinate as part of the model,
i.e. the relation is modeled as a probability distribution in $(x,y)$
space. It renders a Markov-Chain Mote Carlo (MCMC)
realization of the posterior distribution of the linear relation
parameters, which are the zeropoint ($\alpha$), slope ($\beta$), and
dispersion of the Gaussian intrinsic scatter ($\epsilon$). We thus fit
relations of the form

\begin{equation}
 y = \alpha + \beta (x-x_0) + \mathcal{G}(\epsilon)~,\label{eqn:linrel}
\end{equation}

\noindent
where $y=\log(\mbh/M_\odot)$, $x$ are logarithmic luminosities or
masses, i.e. $x=\log(L_H/L_{\odot,H}$) or $x=\log(M/M_\odot)$, and
$\mathcal{G}(\epsilon)$ symbolizes a Gaussian probability distribution
with dispersion $\eps$. The $x$-offset $x_0$ is calculated before
fitting as the mean of the $x$-coordinates of the data in order to
reduce covariance between $\alpha$ and $\beta$. The linear relation
(\ref{eqn:linrel}) implies a powerlaw between $\mbh$ and $L_H$ or
$M_{\ast}$, and $\beta=1$ corresponds to a linear relation between
$\mbh$ and the galaxy property. The presented result for each relation
parameter is the mean and the $68\%$ confidence interval of the
posterior after marginalization over the other parameters. Some
recent papers have suggested a broken power-law fit to BH scaling
relations with bulge luminosity or mass (e.g., Graham \& Scott
2015). This is an interesting possibility, but we do not have a
sufficient sample size in this paper to address that possibility
rigorously. The values quoted in Table \ref{tab:bh_sc} are the
mean values of the parameters in the Markov Chain, and the given
intervals delineate the 68\% confidence interval of the drawn samples
relative to the mean.

\begin{table*}
 \caption{Scaling relation parameters $\log(\mbh)-\log(x)$ with galaxy properties and galaxy subsamples}
 \begin{tabular}{l*{19}{c}}
  \toprule
  $x$ & $x_0$ & \multicolumn{2}{c}{$\alpha$} & \multicolumn{2}{c}{$\beta$} & \multicolumn{2}{c}{$\epsilon$} & \multicolumn{2}{c}{$\Delta a_\mathrm{mega}$} & \multicolumn{2}{c}{$\epsilon_\mathrm{mega}$} & \multicolumn{2}{c}{$\Delta a_\mathrm{late}$} & \multicolumn{2}{c}{$\epsilon_\mathrm{late}$} & \multicolumn{2}{c}{$\Delta a_\mathrm{low}$} & \multicolumn{2}{c}{$\epsilon_\mathrm{low}$} \\
  & & val & err & val & err & val & err & val & err & val & err & val & err & val & err & val & err & val & err \\
  \midrule
$\lbul$ & 10.56 &  8.13 &  0.09 &  0.87 &  0.11 &  0.56 &  0.07 & -0.63 &  0.21 &  0.90 &  0.11 & -0.51 &  0.24 &  0.73 &  0.12 &  0.09 &  0.39 &  0.72 &  0.14 \\
$\lbas$ & 10.64 &  8.12 &  0.09 &  0.99 &  0.13 &  0.57 &  0.07 & -0.80 &  0.22 &  0.93 &  0.13 & -0.58 &  0.24 &  0.76 &  0.13 &  0.04 &  0.37 &  0.76 &  0.13 \\
$\ltot$ & 10.93 &  8.12 &  0.11 &  0.99 &  0.19 &  0.71 &  0.09 & -1.13 &  0.21 &  1.35 &  0.11 & -1.03 &  0.22 &  1.23 &  0.12 &  0.47 &  0.40 &  1.04 &  0.25 \\
 \\
$\mbul$ & 10.66 &  8.12 &  0.08 &  0.88 &  0.10 &  0.52 &  0.06 & -0.58 &  0.21 &  0.79 &  0.12 & -0.49 &  0.22 &  0.70 &  0.12 & -0.22 &  0.31 &  0.71 &  0.12 \\
$\mbas$ & 10.74 &  8.13 &  0.09 &  0.98 &  0.12 &  0.56 &  0.07 & -0.76 &  0.22 &  0.89 &  0.13 & -0.57 &  0.24 &  0.74 &  0.14 &  0.05 &  0.34 &  0.74 &  0.11 \\
$\mtot$ & 11.01 &  8.13 &  0.10 &  1.04 &  0.17 &  0.68 &  0.08 & -1.05 &  0.20 &  1.25 &  0.11 & -0.98 &  0.21 &  1.16 &  0.12 & -0.04 &  0.36 &  0.84 &  0.12 \\

  \bottomrule
 \end{tabular}
\tablecomments{The first column identifies the quantity $x$ that is respectively fitted by a relation $\log(\mbh/\msun) = \alpha+\beta(\log(x/x_\odot)-x_0)+G(\epsilon)$ (Equation \ref{eqn:linrel} in Section \ref{subsec:fitmethod}), where $x_\odot$ is $L_{\odot,H}$ or $M_\odot$, and $G(\epsilon)$ denotes a Gaussian distribution with dispersion $\epsilon$. Given in the second column is the offset $x_0$ (the inverse variance-weighted mean of all $x$), which is subtracted from the $x$ before fitting in order to reduce covariance between $\alpha$ and $\beta$. All other parameters are fit to the data using LINMIX\_ERR and the resulting Markov chain MonteCarlo (MCMC) sample, which is evaluated for the mean (``val'') and the standard deviation (``err'', i.e. the size of the 68\%-confidence interval). These parameters are: the relation zeropoint ($\alpha$), logarithmic slope ($\beta$), and log-scatter in the $y=\log(\mbh/\msun)$-direction ($\epsilon$) for the fit to the entire sample of 44 objects, as well as the respective offset and intrinsic scatter of three subsamples: megamasers (``mega''), spiral galaxies (``late'') and low-mass galaxies (``low''). Low-mass galaxies always constitute half of the sample, i.e. for each relation those 22 galaxies with the lowest $x$ values. The subsamples' offsets are relative to a relation fit to the data minus the subsample; the parameters of these relations are not shown here. The subsample scatter $\epsilon$ is relative to the offset relation with the same $\beta$ as the main relation.
}
\label{tab:bh_sc}
\end{table*} 

\subsection{Subsamples}

When we combine the megamaser disk galaxies with L14, we have a large
enough sample to split galaxies based on various properties, and then
ask where they fall in the BH-bulge mass plane. We will focus
exclusively on the 44 galaxies (L14 and ours) that have dynamical BH
mass measurements as described in \S 4.

We consider the following subsamples.  We investigate the scaling of
the 10 megamaser galaxies, which includes NGC4258 from the L14
sample. We also group all late-type galaxies. Many works
\citep{Hu2008,Greene+08,Gadotti09,Greene+10b,KormendyBenderCornell11}
have suggested that late-type galaxies with pseudobulges may obey
different scaling relations than classical bulges, since secular
evolution may not efficiently fuel BH growth.  In practice, because
there is considerable overlap between the late-type and pseudobulge
samples (7/9 of our megamaser sample), we will only consider one
late-type subsample comprising the eight non-S0 megamasers considered
here and the three additional spiral galaxies from L14. Finally, we
consider a low-mass sample, since if the scaling relations arise from
hierarchical merging via the central limit theorem
\citep[e.g.][]{Peng07,Jahnke+Maccio11} then the scaling relations
would break down at low mass. For this subsample we simply take the 22
lowest-mass or lowest-luminosity galaxies.

We calculate the offset of each subsample from the "primary relation",
which is the relation fitted to the data excluding the respective
subsample. For each Monte-Carlo realization of the primary relation,
the offset $\Delta a$ (Table 8) is the weighted-mean
$y=\log\mbh$-offset of the subsample from the primary relation. The
weights are the $y$- and projected $x$-measurement errors, plus the
intrinsic scatter in the relation, added in quadrature. We take into
account the uncertainty in $\Delta a$ by drawing it from its
(Gaussian) error distribution around the weighted-mean offset. The
intrinsic scatter of the subsample ($\eps$) we subsequently determine
by iteratively varying its value, and adding it in quadrature to the
measurement errors, until $\chi^2=1$ for the given primary relation
slope and subsample offset. We thus calculate $\Delta a$ and $\eps$
for each element of the Markov chain, and the resulting $\Delta a$ and
$\eps$ distributions are evaluated for their mean and
$1\sigma$-uncertainty. The offset for the maser sample is referred to
as $\Delta a_{\rm mega}$, that for the late-type galaxies is $\Delta
a_{\rm late}$, and the low-mass sample is $\Delta a_{\rm low}$.

There is another sample of spiral galaxies with indirect BH masses,
based on reverberation mapping. A subsample of the reverberation
mapped AGN have \hst\ imaging that allows detailed bulge-disk
decompositions \citep{Bentz+09b}. These fall in an inferred $\mbh$
range of $10^7-10^9\,\msun$ that overlaps with the megamaser disks at
the low-mass end. Many of the hosts are also disk
galaxies. \citet{Bentz+09a} find that $\mbh$ is quite tightly
correlated with $\lbul$ (even with no conversion to mass) and 
do not see the long tail to low $\mbh$ that is seen with the
megamaser galaxies. Interestingly, the scatter seen between $\mbh$ and
$\sigma_{\ast}$ is also smaller at $\mbh \approx 10^7\,\msun$ for the
reverberation-mapped sources than the megamaser disk galaxies
\citep[e.g.,][]{Greene+10b,Woo+10}. We discuss this difference in 
\S 6.1.

We also note that there are two other prominent outliers in 
the BH-bulge scaling relations (S0 galaxies
NGC4342 and NGC3998) that have apparently high $\mbh$ for their
stellar mass. In both galaxies, the bulge luminosity is strongly
dependent on the adopted decomposition \citep[for details,
  see][]{Lasker+14a}.  However, even taking the total luminosity as an
upper limit on $\lbul$ still puts their $\mbh$ above the $\mbh-\lbul$
relation. In the case of NGC3998, $\mbh$ would be five times lower if
the gas-dynamical measurement was adopted instead of the
stellar-dynamical model, true for many galaxies with both stellar and
gas dynamical measurements \citep[see][]{Walsh+12}. The BH mass
measurement in NGC4342 is based on stellar dynamics.  Here the large
$\mbh$ for its (bulge) luminosity is similar to an emerging class of
S0 galaxies that appear to have overly massive BHs for their stellar
bulge mass \citep{vdBosch+12,Walsh+15,Walsh+16}. These galaxies tend to be
very compact with large central velocity dispersions and fast
rotation on large scales, live in rich environments, and may have very
different formation histories than the megamaser galaxies studied
here.  However, they contribute to our overall conclusion that there
is very significant scatter in $\mbh$ for bulge masses $< 5 \times
10^{10}\msun$.

\subsection{The $\mbh-M_{\rm bas}$ relation for basic bulges}
\label{subsec:mbh-mbas}

\begin{figure*}
 \centering
 \includegraphics[width=8.5cm]{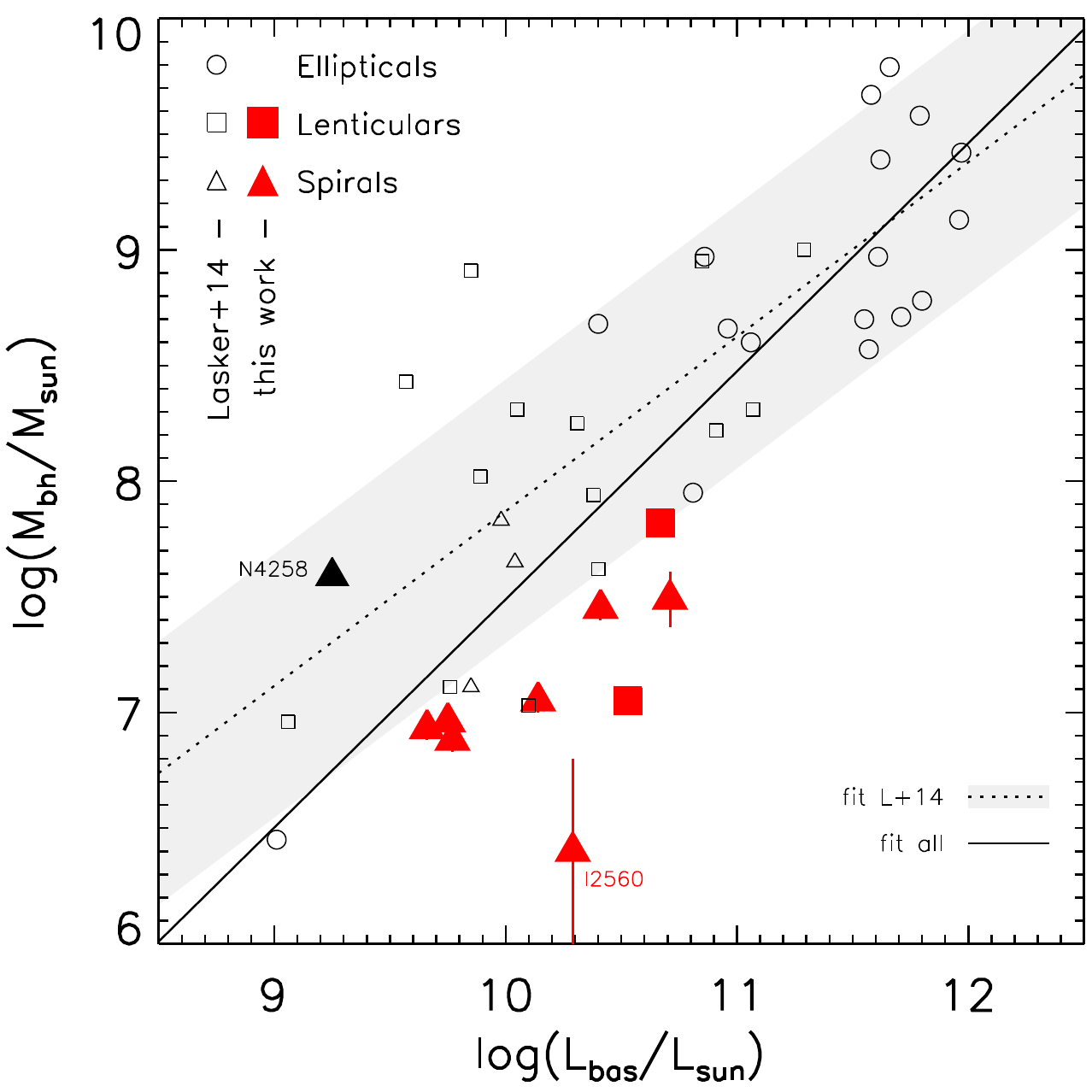}
 \hfill
 \includegraphics[width=8.5cm]{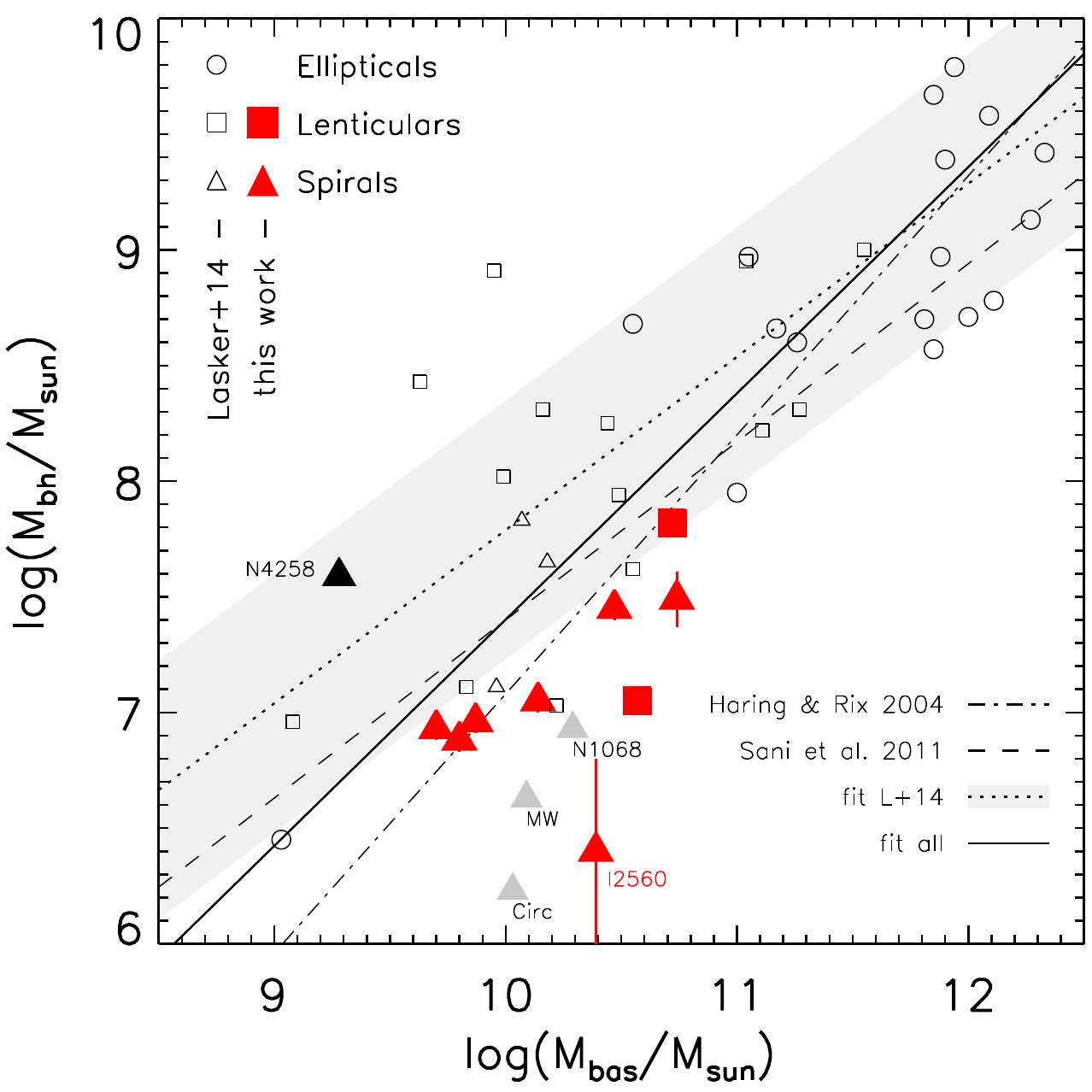}
 \caption{
Correlation of $\log\mbh$ with ``basic'' bulge based on simple
bulge+disk(+point source) decompositions of our megamaser BH hosts. We
show log $\L_{H,\mathrm{bas}}$ (left) and log $M\bas$ (right
panel). The maser disks analyzed in this paper based on HST/WFC3
imaging are indicated by large filled red symbols with error bars. The
errors on the bulge magnitudes are small, as we have not estimated
systematic errors for these basic fits. The filled black symbol
(NGC4258, which also has a megamaser-based $\mbh$), and small open
symbols (with error bars suppressed for clarity) are values from
\protect\cite{Lasker+14a}, with their ``spheroid'' $K$-band luminosity
converted to $L_H$ and $M$ as described in
\S\ref{subsec:colpluslups}. Circles indicate elliptical, squares S0,
and triangles spiral galaxies. The solid line represents the fit to
the combined sample of L14 and the present study. The dotted line
shows the relation fit when restricted to the L14 sample, i.e. with
all megamasers except NGC4258 omitted, and the grey area indicates the
$1\sigma$ intrinsic scatter ($\epsilon$). For comparison, the
$\mbh-M\bul$ relations of \protect\cite{S11} and \protect\cite{HR04}
are overplotted as dashed and dot-dashed lines, respectively, in the
right panel. For illustration, the grey labeled triangles in the right panel show three more
BH hosts with $M\bul$ taken from the literature. These, however, are
excluded from our fit because the Milky Way (MW) bulge mass is highly
uncertain while the megamaser-based $\mbh$ (Circinus, NGC1068) are in
question.}
\label{fig:mbh_l+m_bas}
\end{figure*}

We begin by examining the ``basic'' bulge fit, in which we assume that
the galaxy can be well fit by the combination of a bulge, disk, and
possible point source (Figure \ref{fig:mbh_l+m_bas}, left). These fits
represent an upper limit on the bulge component, and are also a good
analog to fits in the literature to the SDSS
\citep[e.g.,][]{Lackner+Gunn12} and to higher-redshift galaxies
\citep{Bell+12}. The expected range of $H$-band bulge luminosities for
the megamaser disks, based on the scaling relations, is $10^8$ to
$10^{10}~L_{\odot}$. The observed range, in contrast, is much narrower
($3 \times 10^9$ to $4 \times 10^{10}~L_{\odot}$). If we measure the
average offset in $\mbh$ between the megamaser disks and the best-fit
L14 relation ($\Delta a_{\rm mega}$ in Table 8, row 2), we find a mean
offset of $-0.8\pm0.2\dex$ in BH mass from the best-fit relation.

These galaxies tend to have recent or ongoing star
formation in their nuclei, biasing the observed bulge luminosities to
high values compared with the predominantly old stellar populations
that dominate early-type galaxies. We attempt to mitigate these
differences by transforming to stellar mass.  We try to
put both the literature and megamaser galaxies on the same stellar
mass scale to facilitate direct comparison (\S
\ref{subsec:literature_values}). When considering stellar mass rather
than luminosity (Figure \ref{fig:mbh_l+m_bas}, right; 
$\Delta a_{\rm mega}$ in Table 8, row 5), we still see
that the megamaser disks remain offset to smaller BH masses at a given
bulge mass ($-0.8\pm0.2\dex$).

If we instead examine the offset between the best-fit L14 relation,
and the spiral galaxy subset rather than the megamaser subset, the net
offset $\mbh$ declines a bit, $\Delta a_{\rm late}=-0.6\pm0.2$ dex.
The fact that the measured offset $\Delta a_{\rm late}$ is smaller
than $\Delta a_{\rm mega}$ is interesting, and indicates a possible
difference between the distribution of $\mbh$ for megamaser disks and
galaxies with stellar/gas dynamical BH mass measurements.

In summary, when we consider the most general concept of ``bulge'' as
the centrally concentrated component that constitutes a light excess
above the large-scale disk, we find a wide range of BH mass at fixed
bulge mass and a significant offset from the early-type $\mbh-\mbul$
relation for the megamaser disks and late-type galaxies in
general. A qualitatively similar result has been seen before in
late-type galaxies for the subset of pseudobulge galaxies
\citep{Hu2008, Greene+08, S11, Kormendy+Ho13}.

\subsection{The $\mbh-M_{\rm bulge}$ relation for ``classical'' bulge components}
\label{subsec:mbh-mbul}

\begin{figure*}
 \centering
 \includegraphics[width=8.5cm]{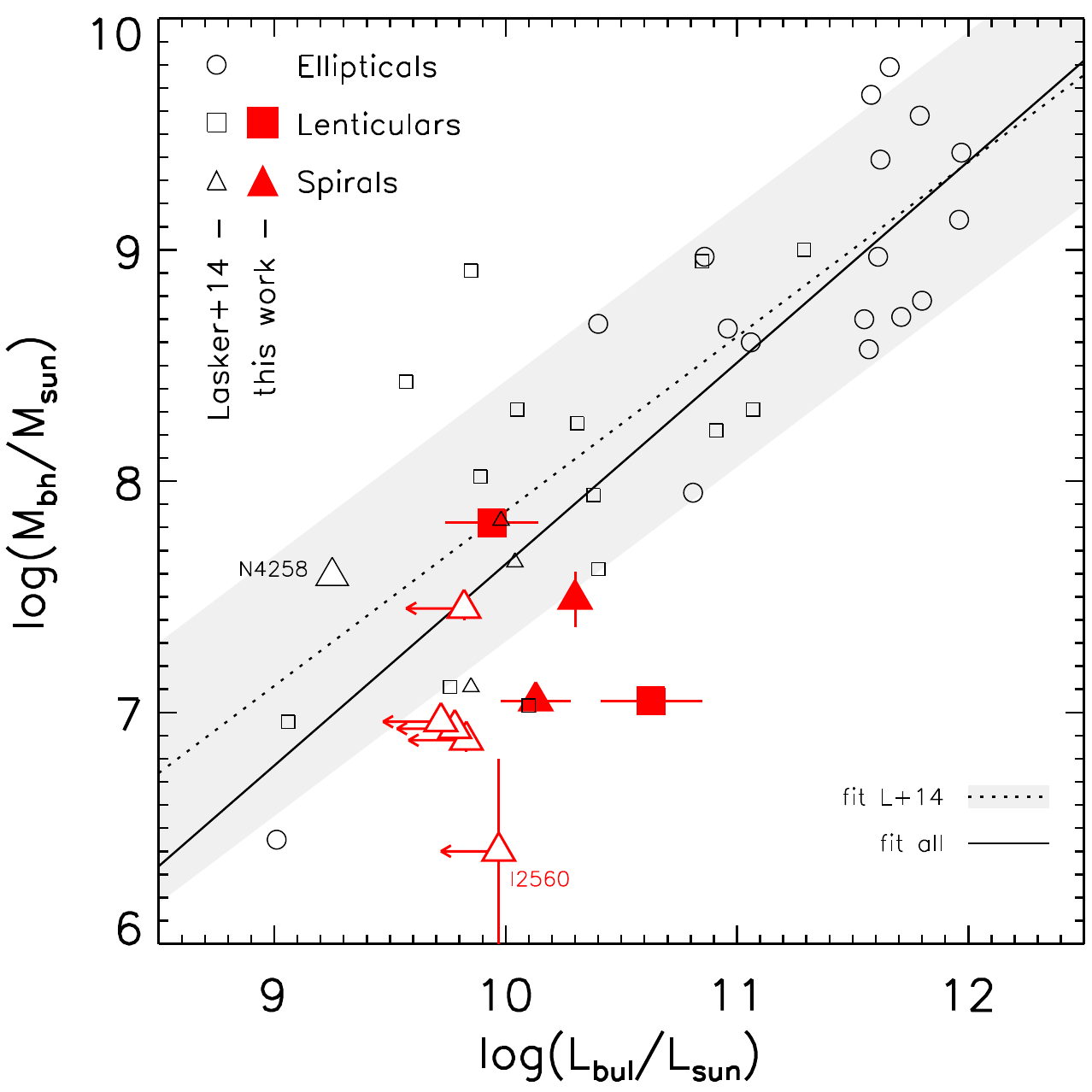}
 \hfill
 \includegraphics[width=8.5cm]{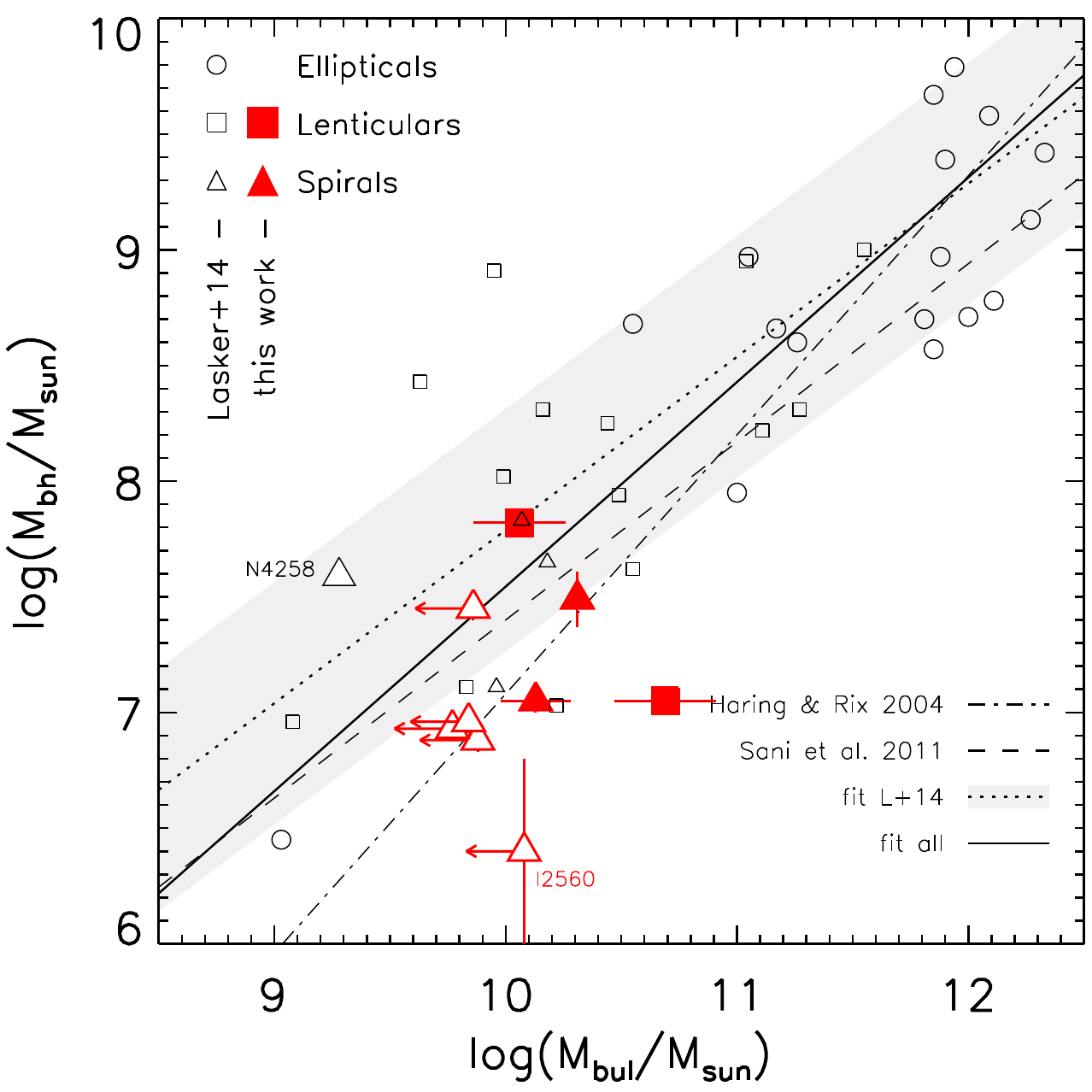}
\caption{Same as Figure \ref{fig:mbh_l+m_bas}, but using the classical bulge
parameters from our adopted multi-component decompositions for the
megamaser hosts (filled red symbols with error
bars). In cases where we do not identify a classical bulge, we plot an 
open symbol. The megamaser disk galaxy NGC4258 is 
indicated as an open triangle.
The fits and the literature sample
\protect\citep{Lasker+14a} are the same as in Figure
\ref{fig:mbh_l+m_bas}. Using the more detailed decompositions 
moves the megamasers
closer to the relation of the L+14 sample, but
only slightly so. Similarly, the conversion to mass reduces the
scatter marginally. The megamaser galaxies also appear to have lower
$\mbh$ at a given $L\bul$ or $M\bul$ than the general BH host
population, while their scatter is similar to other galaxies in the
low-mass regime they occupy.}
\label{fig:mbh_l+m_ref}
\end{figure*}

In \S 4 we described in detail our attempt to isolate a classical
bulge component in the maser disk galaxies. Considering the colors,
shapes, and structures of the putative ``classical'' bulge components,
we identified a classical bulge in four cases (NGC1194, NGC2960,
NGC3393, and UGC3789).

We compare the BH and stellar mass for the confirmed classical bulges
(solid symbols) in Figure \ref{fig:mbh_l+m_ref}. Obviously, these
classical bulge components will, by construction, contain less stellar
mass than the basic bulges (typically by $\sim 0.2\dex$), nominally
improving the agreement with the elliptical galaxy $\mbh-\mbul$
relation. We indeed see better agreement overall when plotting $\mbh$
against classical rather than basic bulge measurement: the offset in
$\mbh$ is $\Delta a_{\rm late}=-0.6\pm0.2\dex$. However, if we focus
only on those galaxies where we believe there is a secure classical
bulge component (filled symbols in Fig. \ref{fig:mbh_l+m_ref}), we see
that there is still a net offset towards lower $\mbh$ at a given bulge
mass. Thus, we tentatively conclude that simply identifying more
robust classical bulges in these galaxies will not eradicate the trend
towards higher scatter and lower $\mbh$ at a given bulge mass. 
As noted by \cite{S11} and \cite{Lasker+14a}, it is still difficult
to definitively rule out that observational issues (e.g., hidden 
nuclear star clusters or other small-scale components) are contaminating 
our measurements, but as the number of galaxies with \hst\ data 
and high-fidelity dynamical BH masses increase, it becomes more and more 
clear that there is simply a very wide range of $\mbh$ in these low-mass 
galaxies.

\subsection{$\mbh-M_{\rm tot}$ relation}
\label{subsec:mbh-mtot}

\begin{figure*}
 \centering
 \includegraphics[width=8.5cm]{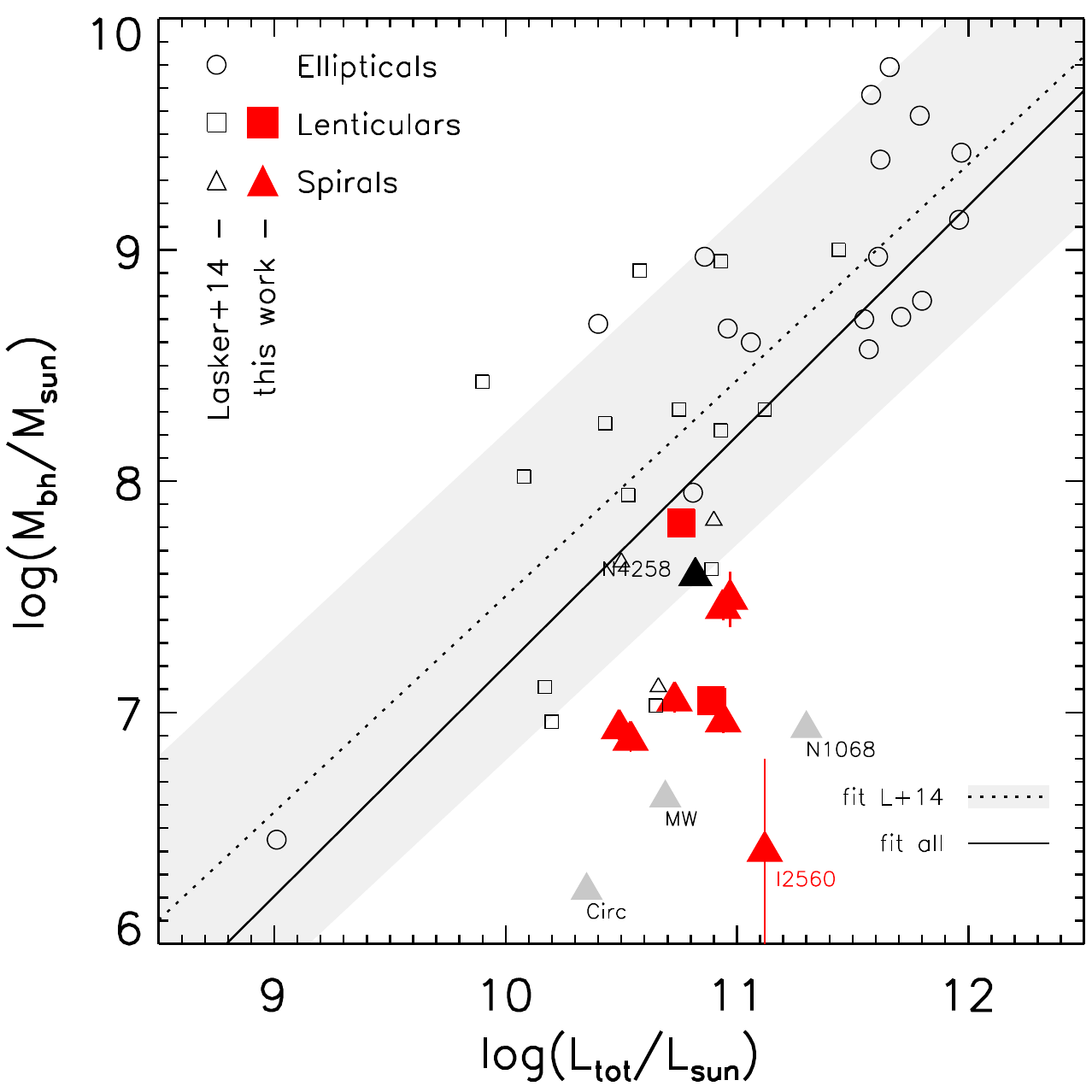}
 \hfill
 \includegraphics[width=8.5cm]{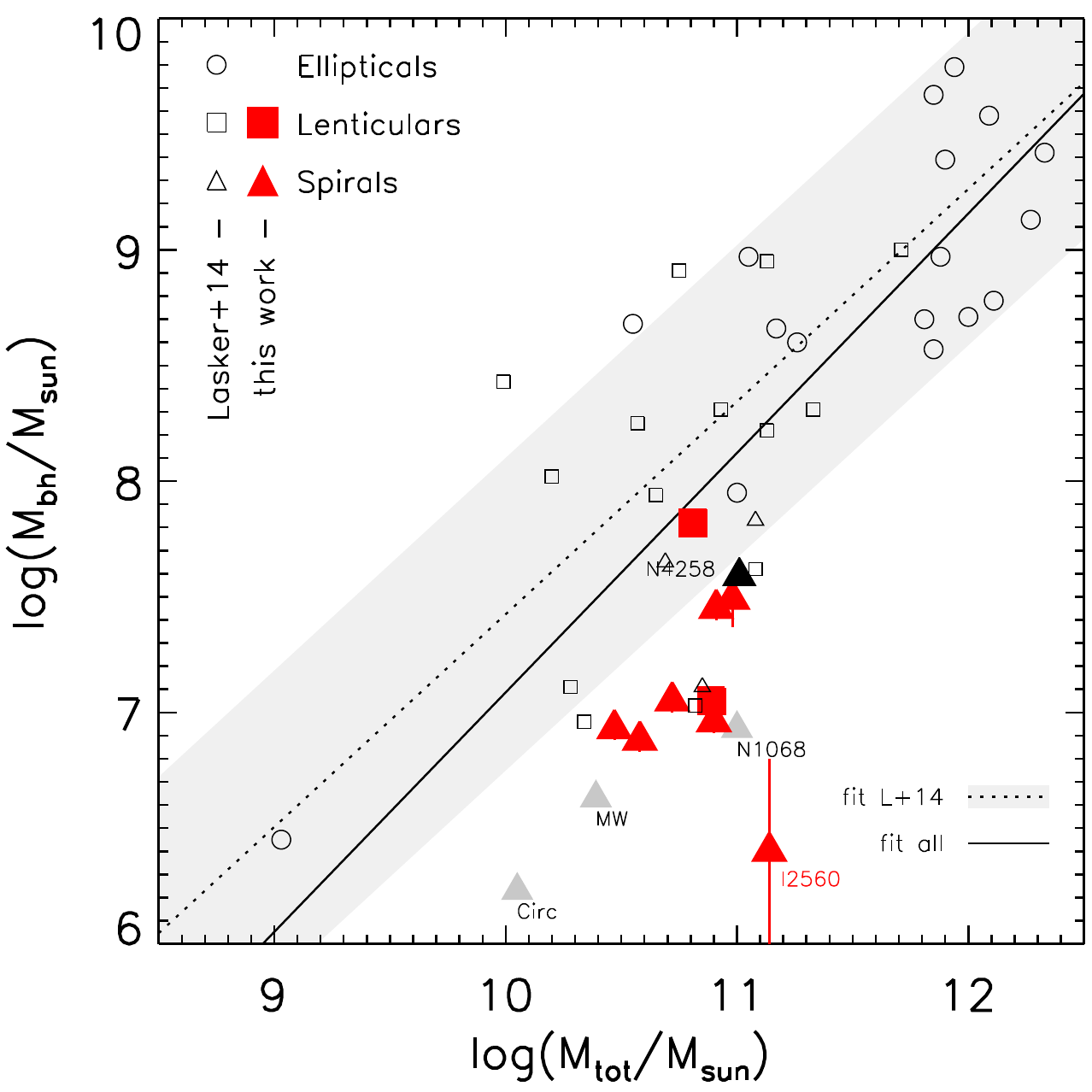}
\caption{Correlation of log $\mbh$ with total $\log L_{H,\mathrm{tot}}$ (left
panel) and $\log M\tot$ (right panel), for the megamaser BH hosts
(large filled symbols, red for this study, NGC4258 in black) and for
the L14 sample of BH hosts with $\mbh$ from stellar or gas kinematics
(small symbols). Symbols shapes (Hubble type) and lines (relation
fits) are defined as in Figures \ref{fig:mbh_l+m_bas} and
\ref{fig:mbh_l+m_ref}. As for the relation with bulges, the megamaser
hosts reside at lower $\mbh$ than predicted by the fit (dotted line)
to the predominantly early-type BH host galaxy sample (open symbols)
of \protect\cite{Lasker+14a}.
}
\label{fig:mbh_l+m_tot}
\end{figure*}

There has been considerable interest in recent years in the relation
between total stellar mass and $\mbh$, with various claims that total
stellar mass also should show a relationship with $\mbh$
\citep{Peng07}, possibly even tighter than bulge mass
\citep{Jahnke+Maccio11}, and one that does not evolve with redshift
\citep{Cisternas+11,Jahnke+09,Bennert+10,Bennert+11b}. We examine the
$\mbh-M_{\rm tot}$ relation using the L14 and our megamaser samples in
Figure \ref{fig:mbh_l+m_tot}. Focusing on the stellar mass range where
the megamasers are found, we see here the most striking mismatch
between the range in $\mbh$ ($2.5\dex$) and the range in stellar mass
($0.4\dex$). Thus, at a fixed stellar mass, galaxies may contain BHs
with a wide range of mass. We do not see strong evidence that the
total stellar mass to $\mbh$ relation is tighter than others in the
literature, and we see virtually no correlation at all below
$M_*\approx 10^{11}\,\msun$. Regardless, $\eps(\mbh-\mtot)$ is still marginally consistent with $\eps(\mbh-\mbul)$ \citep[cf.][]{Lasker+14b}. We measure an offset between the
megamasers and the best-fit L14 relation of $\Delta a = -0.8 \pm 0.2$ dex, similar to the apparent offset between active and quiescent BHs observed by \cite{Reines+Volonteri15}.

\section{Discussion and Summary}
\label{sec:disc+sum}

With the megamaser disk galaxies, we have a sample of $\sim L^*$
spiral galaxies with very precise BH mass
measurements. The masers allow us to explore BH demographics in spiral
galaxies more robustly than any other dynamical method. Firstly, we
can spatially resolve the spheres of influence of much lower-mass BHs,
allowing us to probe the full range of $\mbh$ at fixed galaxy
property. Secondly, the maser disks are not impacted by the dust and
mixed stellar stellar populations that challenge stellar and gas
dynamical techniques \citep[although see also][]{denBrok+15}.

\subsection{The Role of Bias in Different BH Samples}

The megamaser disk galaxies span $1.5\dex$ in BH mass
but only $0.6\dex$ in galaxy mass. Furthermore, the
ratio of $\mbh / M_*$ is, on average, considerably lower than what
is seen in more massive elliptical and S0 galaxies. This long tail to
low $\mbh$ at fixed stellar mass is seen most conclusively in 
the maser disk samples, likely due to the
difficulties of resolving the gravitational sphere of influence for
low-mass BHs \citep{Batcheldor10,Gultekin+11a,vdBosch+15}. 
In the stellar mass range probed by the maser disks 
($M_{\ast} < 10^{11}\msun$), we see a hint 
in this paper that stellar dynamical, gas dynamical, 
and reverberation-mapped samples do not truly sample the full range 
of $\mbh$ at a given $M_\ast$. In particular, only the megamaser disk 
galaxies extend to the lowest $\mbh$ probed at a fixed galaxy 
mass. In the case of 
the stellar and gas dynamical measurements, it is not 
surprising that the BHs do not sample the low-mass regime, since 
we cannot resolve their sphere of influence.  Thus, we argue 
that only the megamaser disks reveal the true distribution 
of $\mbh$ at a given galaxy property.

It is more difficult to pin down the origin of the difference between
the maser and the reverberation mapped sources. One possibility is a
bias in the reverberation-based BH masses, which after all have been
calibrated to follow the inactive \msigma\ relation. Alternatively,
the reverberation sources may also have a bias toward higher $\mbh$ at
a given galaxy mass, due to the preferential selection of the most
luminous sources with correspondingly high mass BHs. Until we
come to a full understanding of this issue, it will be difficult to
fully compare the scaling relations for single-epoch virial BH
masses with dynamical BH masses (e.g., Graham \& Scott 2015; Reines
\& Volonteri 2015).

Alternatively, the megamaser disk sample could have a bias towards
lower $\mbh$ at fixed galaxy property due to their selection as active
galaxies if, for some reason, the megamaser disks pick out galaxies
that are preferentially growing towards the end state of the
\msigma\ relation.  The megamaser disks are nearly the only active
galaxies with dynamical BH masses, so it is worth considering the
possibility that masers select a non-representative sample.  We argue
against that possibility, repeating the arguments in 
\citet{Greene+10b}. The galaxies are found,
on average, a factor of four below the relation between $\mbh$ and
$L_{\rm bulge}$. To erase this offset, at their current Eddington
ratios of $\sim 10\%$ \citep{Greene+10b}, would require $\sim 1$~Gyr
of steady BH growth. On the other hand, typical lifetimes of AGN are
likely shorter than this \citep{Martini+Weinberg01} while pseudobulge
growth times are much longer than this \citep{Kormendy+Kennicutt04},
so how the megamaser disks would know to grow at this particular
moment is difficult to understand. Furthermore, if such a bias
impacted the maser galaxies, we would expect to see the same effect in
the reverberation-mapped sources, which we do not.

In principle, it is also possible that megamaser disk galaxies are
biased against the most massive BHs \citep{Reines+Volonteri15,vdBosch+16}. However, if there were a large sample of spiral
galaxies with very massive BHs ($\gtrsim 10^8\,\msun$) we would likely
know about them already from stellar- or gas-dynamical $\mbh$ measurements.

\subsection{Differences in Scaling Relations}

Barring such biases, megamaser disks enable us to probe the underlying
distribution of BH mass for spiral galaxies. The observations
presented here apply (at minimum) to all BHs in spiral galaxies with
$M_{\ast} < 10^{11}\msun$. That is, at a fixed galaxy property, there
is a large range of $\mbh$, extending systematically below the relations
defined by the early-type, massive galaxies.

We consider two explanations for the differences in scaling relations
between the masers and early-type galaxies. One is that the scaling
relations vary with galaxy morphology. In this picture, the formation
history of the galaxy (e.g., the formation of a massive bulge) is tied
to the fueling and feedback processes of the growing BH. For instance,
galaxies that build their central bulges primarily with
secular processes may never efficiently fuel their BHs
\citep[e.g.,][]{Hu2008,Greene+08,Gadotti08}. This would result in the
observed offset to lower masses, along with significant scatter,
preferentially among spirals.  To test these possibilities, we
urgently need BH mass measurements in more bulge-dominated low-mass
galaxies like M32. We also note the intruiging possibility raised by 
\cite{Saglia+16} that at high enough mass density, the BHs in pseudobulges obey 
the same scaling relations

The other possibility is that the BH scaling relations are driven by
merging via the central limit theorem, as advocated by \cite{Peng07}
and \cite{Jahnke+Maccio11}.  In this picture, galaxies with fewer
mergers do not converge to a tight scaling relation, leading to a
strong dependence of scatter on galaxy (or halo) mass.  If we could
measure the scatter in the scaling relations as a function of mass and
morphology over the full range of $\mbh$, we could determine which of
these scenarios is preferred. Unfortunately, we still have very few
measurements at low mass, and vanishingly few in low-mass early-type
galaxies (M32, \citealt{vdBosch+deZeeuw10}; NGC404, \citealt{Seth+10},
Nguyen et al. in prep.; NGC 4395 \citealt{denBrok+15}). It remains
very challenging to distinguish these two possibilities.

Finally, there is the possibility that the scaling relations are
completely artificial and actually just define an upper envelope that
arises due to problems with the stellar and gas-dynamical methods.
Others have considered this possibility
\citep{Batcheldor10,Gultekin+11a, vdBosch+15}. In the coming decade,
ALMA \citep{Davis14} and 30m-class telescopes \citep{Do+14} will
provide an order of magnitude increase in angular resolution, allowing
us to test this possibility. What we really need are megamaser disks
in early-type galaxies. Searches thus far have not been successful
\citep{vdBosch+16}, but they are worth continuing.

Despite significantly increasing the sample of $L^*$ spiral galaxies
with dynamical $\mbh$ measurements, our samples at low mass and
late-type morphology remain small. Thus, our conclusions are not yet
definitive. While we did our best to identify
classical bulges with photometric indicators, a combination of
photometry and kinematics would undoubtedly work better
\citep[e.g.,][]{Erwin+15}. Luckily, there are additional
megamaser disk galaxies with Keplerian rotation curves and 
secure $\mbh$ being observed with \hst\ in Cycle 22 (P.I. Greene). We 
are also securing AO-assisted integral-field observations of the 
stellar and gas kinematics in five of these objects with SINFONI 
on the VLT (P.I. Greene). The combination of these two data sets should 
prove powerful in setting our results on firmer ground. 


\section*{Acknowledgments}
\label{sec:acknowledgments}

We thank Stefano Zibetti for providing his $(g-i,i-H)$ mass-to-light ratio tables. RL thanks Arjen van der Wel for advice on reconstructing the WFC3 PSF, as well as Remco van den Bosch and Akin Yildirim for useful discussions. We are also grateful to Chien Y. Peng for his substantial advice on our analysis, and to Dimitri Gadotti, Alister Graham, Kayhan G\"ultekin, John Kormendy, Amy Reines and Marta Volonteri for their comments on the manuscript.
JEG was partially supported by NSF grant AAG:1310405.
Based on observations obtained with the Apache Point Observatory 3.5-meter telescope, which is owned and operated by the Astrophysical Research Consortium; on observations at Kitt Peak National Observatory, National Optical Astronomy Observatory (NOAO Prop. ID: 2011A-0170; PI: J. Greene), which is operated by the Association of Universities for Research in Astronomy (AURA) under a cooperative agreement with the National Science Foundation; and the Las Campanas Observatory 100-inch du Punt telescope, operated by the Carnegie Insitution for Science (CIS).
Funding for the NASA-Sloan Atlas has been provided by the NASA Astrophysics Data Analysis Program (08-ADP08-0072) and the NSF (AST-1211644). This research has made use of the NASA/IPAC Extragalactic Database (NED) which is operated by the Jet Propulsion Laboratory, California Institute of Technology, under contract with the National Aeronautics and Space Administration.




\appendix

\section{Detailed notes on galaxy decompositions}
\label{sec:detailed_decomp}

\subsection{IC 2560}

\begin{figure*}
  \centering
  \includegraphics[width=15cm]{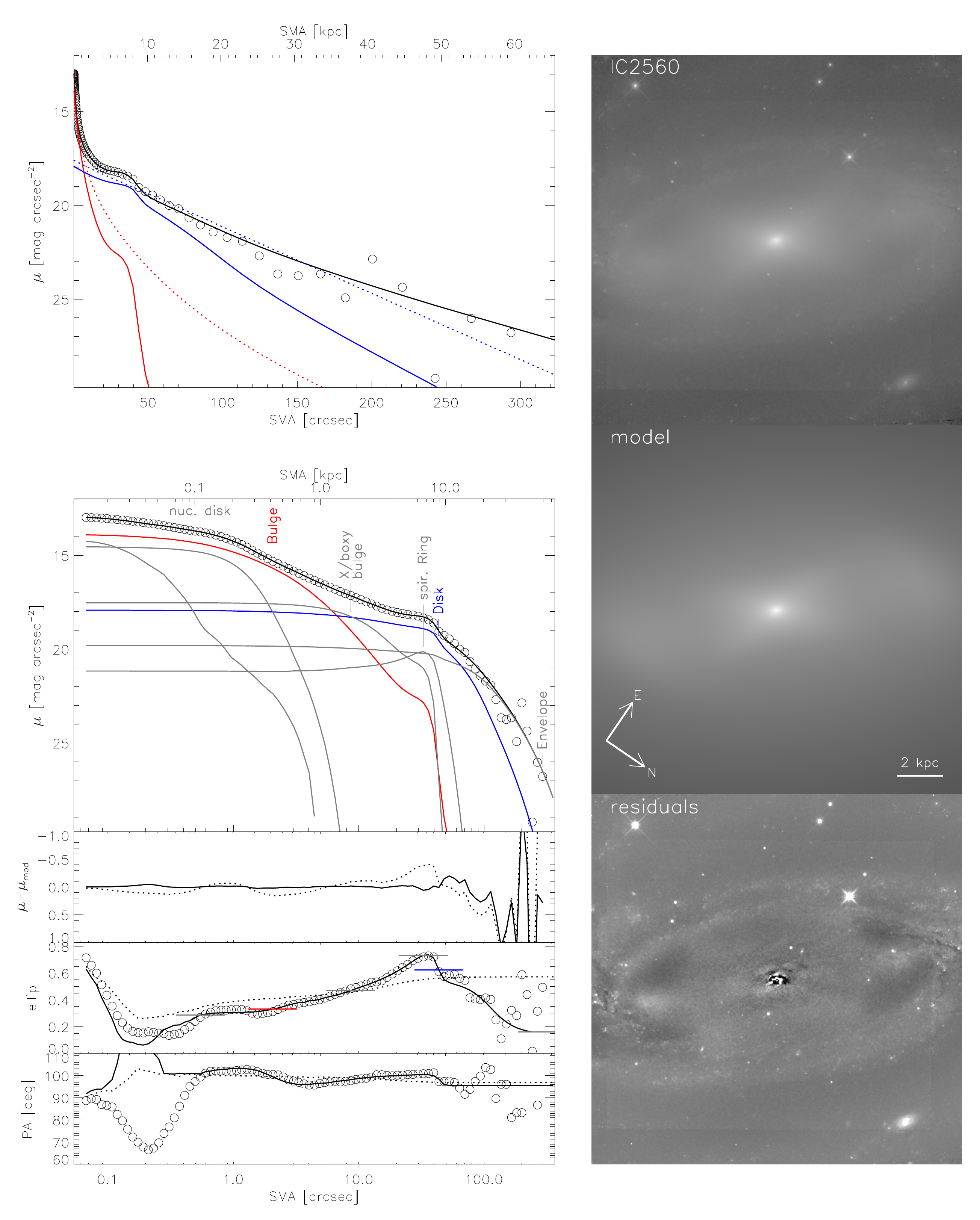}
  \caption{
IC2560 photometric data and model. \textit{Left panels:} semi-major
axis (SMA) profiles of $H$-band surface brightness ($\mu$), data-model
residuals ($\mu-\mu_\mathrm{mod}$), ellipticity (ellip), and
east-of-north position angle (PA). Open circles show the observed
data, and solid lines the full model (thick black), its bulge (red),
main exponential disk (blue), and all other model components (thin
grey lines). For comparison with our best-fit model, shown by
the dashed lines are total (black), bulge (red) and disk (blue)
profiles of the ``basic model'', which includes only a bulge and disk
(for IC2560, the point source could not be fitted in the basic model
as the best-fit bulge+disk model overpredicts the central flux). The
top-left panel plots $mu$ against linear SMA and shows the basic model
bulge and disk, while the lower panels use a logarithmic SMA scaling
and omit the basic model components for better visibility of the full
model's multiple components. See Figure \ref{fig:IC2560-2}, for the
basic model components' $\mu-\log(\text{SMA})$ profiles. The names and
ellipticities (horizontal bars) of the full model's components are
indicated at the SMA distances where the components' contribution to
the total flux is maximal. In IC2560's full model, the envelope
accounts for the strongly increased flattening at $\gtrsim 20\kpc$,
the spiral arms are tightly wound and hence modeled by a ring with
inner truncation, and the X-shaped pseudobulge by a \sersic\ with
best-fit index $n=0.5$ and a 4th-order isophote harmonic. {\it Right
  panels, from top to bottom:} the image data and full model on a
logarithmic greyscale, and full model residuals on a linear
greyscale. Evident is a residual spiral structure in the innermost
regions, presumably from an unmodeled nuclear disk.}
  \label{fig:IC2560-1}
\end{figure*}

\begin{figure*}
  \centering
  \includegraphics[width=17cm]{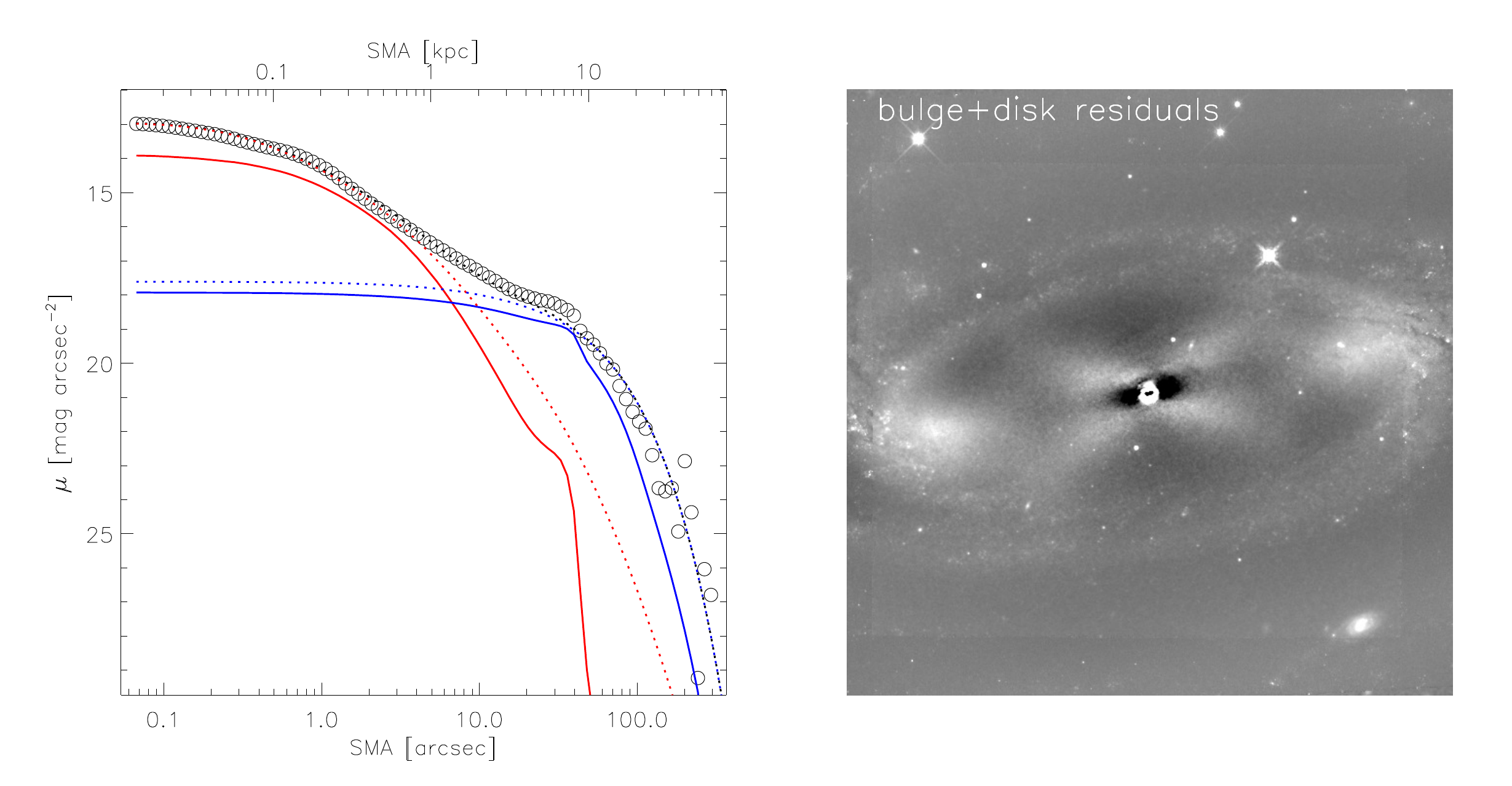}
  \caption{
IC2560 photometric data and model, continued from Figure
\ref{fig:IC2560-1}. {\it Left panel:} logarithmic semimajor-axis (SMA)
$H$-band surface brightness ($\mu$) of the data (open circles), full
model (solid lines) and basic bulge+disk model (dashed). The black
lines represent the total model $\mu$, red the bulge, and blue the
disk contribution. The other components of the full model are not
shown (but see Figure \ref{fig:IC2560-1}. {\it Right panel:}
data-model residuals of the basic (bulge+disk) model. The bright onset
of the spiral arms is unaccounted for, as well as the X-shaped
pseudobulge. The central brightness is overpredicted by the basic
model, and therefore the point source cannot be fitted before the
mentioned components are included in a full model.}
  \label{fig:IC2560-2}
\end{figure*}

IC2560 (Figures \ref{fig:IC2560-1} and \ref{fig:IC2560-2}) contains a
large-scale disk with well-defined spiral arms at intermediate
radii. The disk is detectable out to $\sim 150''$ ($30\kpc$) along the
major axis (with PA$ = -50\deg$) and has an axis ratio of $\sim0.6$;
it grows rounder at large radius. The tightly-wound spiral arms range
from 6-12 kpc ($30-60"$) and broadly resemble a ring. 
The central X-shaped bulge/bar dominates the light inside
$\sim10''$ ($2~\kpc$). Finally, the surface-brightness
profile exhibits a ``knee'' at $1''$ ($0.2~\kpc$) that corresponds to 
a nuclear disk. 

The spiral ring and X-shaped bulge/bar are apparent in the residuals 
to the basic bulge+disk model (Figure \ref{fig:IC2560-2}). Thus, 
in our more complex fit, we introduce one component
for the ring and one for the X-shaped bar/bulge. We also include a
large-scale (extended) envelope, with a best-fit exponential scale
radius of $47''$ ($9.5~\kpc$) to accommodate the obviously rounder
outer component. Our best-fit model (Figure
\ref{fig:IC2560-1}) also includes a faint nuclear disk that is exposed in the residual image and indicated by profile inflections around
$1''$. Although the formal $\chi^2$ improves only marginally when the disk
is added, we see spiral structure associated with the nuclear disk in the 
residual image, and the disk model
perfectly fits the apparent knee in the $\mu$-profile at $\text{SMA}\sim1''$
($0.2\kpc$). The bulge component changes when the nuclear disk is
included, growing by $1 \arcsec$ (a factor of two) and $0.34\mg$, as
it is no longer trying to fit the very compact nuclear disk. 

In the best-fit model, we fix the \sersic\ index of the ring, envelope
and nuclear disk to 1, 1, and 0.5 respectively.  The ring is modified
by an inner truncation. The X-shape is modeled by a 4th-order Fourier
mode. Our estimate for the systematic (modeling) error of the
classical bulge magnitude is $0.4\mg$, which we derive by considering
the difference between the best-fit values from our reference model
and several alternative models: allowing a free \sersic\ index for
first the envelope (bulge magnitude unchanged,
$n_\mathrm{env}\rightarrow0.6$), then the nuclear disk ($+0.3\mg$,
$n_\mathrm{nuc.disk}\rightarrow0.8$), and finally the main disk
($-0.4\mg$, $n_\mathrm{d}\rightarrow0.2$). We also construct
  alternative models by omitting from the model the nuclear disk
(bulge $+0.3\mg$), or omit the X-shaped pseudobulge ($-0.6\mg$).

Our high resolution data allows us to cleanly separate the
  X-shaped bar/bulge from from a nuclear disk and the small, round
  (possibly classical) bulge. In the low-resolution and shallower
  \emph{Spitzer} data of \cite{S11}, they are clearly fitting the disk
  and envelope together, since their disk is rounder than their bulge
  component. Likewise, their bulge component contains multiple
  components. In net, they find a B/T of about 0.5, i.e. significantly
  higher than even the B/T of 0.18 in our basic model.

\subsection{NGC 1194}

\begin{figure*}
  \centering
  \includegraphics[width=15cm]{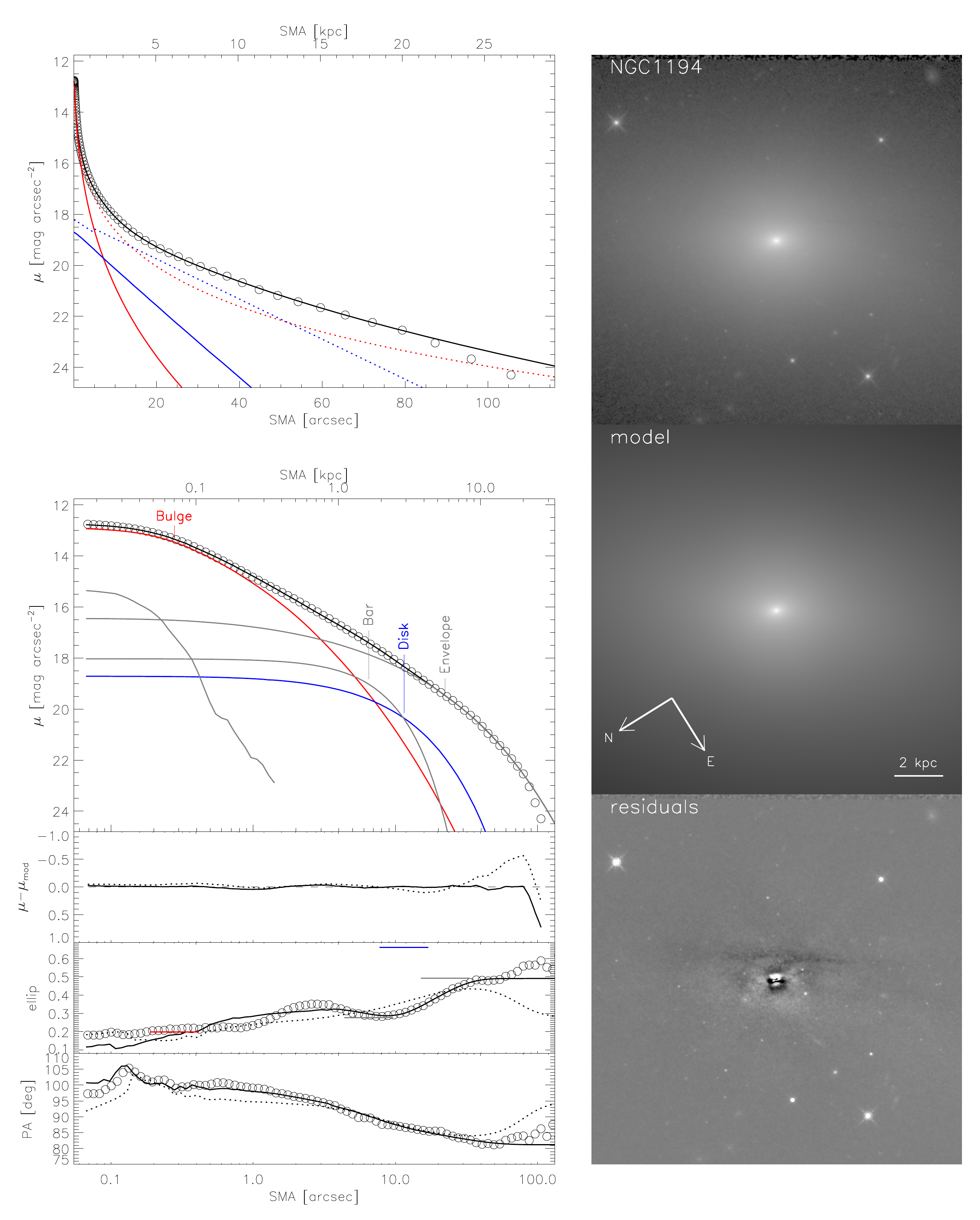}
  \caption{NGC1194 photometric data and model, with layout as in Figure
  \ref{fig:IC2560-1}. \textit{Left panels:} semi-major axis (SMA)
  profiles of $H$-band surface brightness ($\mu$), data-model
  residuals ($\mu-\mu_\mathrm{mod}$), ellipticity (ellip), and
  east-of-north position angle (PA). Open circles: data, solid lines:
  full model, dashed lines: basic (bulge+disk) model. Thick black:
  total model image profiles, red: bulge, blue: disk, and thin grey:
  all other components. Only select profiles are shown in the
  $\mu-\text{SMA}$ (top panel) and $\mu-\log\text{SMA}$ (second from
  top) plots (see also Figure \ref{fig:NGC1194-2}). Ellipticities of
  individual components are indicated by horizontal bars. {\it Right
  panels:} Images of the data, model and residuals. In the full model,
  the envelope component is required to allow the point source to be
  fitted; otherwise, the bulge is too bright in the center with a
  comparatively high ($n\sim7$) \sersic\ index. It also provides the
  higher flattening in the outer parts compared to the bulge, which
  otherwise (in the basic model) dominates the
  light at $\gtrsim 10\kpc$. The intermediate-scale ($\sim 2\kpc$)
  component is tentatively termed ``bar'' here for its compact profile
  (\sersic\ $n\sim0.8)$ and $\sim90\deg$ PA offset from the disk major
  axis. It is strong enough to be required for fitting the disk and
  envelope separately.}  \label{fig:NGC1194-1} \end{figure*}
\begin{figure*}
  \centering
  \includegraphics[width=17cm]{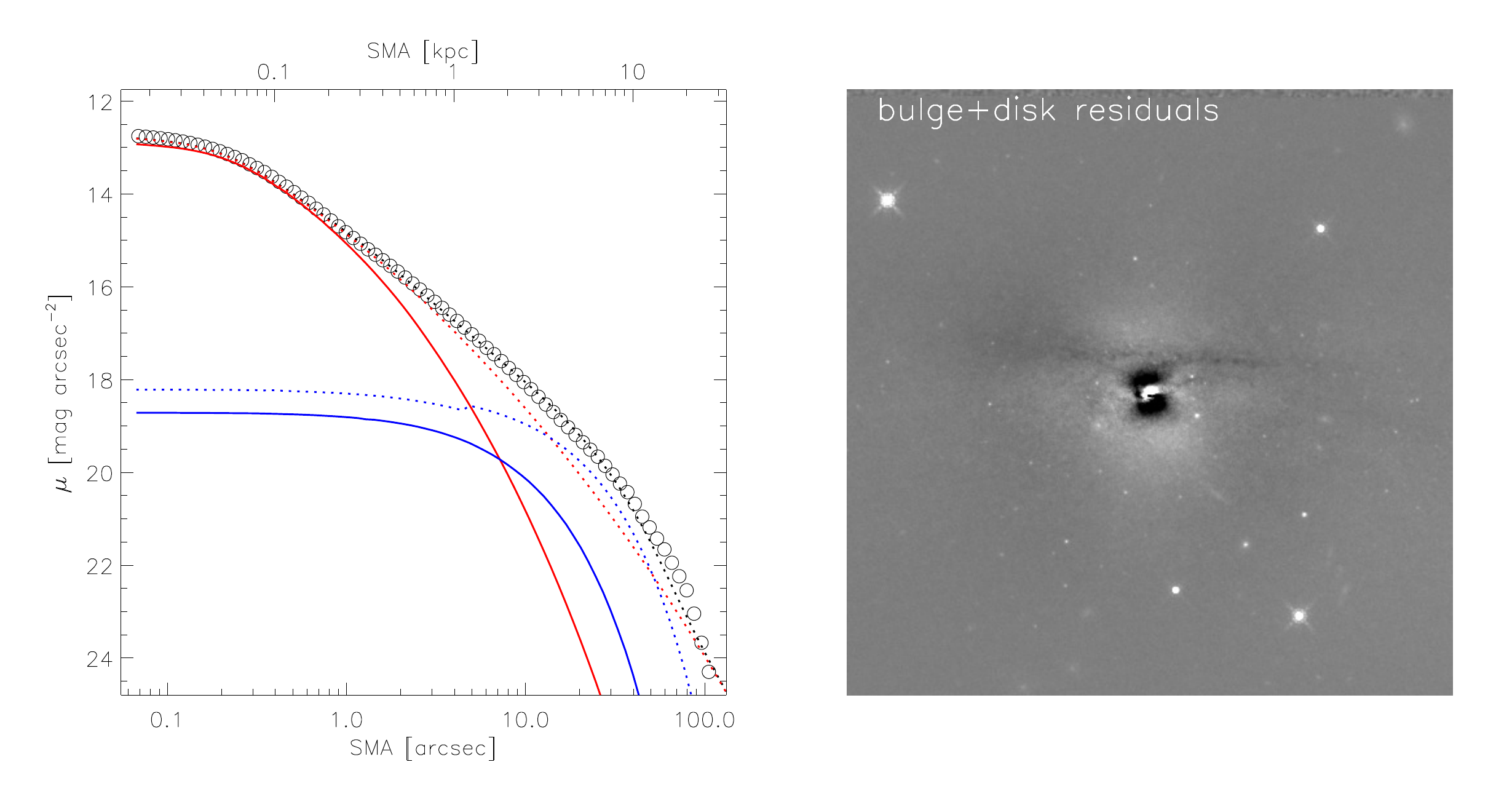}
  \caption{NGC1194 photometric data and model, continued from Figure \ref{fig:NGC1194-1}. {\it Left panel:} SMA surface brightness ($\mu$) of the data (open circles), full model (solid lines) and basic bulge+disk model (dashed), separately for total light (black), bulge (red) and disk (blue). {\it Right panel:} image of the basic model residuals. In this basic model, which in particular does not include the envelope component, the bulge outer profile is too extended and does not follow the downturn (``knee'') at $\sim 80\arcsec$ ($\sim 20\kpc$). The excess brightness of the compact ``bar'', which is slightly elongated along the minor axis (here: vertical orientation) can also clearly be spotted in the basic model residuals, as well as the dust lane parallel to the major axis.}
  \label{fig:NGC1194-2}
\end{figure*}

NGC1194 (Figures \ref{fig:NGC1194-1} and \ref{fig:NGC1194-2}) appears
to be an S0 seen at relatively high inclination. The innermost region
($\sim1''$ nucleus) is round, bright, and distinct from
the flatter ($q\sim0.5$) outer regions, which can be visually traced 
to about $90''$ ($20\kpc$). The ellipticity profile features a local
peak at $\sim3''$ ($\sim0.7\kpc$), followed by a trough at $\sim8''$
($2\kpc$). Subtracting the basic bulge+disk model additionally reveals
a dust lane that is parallel to the major axis but offset 
by $\sim4''$ to the south-west, as well as a minor-axis boxy light excess 
at $\sim 6'' / 1.5\kpc$. 

Turning to the basic model first, we see that there is no need for an
AGN component as the central light is already slightly overestimated 
by the bulge+disk model.  Out to $\sim 1''$ a major-axis excess in residual
light may indicate an edge-on nuclear disk, while on larger
scales (''$\sim6''$; Figure \ref{fig:NGC1194-2}) we may be seeing the
residuals of an end-on bar. There is a local ellipticity peak
$\sim40''$ likely pointing to a disk component. At large radius, the
basic model fits the extended light profile with the bulge (leading to
a very high \sersic\ index $n=6.8$), but the best-fit
bulge is too round to fit the outer component properly and leads to a
over-extended profile at large radii compared with the profile in the
data.

Our adopted model includes a bar-like component and an "envelope", in
addition to a bulge, disk, and point source. The envelope, which has a
best-fit \sersic\ profile with $n=2$, reduces the flux excess and axis
ratio at the largest scales ($\gtrsim100''$) compared to the
  basic bulge+disk model, as well as reducing the bulge
\sersic\ index ($n=3.2$). The envelope is intermediate in flattening
between bulge and disk: $q=0.5$, versus $0.7$ and $0.4$,
respectively. This outer component is as flattened as the inner disk,
and carries a large share of the total flux ($\sim 75\%$), so we
surmise that it probably represents a large-scale thick
disk. Supporting this interpretation, there is also a large HI-disk in
NGC1194 \citep{Sun+13}.  The bar-like component fits the apparent
minor axis excess on $\sim 10''$ scales. An additional
$0.3\arcsec$($80\pc$)-sized nuclear component can be fit and slightly
improves residuals, but we refrain from including it in our adopted
model due to the complicated and dust obscured center. However, we
retain such a model with nuclear component as an alternative that
serves to estimate the systematic modeling errors of the model
parameters.

Turning to the systematic errors, we find that the basic bulge is
$1.8\mg$ brighter than the bulge component in our adopted model. If we
remove just the envelope from the adopted model, we obtain a $1.6\mg$
brighter bulge. Interestingly, if we {\it add} to the adopted model
the putative nuclear component, while constraining the envelope
\sersic\ index to $n=1$ to reduce degeneracy, we again get a
significantly {\it brighter} bulge than in our adopted model (by
$0.8\mg$). Conversely, the bulge parameters barely change when we fit
the large-scale disk with a \sersic\ profile ($0.1\mg$ fainter bulge,
and $n_\mathrm{disk} \rightarrow 0.8$). Finally, we explored
  whether masking of dust features (central, and particularly the lane
  parallel to the major axis) alters our results and found virtually
  no difference in parameters when applying the mask. In summary,
these alternative models indicate a systematic bulge magnitude
uncertainty of $0.8\mg$.

\subsection{NGC 2273}

\begin{figure*}
  \centering
  \includegraphics[width=15cm]{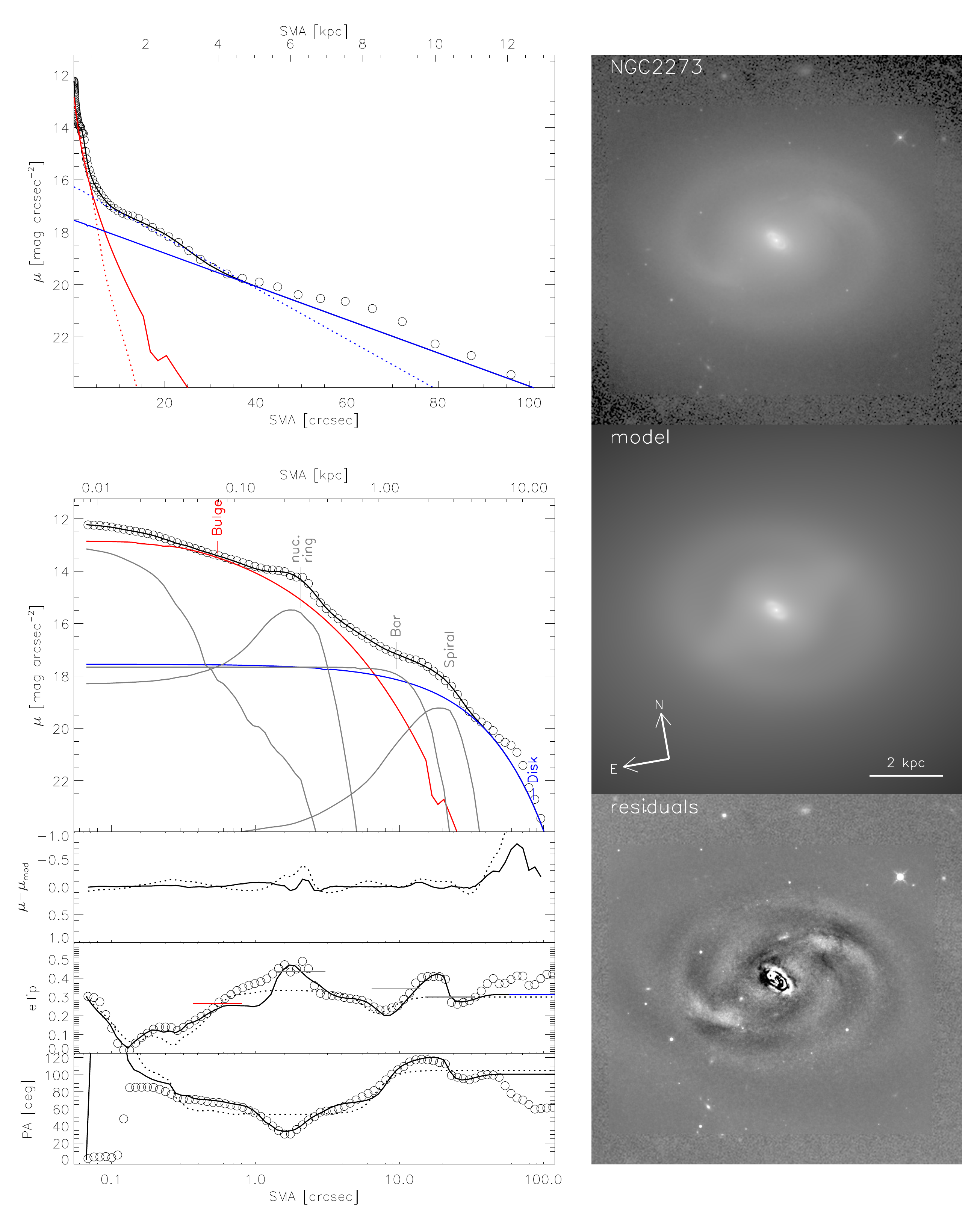}
  \caption{NGC2273 photometric data and model, with layout as in Figure
\ref{fig:IC2560-1}. \textit{Left panels:} semi-major axis (SMA)
profiles of $H$-band surface brightness ($\mu$), data-model residuals
($\mu-\mu_\mathrm{mod}$), ellipticity (ellip), and east-of-north
position angle (PA). Open circles: data, solid lines: full model,
dashed lines: basic (bulge+disk) model. Thick black: total model image
profiles, red: bulge, blue: disk, and thin grey: all other
components. Only select profiles are shown in the $\mu-\text{SMA}$
(top panel) and $\mu-\log\text{SMA}$ (second from top) plots (see also
Figure \ref{fig:NGC2273-2}). Ellipticities of individual components
are indicated by horizontal bars. {\it Right panels:} Images of the
data, model, and residuals. The full model traces the data much better
than the basic model, in particular regarding ellipticity and PA. It
models the spiral arms by a ring with inner truncation, and dispenses
with modeling the rotation of the arms as they are too tightly wound
for a stable fit. The other prominent features 
are the bar (\sersic\ profile with best-fit $n \sim 0.2$) and
the nuclear disk and ring, which are both modeled by one component
with a Gaussian profile and an inner truncation applied. The outer disk
and (faint) spiral structure are not shown on the image area and are not
separately modeled as their degeneracy with the main
exponential disk is large.}
  \label{fig:NGC2273-1}
\end{figure*}

\begin{figure*}
  \centering
  \includegraphics[width=17cm]{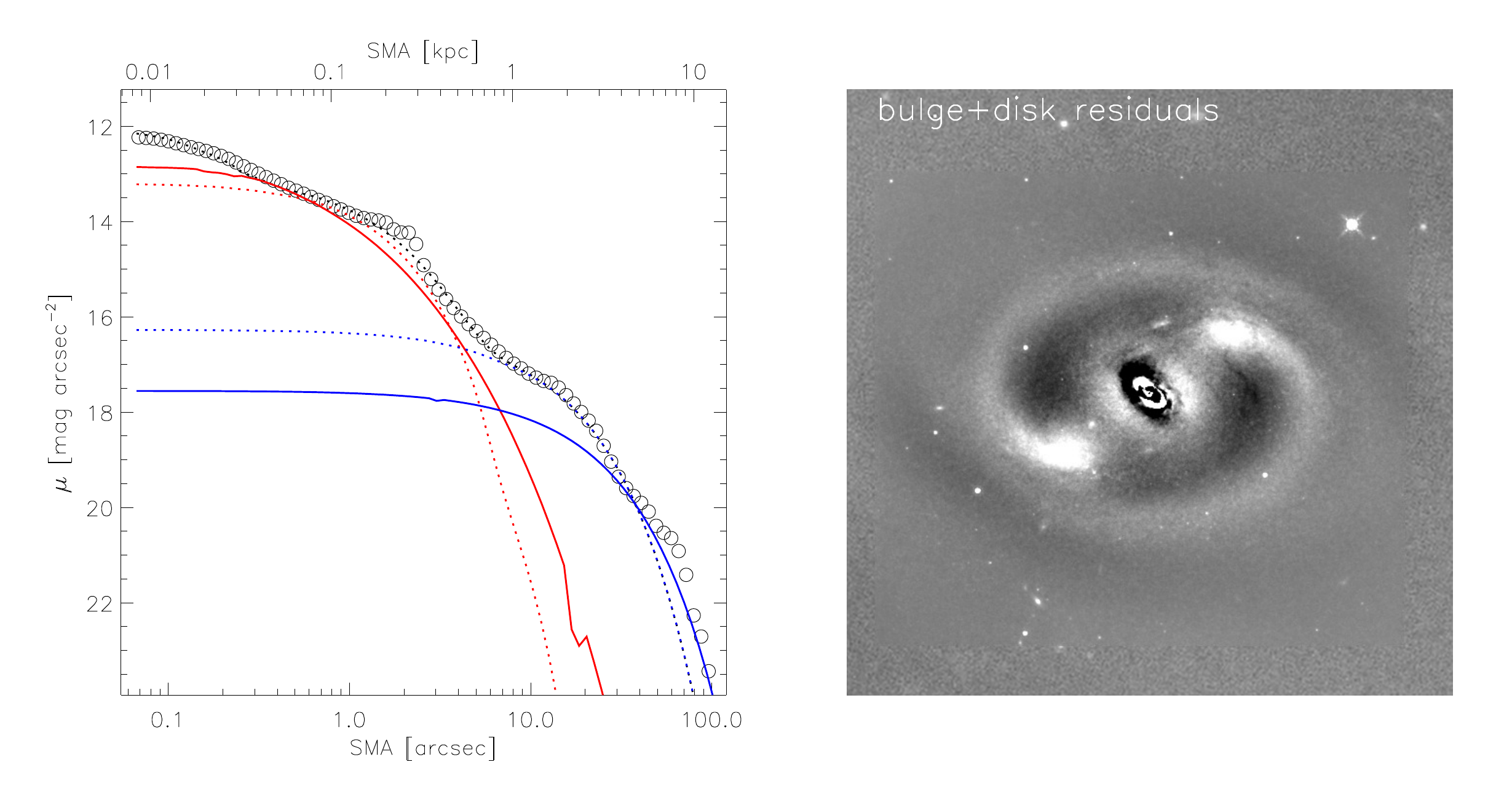}
  \caption{NGC2273 photometric data and model, continued from Figure \ref{fig:NGC2273-1}. {\it Left panel:} SMA surface brightness ($\mu$) of the data (open circles), full model (solid lines) and basic bulge+disk model (dashed), separately for total light (black), bulge (red) and disk (blue). {\it Right panel:} image of basic model residuals. In both the profiles and residuals it is clear that the basic model cannot account for the nuclear disk (near $\text{SMA} \sim 2\arcsec$, or $\sim 250\pc$) and cannot distinguish the extended profile of the main disk from the compact profile of the spiral arms (the ``knee'' at $\sim 20\arcsec\,/\,2\kpc$)}
  \label{fig:NGC2273-2}
\end{figure*}

NGC2273 (Figures \ref{fig:NGC2273-1} and \ref{fig:NGC2273-2}) is a
nearby ($26~\Mpc$) spiral galaxy with several prominent rings and a
central disk.  Due to the proximity of the galaxy, the inner structure
is well-resolved and the galaxy extends $\sim 100$\arcsec\ on the
sky. At the largest radii ($\text{SMA}\gtrsim60'' / 7.5~\kpc$), the
light distribution is flattened ($q\sim0.6$) with a major-axis PA of
about $140\deg$ E of N. The ends of the spiral arms are also
visible. Further inwards, the arms dominate the light.  They are
tightly wound, forming a broad ring at $10-20''$ ($1.3-2.5\kpc$) along
the SMA.  The ellipticity of the ring is lower than that of the outer
disk, and the PA is misaligned by about $+40~\deg$ from the outer
disk. The spiral arms emerge from opposite sides of an apparent bar,
which thickens to become lens-like towards the center.  Located well
inside this bar/bulge region, at $\text{SMA}=2''$ ($250~\pc$), is a
bright and slightly asymmetric nuclear ring and disk
\citep[][]{Mulchaey+97,Erwin+Sparke03,Gu+03} with similar PA and axis
ratio as the outer disk \citep[see
also][]{Petitpas+Wilson02,Barbosa+06,Falcon-Barroso+06}. In the galaxy
center, near the resolution limit, the brightness rises steeply.  This
feature appears round and is $~1''$ ($130~\pc$) in diameter.  It is
either an inner bulge or a barely resolved star cluster.

Starting with a basic bulge+disk+psf model (Figure 1), the best-fit bulge component traces the light of the nuclear ring,
while the disk component broadly accounts for the main spiral arms.
We improve on the basic model by adding a nuclear ring, bar, and
spiral arms (Figure 2). We find a Gaussian profile with inner
truncation for the nuclear ring and the spiral component. We found
 that this choice fits the data better than an
exponential, and avoids the degeneracy of the more general
\sersic\ profile in the presence of the simultaneously
  adjustable truncation parameters. The bar component is a
\sersic\ with $n\sim0.2$ and boxy isophotes (Fourier amplitude
$a_4=-0.1$) as expected.  The bulge and nuclear disk components are
oriented along the (outer) major axis ($\sim50\deg$ E of N), while the
bar component is rotated by a relative $+80\deg$. The large-scale disk
provides a good fit beyond $\gtrsim20'' (2.5~\kpc$) aside from the
spiral arms themselves, which cannot be robustly modeled.

We bracket systematic uncertainties in our reference model with three
additional models. Replacing the exponential disk by a
\sersic\ profile leads to a $0.2\mg$ reduction in the classical bulge
light, and the disk index of $n=1.5$ shows that the corresponding
component indeed traces the exponential part of the profile. We test
the effect of removing the bar, which was visually confluent with the
bulge, and obtain a $0.6\mg$ increase in bulge flux. In this
  modification, the classical bulge $R_e$ increases by a factor of
two, becomes steeper ($n=3.8$ versus the reference $n=2.1$), and
effectively accounts for most of the light inside the spiral
ring. However, this model without a bar results in strongly increased
residuals, and we thus prefer to include the bar. Finally, we test
using a \sersic\ profile for the nuclear ring and model a nuclear disk
instead of the ring.  We find virtually unchanged classical bulge
parameters, but significantly elevated residuals in the latter
  case. The resulting low \sersic\ index (0.1) of the disk also
suggests that an inner truncation is appropriate to model the nuclear
disk. In conclusion, we arrive at a conservative systematic uncertainy
of $0.4\mg$ for the classical bulge magnitude of NGC2273.

\subsection{NGC 2960}

\begin{figure*}
  \centering
  \includegraphics[width=15cm]{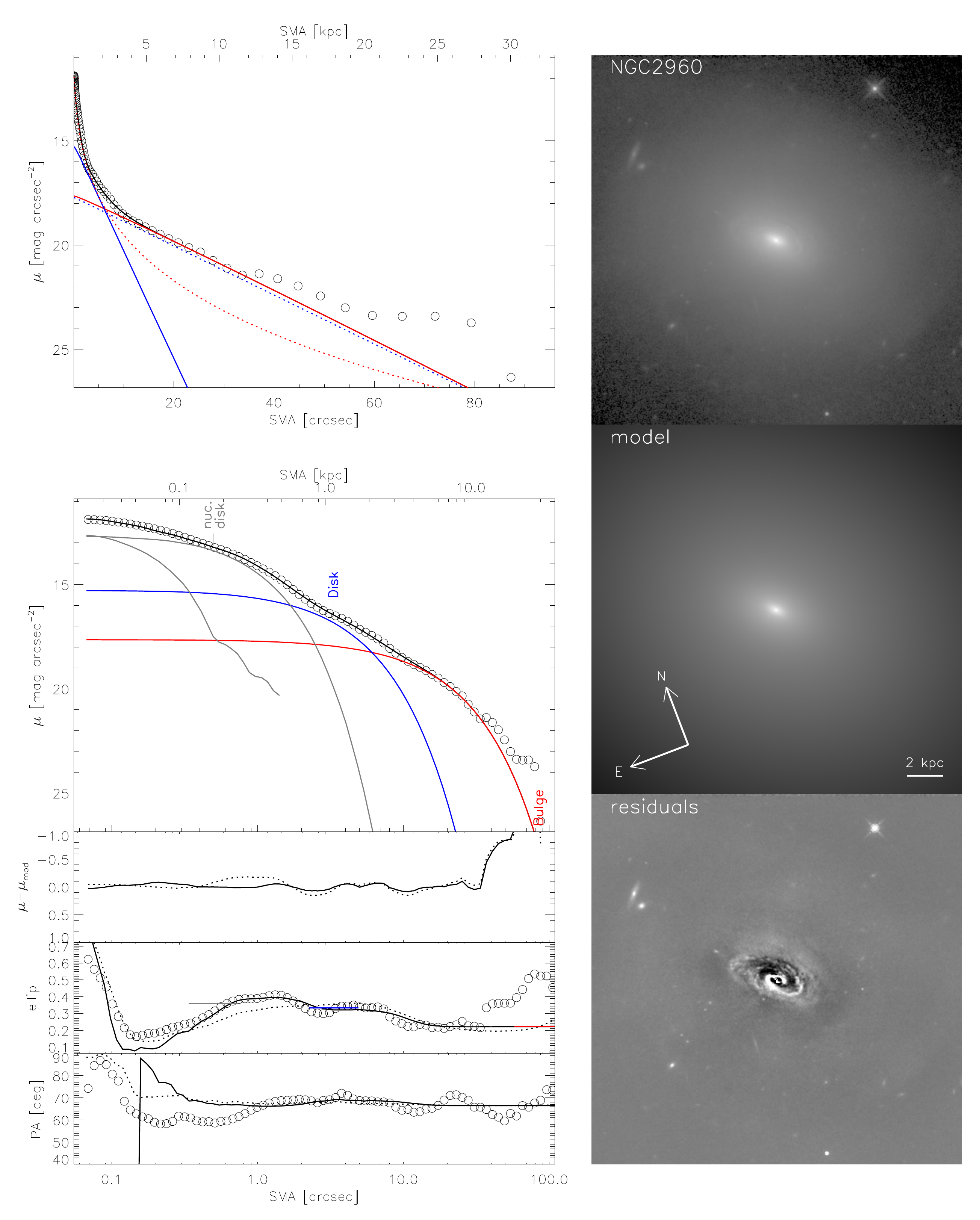}
  \caption{NGC2960 photometric data and model, with layout as in Figure
\ref{fig:IC2560-1}. \textit{Left panels:} semi-major axis (SMA)
profiles of $H$-band surface brightness ($\mu$), data-model residuals
($\mu-\mu_\mathrm{mod}$), ellipticity (ellip), and east-of-north
position angle (PA). Open circles: data, solid lines: full model,
dashed lines: basic (bulge+disk) model. Thick black: total model image
profiles, red: bulge, blue: disk, and thin grey: all other
components. Only select profiles are shown in the $\mu-\text{SMA}$
(top panel) and $\mu-\log\text{SMA}$ (second from top) plots (see also
Figure \ref{fig:NGC2960-2}). Ellipticities of individual components
are indicated by horizontal bars. {\it Right panels:} Images of the
data, model and residuals. The steepening of the profile inside
$\sim13\arcsec$ ($4.5~\kpc$) apparently signals a bulge, but the
increased flattening inside $\sim
13\arcsec$ ($4.5~\kpc$), the flocculent spiral arms, the dust lanes
and two-part inner profile (inflection at $\sim 1\kpc$)
originate in a nuclear ($\sim 100\pc$-scale) and kpc-scale
star-forming disk that dominate the light in this region. We identify
the smooth and round part of the galaxy outside of these two disks as
a likely bulge, and surmise that its near-exponential profile
(i.e. its relatively low \sersic\ index) is a result of a recent
merger, which is likely also responsible for the profile distortions
and light excess at the largest observable radii ($\gtrsim
50\arcsec\,/\,6\kpc$).  } 
\label{fig:NGC2960-1} 
\end{figure*}

\begin{figure*}
  \centering
  \includegraphics[width=17cm]{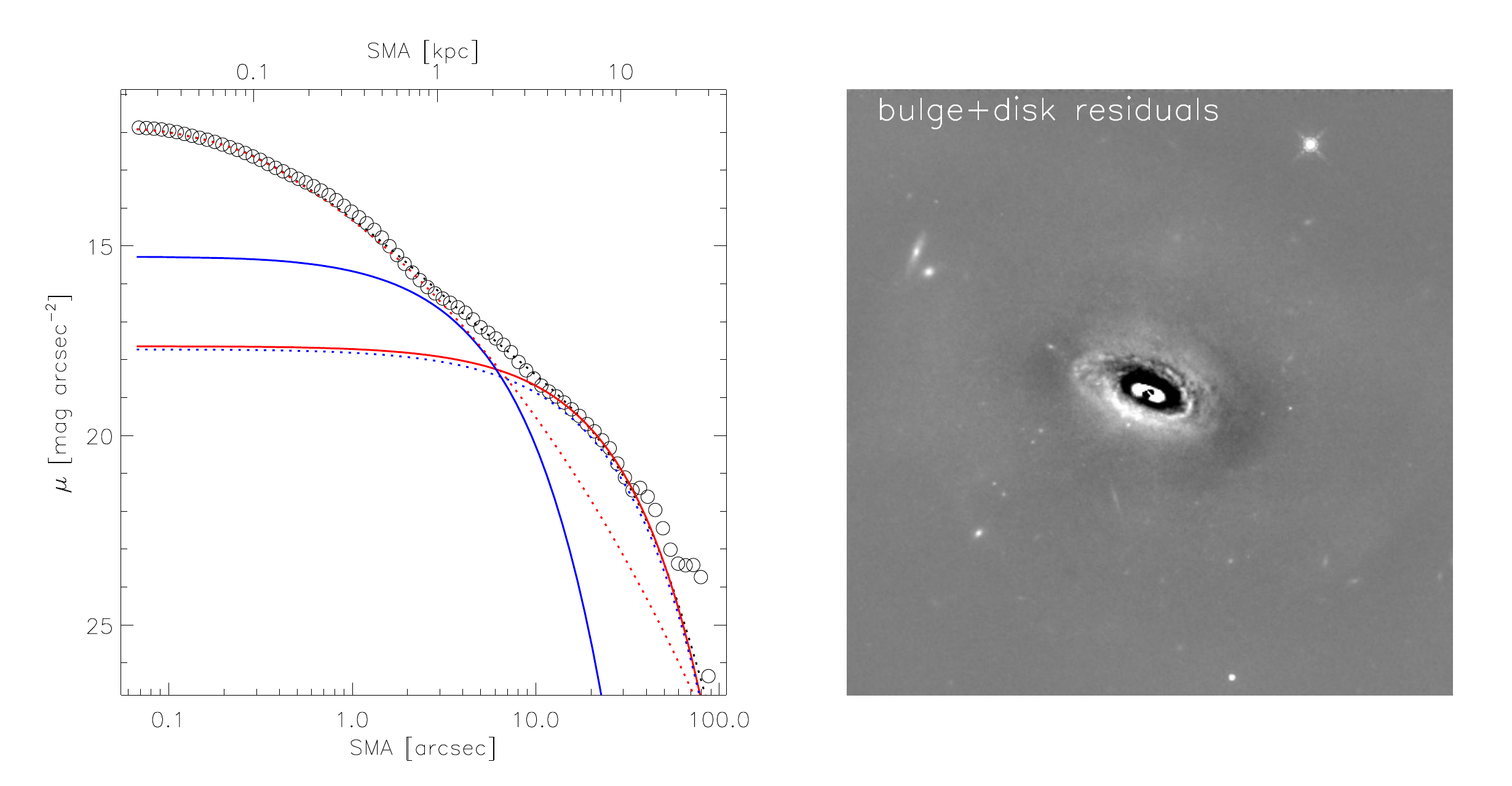}
  \caption{NGC2960 photometric data and model, continued from Figure
\ref{fig:NGC2960-1}. {\it Left panel:} SMA surface brightness ($\mu$)
of the data (open circles), full model (solid lines) and basic
bulge+disk model (dashed), separately for total light (black), bulge
(red) and disk (blue). {\it Right panel:} image of basic model
residuals. The nuclear and kpc-scale starforming disk are clearly
evident in the basic model residuals.}
  \label{fig:NGC2960-2}
\end{figure*}

NGC2960 (Figures \ref{fig:NGC2960-1} and \ref{fig:NGC2960-2}) is a
bulge-dominated galaxy with a kpc-scale embedded disk. Its
semi-major axis is oriented $150\deg$ East of North, and the
average flattening is rather low, $q\sim 0.7$. The radial surface
brightness profile appears to exhibit a bulge-disk structure of a
typical lenticular galaxy, with a prominent steepening of the profile
inside $\sim13''$ ($4.5~\kpc$) and an exponential decline
outside. However, two salient features distinguish NGC2960 visually
from a simple lenticular: the increase of ellipticity towards the
center, and an inner ($\text{SMA} \sim 8'' / 2.7~\kpc$) flocculent and
dusty disk. The residual image of a basic
bulge+disk model (Figure \ref{fig:NGC2960-1}) reinforces the
impression of a dusty disk, and reveals embedded asymmetric 
spiral structure. The residual image also shows that a highly
inclined nuclear disk (residuals at $\text{SMA}\lesssim1.5'' / 500~\pc$) is
responsible for the low ellipticity at the smallest radii.  Both
the nuclear disk and the kpc-scale disk are indicated by
separate ellipticity peaks ($e \sim 0.4$ respectively) at radii
corresponding to their visual dominance. 

By contrast, the large-scale component is round ($q \sim 0.8$) and the
round, smooth and dominant outer component morphologically resembles
an elliptical galaxy. We thus might interpret this large-scale
component as the bulge, despite its exponential profile. At the
largest radii at which the light distribution can be traced, from
$80''$ ($30~\kpc$) to $30''$ ($10~\kpc$), the profile again flattens
and is marked by tidal or shell-like features that can be seen even in
the science frames. These features indicate that NGC2960 has undergone
a recent interaction.  The merger may have distorted the radial
brightness profile of the bulge, resulting in a mostly exponential
shape with a marked upturn only at the largest radii.

The basic bulge+disk model includes a flat ``bulge'' that
over-predicts the central flux. To extract the bulge parameters more
reliably, and to account for the disks, we add an additional nuclear
component with a \sersic\ profile (Figure \ref{fig:NGC2960-1}). Both
inner components converge to relatively compact profiles ($n=1.4$ and
$0.5$). The comparison of this 4-component model with the data profile
makes it clear that both inner components trace the light of nuclear
and kpc-scale disk, and that the inner region is reasonably fit by
these two components alone. We have also explored models with an
additional central bulge, resulting in five components (including the
central point source). However, the residuals barely improve, and the
corresponding ``bulge'' component is still significantly flatter
($q=0.6$) than the large-scale outer profile. We thus do not include
this additional component. We have also accounted for the outermost,
very flattened light by a separate component, but find that its
inclusion does not affect the parameters of the other components.

Given these many complications, we do not find a clear classical bulge
component in NGC2960; instead the center is fit by the sum of the two
disks. We judge that the third, large-scale component corresponds best
to a classical bulge, even with a low best-fit \sersic\ $n\sim 1$. For
comparison, \cite{Vika+12} have fitted a two component model here, and
interpreted the inner component as the bulge, which, given its size,
presumably fits both the nuclear and the kpc-scale disk simultaneously
and is hence more flattened than the outer component (the "disk" in
their interpretation). The \citealt{Vika+12} model hence corresponds
to our basic model, which does not account for separation of the two
inner components and their clear disk morphology. We speculate that
the kpc-scale disk was recently accreted and thus still contains gas
and spiral structure. If we take one or both of the inner disks as the
``bulge'' instead, then the magnitude drops by $0.4-1.4\mg$.  Other
modifications, like allowing a \sersic\ instead of exponential profile
for either disk, or including an envelope, have comparatively minor
effects on the bulge parameters. The alternative models provide us
with a $0.6\mg$ estimate of the systematic bulge magnitude
uncertainty.

\subsection{NGC 3393}

\begin{figure*}
  \centering
  \includegraphics[width=15cm]{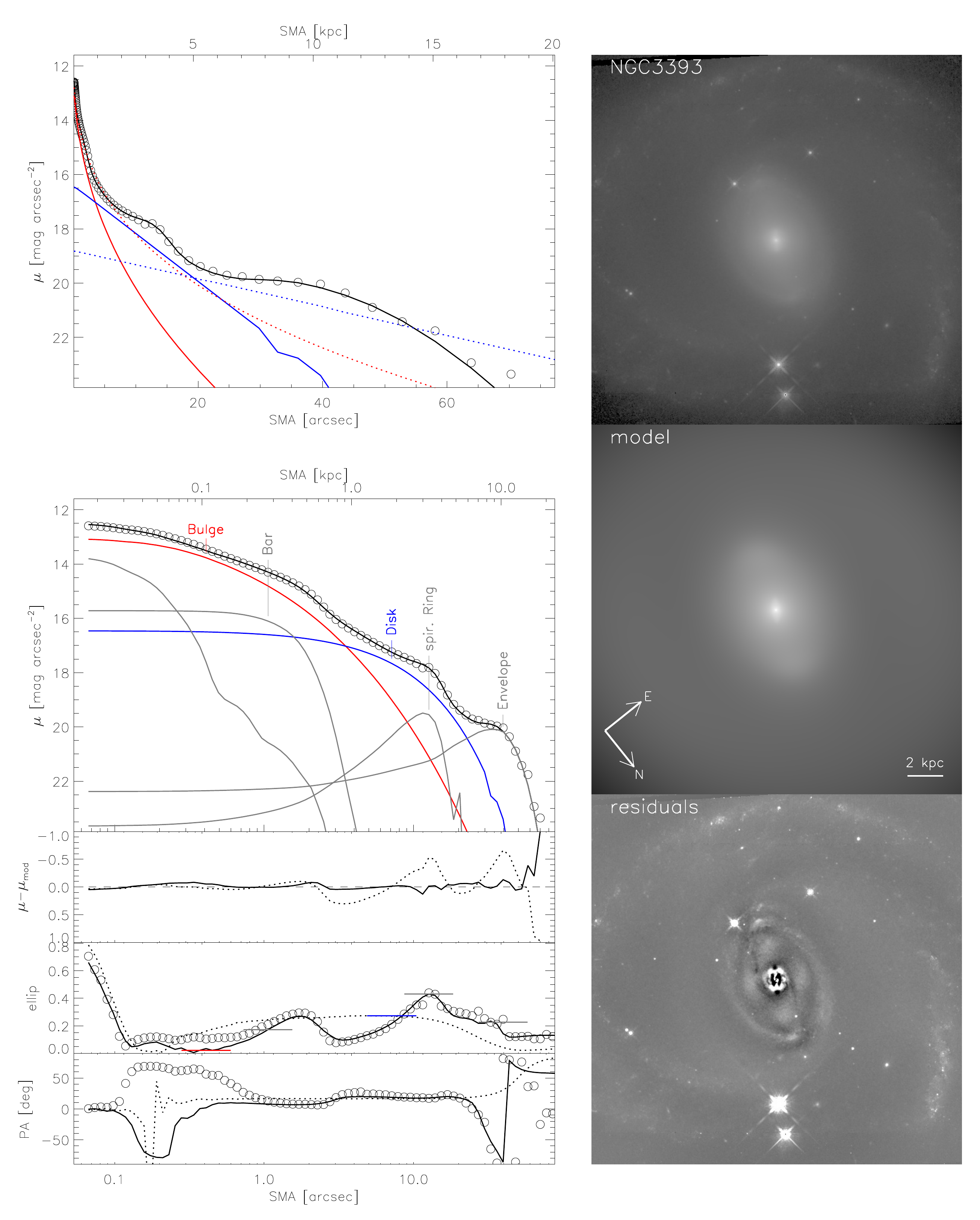}
  \caption{NGC3393 photometric data and model, with layout as in Figure
\ref{fig:IC2560-1}. \textit{Left panels:} semi-major axis (SMA)
profiles of $H$-band surface brightness ($\mu$), data-model residuals
($\mu-\mu_\mathrm{mod}$), ellipticity (ellip), and east-of-north
position angle (PA). Open circles: data, solid lines: full model,
dashed lines: basic (bulge+disk) model. Thick black: total model image
profiles, red: bulge, blue: disk, and thin grey: all other
components. Only select profiles are shown in the $\mu-\text{SMA}$
(top panel) and $\mu-\log\text{SMA}$ (second from top) plots (see also
Figure \ref{fig:NGC3393-2}). Ellipticities of individual components
are indicated by horizontal bars. {\it Right panels:} Images of the
data, model and residuals. In addition to bulge, disk and central
point source, the full model features all major visible structures:
the nuclear bar ($\sim500\pc$ scale) oriented $\sim\,-35\deg$ E of N
(roughly vertical on the shown image); the bright elongated ring,
which appears to delineate the boundary of a $\sim 3\kpc$ bar; and an
outer round ring (flattening near zero) which appears to consist of
weakly defined and tightly wound low-surface brightness spiral arms
and marks the boundary of the visible disk (seen in the corners of the
field shown here). Both rings have a Gaussian profile and an inner
truncation. The residuals suggest an additional nuclear ring that
touches the ends nuclear bar, but it is not separately modeled due to
ensuing excessive model degeneracy.}
  \label{fig:NGC3393-1}
\end{figure*}

\begin{figure*}
  \centering
  \includegraphics[width=17cm]{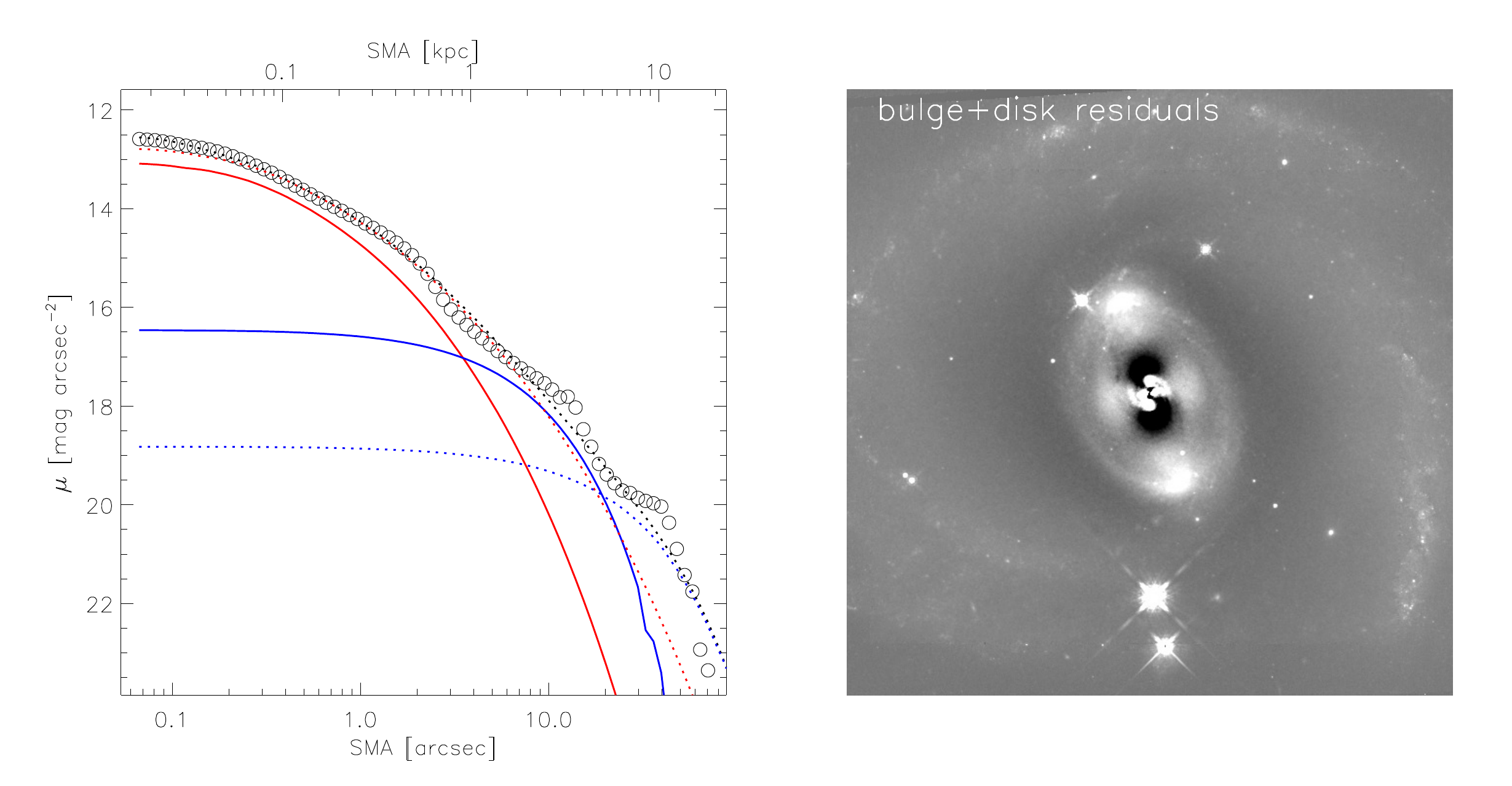}
  \caption{NGC3393 photometric data and model, continued from Figure
\ref{fig:NGC3393-1}. {\it Left panel:} SMA surface brightness ($\mu$)
of the data (open circles), full model (solid lines) and basic
bulge+disk model (dashed), separately for total light (black), bulge
(red) and disk (blue). {\it Right panel:} image of basic model
residuals. The inability of the basic model to represent the galaxy
light is evident, and the basic model residual image exposes both the
intermediate ring (elongated) and the outer ring (round, faint spiral
structure). The nuclear bar is not clearly visible here, but indicated
still by the central isophotal twist.}
  \label{fig:NGC3393-2}
\end{figure*}

NGC3393 (Figures \ref{fig:NGC3393-1} and \ref{fig:NGC3393-2}) is a
late-type galaxy that contains two prominent rings.  It is probably
seen face-on based on the round outer disk, which extends to $\sim
80''$ ($21~\kpc$). The large-scale disk is slightly lopsided towards
$-50\deg$ E of N, and features two asymmetric spiral arms with
prominent star-forming regions. The arms barely connect to the galaxy
center and broadly resemble a ring, which is clearly indicated by the
peak in the SB profile at $40'' / 10\kpc$. The inner ring at
intermediate radii ($\text{SMA}=13\arcsec\,/\,3.5~\kpc$) is much more
elongated and has a lower axis ratio than the outer disk. The ring
might be the star-forming boundary of a large-scale bar. The innermost
region ($\lesssim 2'' / 0.5~\kpc$) appears to be dominated by the
bulge, i.e. a steep increase in surface brightness and round
isophotes.  It harbors yet another bar-like light distribution, with a
PA $140\deg$ E of N, which is misaligned by $-20\deg$ from the major
axis defined by the inner ring. 

Fitting and removing the basic model (bulge+disk+nucleus, see Figure
\ref{fig:NGC3393-2}) exposes residuals from the bar and each ring.
The basic fit yields an extended ($6.4'' / 1.6~\kpc$ and $n=3.5$)
``bulge'' component that effectively accounts for all the light at and
inside the inner ring and is more flattened ($q=0.7$) than the outer
disk.

We improve the NGC3393 model by adding separate components for the
central bar and both rings. With $n=0.25$ and $q=0.4$, the bar is
compact and flattened as expected. We model the rings with Gaussian
profiles ($n=0.5$) but use an inner truncation so that the ring light
does not eat away the disk. This, our best-fit model, has classical
bulge parameters $m=11.2$ mag, $R_e=2.4''\,(0.6~\kpc)$, $n=2.6$ and
$q=0.97$. This is $1\mg$ fainter than the bulge in the basic
bulge+disk model, and $0.2\mg$ brighter than for the intermediate
model without truncations. We adopt this difference as a rough
estimate of the systematic bulge magnitude uncertainty.

\subsection{NGC 4388}

\begin{figure*}
  \centering
  \includegraphics[width=15cm]{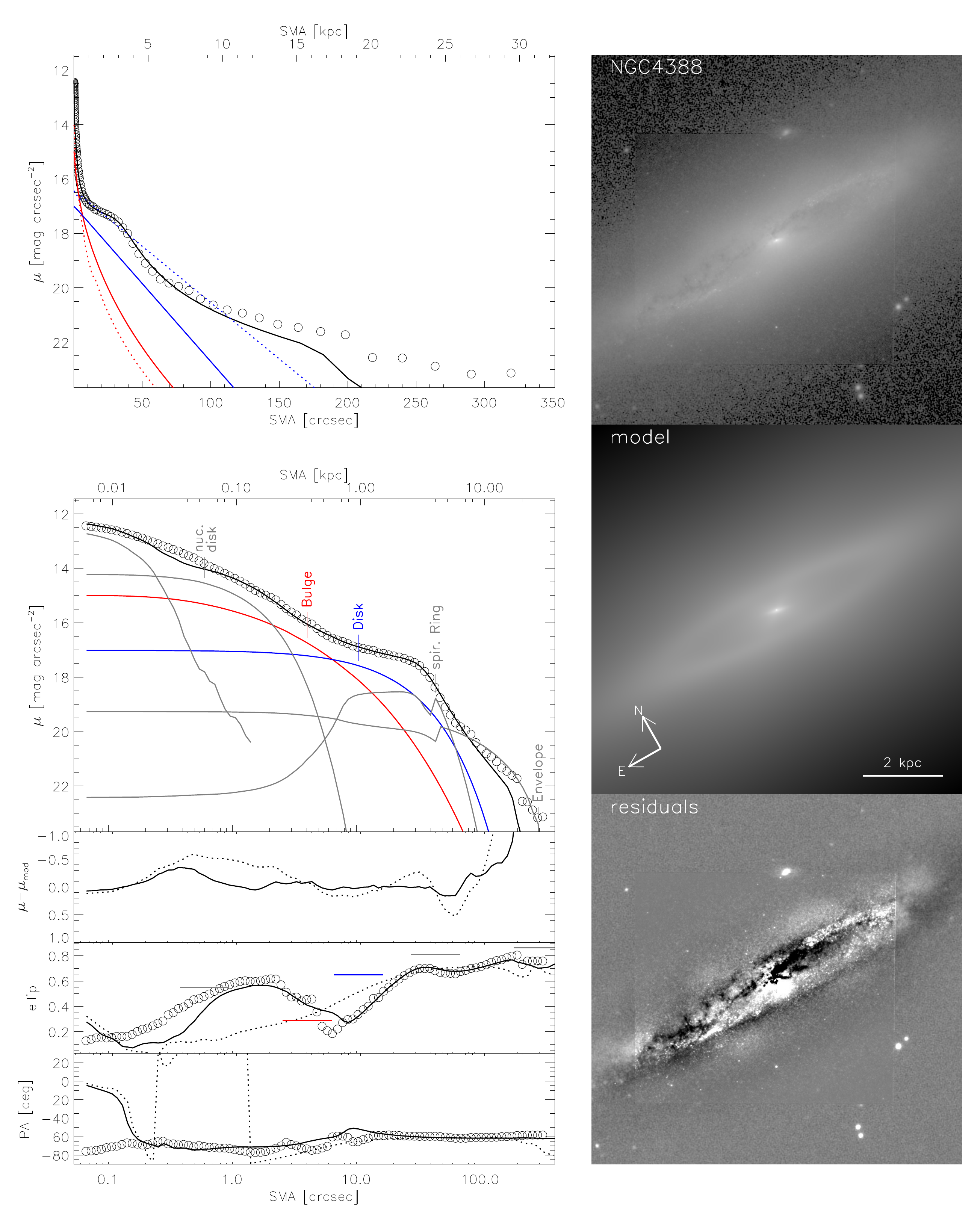}
  \caption{NGC4388 photometric data and model, with layout as in Figure
\ref{fig:IC2560-1}. \textit{Left panels:} semi-major axis (SMA)
profiles of $H$-band surface brightness ($\mu$), data-model residuals
($\mu-\mu_\mathrm{mod}$), ellipticity (ellip), and east-of-north
position angle (PA). Open circles: data, solid lines: full model,
dashed lines: basic (bulge+disk) model. Thick black: total model image
profiles, red: bulge, blue: disk, and thin grey: all other
components. Only select profiles are shown in the $\mu-\text{SMA}$
(top panel) and $\mu-\log\text{SMA}$ (second from top) plots (see also
Figure \ref{fig:NGC4388-2}). Ellipticities of individual components
are indicated by horizontal bars. {\it Right panels:} Images of the
data, model and residuals. The full model accounts for the nuclear
disk, spiral arms, and outer disk (envelope) with exponential profiles,
where the spiral arms component is modeled as a ring. 
The innermost profile is partially underpredicted, but
corresponding models accounting for it by an additional component
proved too degenerate, and dust in the center prevents a more
accurate interpretation of the inner structure based on our image.}
  \label{fig:NGC4388-1}
\end{figure*}

\begin{figure*}
  \centering
  \includegraphics[width=17cm]{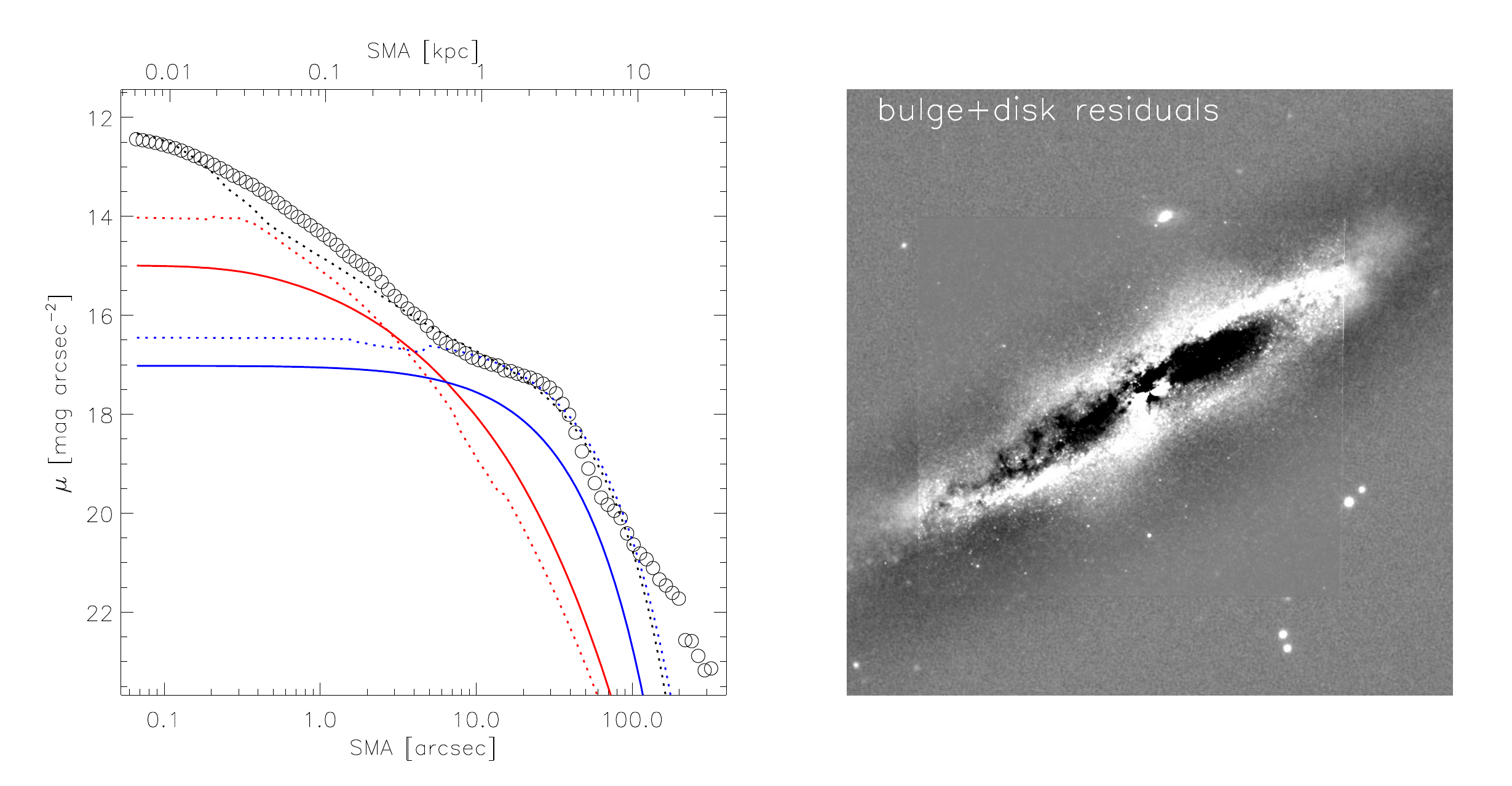}
  \caption{NGC4388 photometric data and model, continued from Figure
\ref{fig:NGC4388-1}. {\it Left panel:} SMA surface brightness ($\mu$)
of the data (open circles), full model (solid lines) and basic
bulge+disk model (dashed), separately for total light (black), bulge
(red) and disk (blue). {\it Right panel:} image of basic model
residuals. The basic model residuals clearly expose the spiral arms,
which wind tightly to nearly form a ring, as well as the thin
(inclined) central nuclear disk and intervening dust lanes at all
radii.}
  \label{fig:NGC4388-2}
\end{figure*}

NGC4388 (Figures \ref{fig:NGC4388-1} and \ref{fig:NGC4388-2}) is a
spiral galaxy seen at high inclination (disk
$q=0.35~\rightarrow~i\gtrsim 70\deg$).  The tightly-wound 
spiral arms are visually defined between
$\text{SMA}\sim30$ and $50\arcsec$ ($2.8$ and $4.6~\kpc$). They form a
ring that can be identified in the surface brightness and ellipticity
profiles. In the center of NGC4388 there is a
bright, central ($\lesssim 2.5" / 200~\pc$) disk, seen almost edge-on
with very low axis ratio, and at a $\sim-15\deg$ misalignment from the
large-scale major axis. The entire region interior to the ring
contains several dust lanes. Outside of the spiral/ring, the profile
has two nearly exponential parts, with a change to a larger scale
radius (factor of several) and ellipticity (by $\sim0.2$) occurring at
$\sim 80"$ ($7~\kpc$). We identify the inner exponential as
the main disk, while the outer exponential ``envelope'' as
a very extended disk, due to its high flattening ($q\sim 0.1$) and
$\sim100"$ ($9~\kpc$) scale radius.

In addition to the bulge, disk and nucleus (point source), we account
for the nuclear disk, spiral arms, and envelope with exponential
profiles. The spiral arms are modeled as a ring with an inner
truncation.  To reduce degeneracy, we chose an exponential instead of
a \sersic\ profile for the spiral ring, and force the bulge, disk and
envelope to share a common PA. The best-fit bulge of our adopted
reference model is very round ($q\sim0.8$, compared to the disk
$q\sim0.3$), has an intermediate \sersic\ index $n=2.2$, and boxy
isophotes. The nuclear disk dominates the surface brightness in the
center, and the PA of the model component converges to the observed
value, which provides confirmation that this component is suitably
accounted for in our adopted model \citep[see also discussion of the
nuclear disk from the kinematics as observed by SINFONI
in][]{Greene+14}.  The outer envelope also makes a significant
difference to the fit in this case, changing the bulge magnitude by
$-0.5\mg$.

Of the various models we fitted, we mention here the model with a
\sersic\ profile (instead of exponential) for the nuclear disk, a
\sersic\ for the main disk, and a model without the envelope
component.  These result in classical bulge magnitude changes of
$+0.7$, $-0.3$ and $-0.5\mg$, respectively, and lead us to a
systematic bulge magnitude uncertainty estimate of $0.5\mg$. For
reference, the classical bulge in the best-fit bulge+disk+point-source
model is $0.3\mg$ fainter than in our adopted 6-component model.

\subsection{NGC 6264}

\begin{figure*}
  \centering
  \includegraphics[width=15cm]{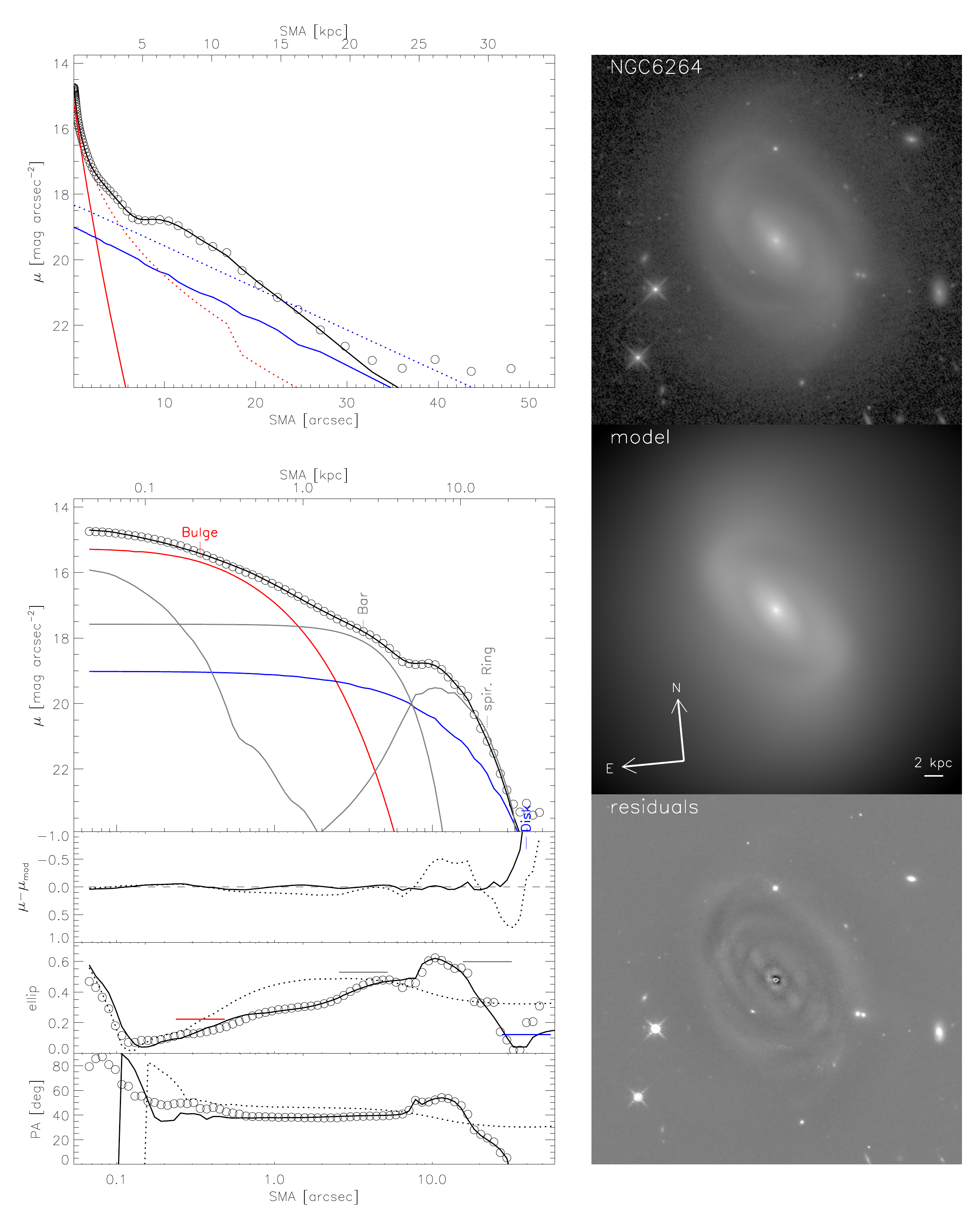}
  \caption{NGC6264 photometric data and model, with layout as in Figure
\ref{fig:IC2560-1}. \textit{Left panels:} semi-major axis (SMA)
profiles of $H$-band surface brightness ($\mu$), data-model residuals
($\mu-\mu_\mathrm{mod}$), ellipticity (ellip), and east-of-north
position angle (PA). Open circles: data, solid lines: full model,
dashed lines: basic (bulge+disk) model. Thick black: total model image
profiles, red: bulge, blue: disk, and thin grey: all other
components. Only select profiles are shown in the $\mu-\text{SMA}$
(top panel) and $\mu-\log\text{SMA}$ (second from top) plots (see also
Figure \ref{fig:NGC6264-2}). Ellipticities of individual components
are indicated by horizontal bars. {\it Right panels:} Images of the
data, model and residuals. The non-axisymmetric structure of NGC6264
is produced by the bar and spiral arms, which we model by a compact
(best-fit $n=0.47$) \sersic\ and an exponential profile with both
inner truncation and coordinate rotation, respectively. The bulge and
bar profiles largely overlap and thus are somewhat degenerate, however
their very different axis ratios and the steep profile in the center
justify two seperate components here.}
  \label{fig:NGC6264-1}
\end{figure*}

\begin{figure*}
  \centering
  \includegraphics[width=17cm]{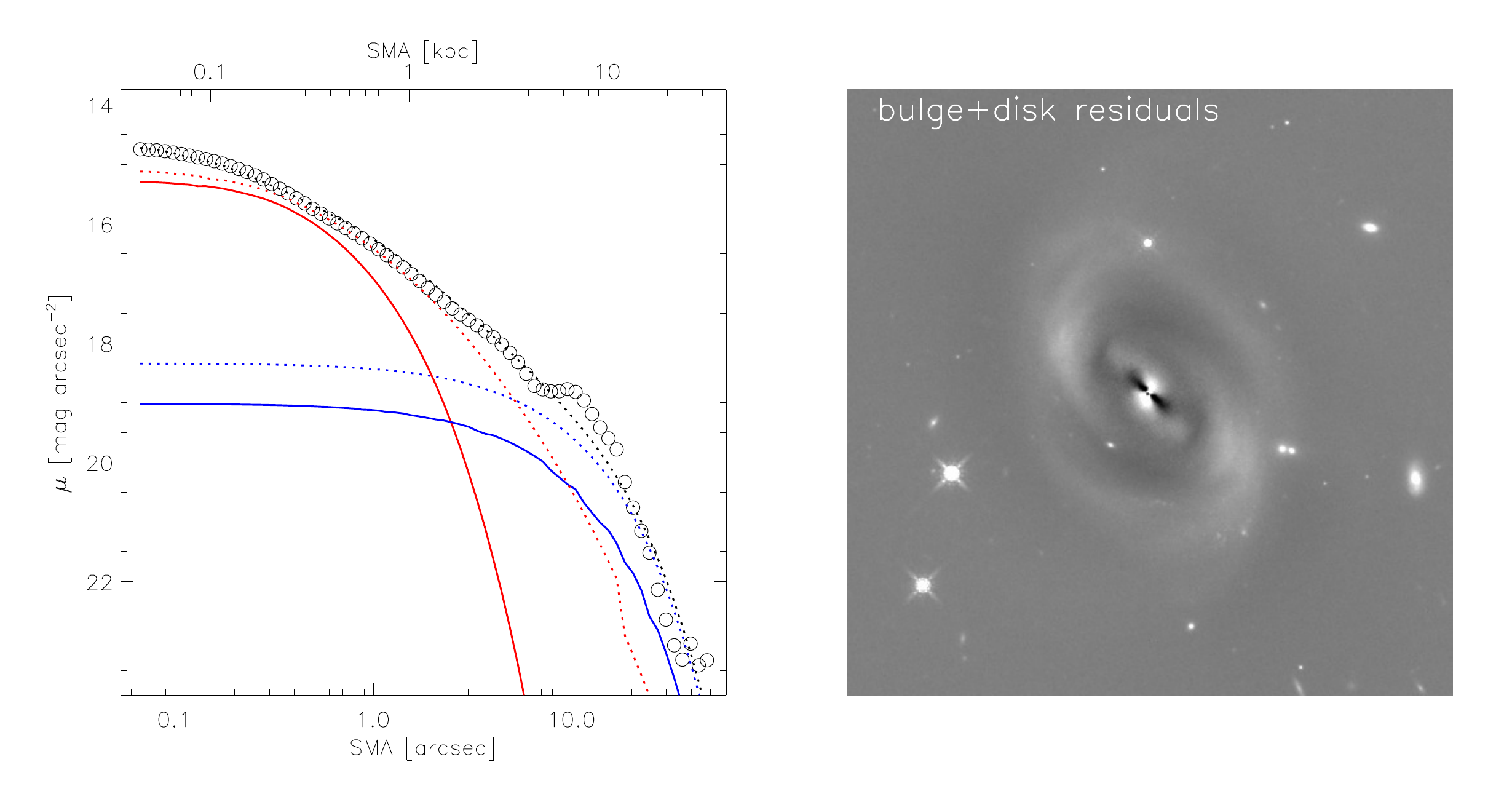}
  \caption{NGC6264 photometric data and model, continued from Figure
\ref{fig:NGC6264-1}. {\it Left panel:} SMA surface brightness ($\mu$)
of the data (open circles), full model (solid lines) and basic
bulge+disk model (dashed), separately for total light (black), bulge
(red) and disk (blue). {\it Right panel:} image of basic model
residuals, which clearly exhibits the central bulge as distinct from
the bar.}
  \label{fig:NGC6264-2}
\end{figure*}

NGC6264 (Figures \ref{fig:NGC6264-1} and \ref{fig:NGC6264-2}) is
dominated by a bar and a pair of smooth spiral arms. A distinct bulge
is not immediately visible, but there is a small ($\sim0.3'' /
200\pc$) and round central light concentration embedded within the
conspicuous bar. The bar is apparent also in the local maximum around
$5"$ ($3~\kpc$) in the surface brightness and ellipticity profiles at
constant PA. Within $\sim5''$ ($350~\pc$), the PA profile shows a
$\sim10\deg$ step, and grows considerably rounder towards smaller
radii, suggesting that we are seeing an underlying bulge.

The spiral arms are smooth and show little signs of star formation,
perhaps due to the decreased spatial resolution ($\sim100~\pc$ at the
distance of $136~\Mpc$). The tightly wound spiral arms emerge from the
ends of the bar, then become clearly defined at larger radii and
obtain full strength around $\text{SMA}\sim10"$ ($7~\kpc$). This
region is marked by a local maximum in the surface brightness,
ellipticity and PA profiles, followed by a rapid brightness drop
towards larger radii. At $\sim30''$ ($20~\kpc$) the surface brightness
profile transitions into a faint ($\sim23\magarcsec$) floor that we
interpret as an envelope or halo that is traceable out to $50''$
($35~\kpc$).

When fitted with a basic model (bulge, disk and point source), the
``bulge'' component is more flattened than the ``disk'' and appears to
predominantly fit the light of the bar. Round residuals near the
center expose an underlying rounder light distribution, i.e. the
probable bulge. We add another \sersic\ component to directly model
the bar. The residuals in this second fit are greaty improved, and we
find a small bulge component ($R_e=1'' / 0.7~\kpc$, $q=0.7$), and a
larger flat ($R_e=4'' / 2.6~\kpc$, $q=0.4$) and compact (\sersic\
$n=0.5$) component, as expected for a bulge and a bar. Using an
exponential profile with inner truncation and powerlaw rotation, we
model the spiral arms and the underlying exponential disk
component. The adopted 4+1-component model (bulge, disk, bar, spiral,
plus the nuclear point source) is a good fit aside from a possible
outer envelope (or halo), which was too faint to be fit robustly. 
We choose not to account for the latter by
another component, as it converges to a very large 
($R_s=100" / 70~\kpc$) scale, suggesting a degeneracy with the 
background uncertainty.

We experimented with alternatives to this adopted model, and quote the
change of bulge magnitude incurred by some of these
modifications. Using a \sersic\ instead of exponential profile for the
disk, or adding an exponential component for the envelope (halo),
results in a $0.4$ and $0.2\mg$ fainter bulge, respectively. Modeling
bar and bulge with only one \sersic\ profile, but retaining the spiral
component apart from disk and AGN leads to a much
brighter bulge ($-1.9\mg$), with classical bulge parameters that are close to
those of the basic model parameters and differ by $-1.4\mg$ in
$m_\mathrm{bul}$ from the reference model. As a conservative estimate,
we thus assign a bulge magnitude uncertainty of $0.7\mg$. \\ 

\subsection{NGC 6323}

\begin{figure*}
  \centering
  \includegraphics[width=15cm]{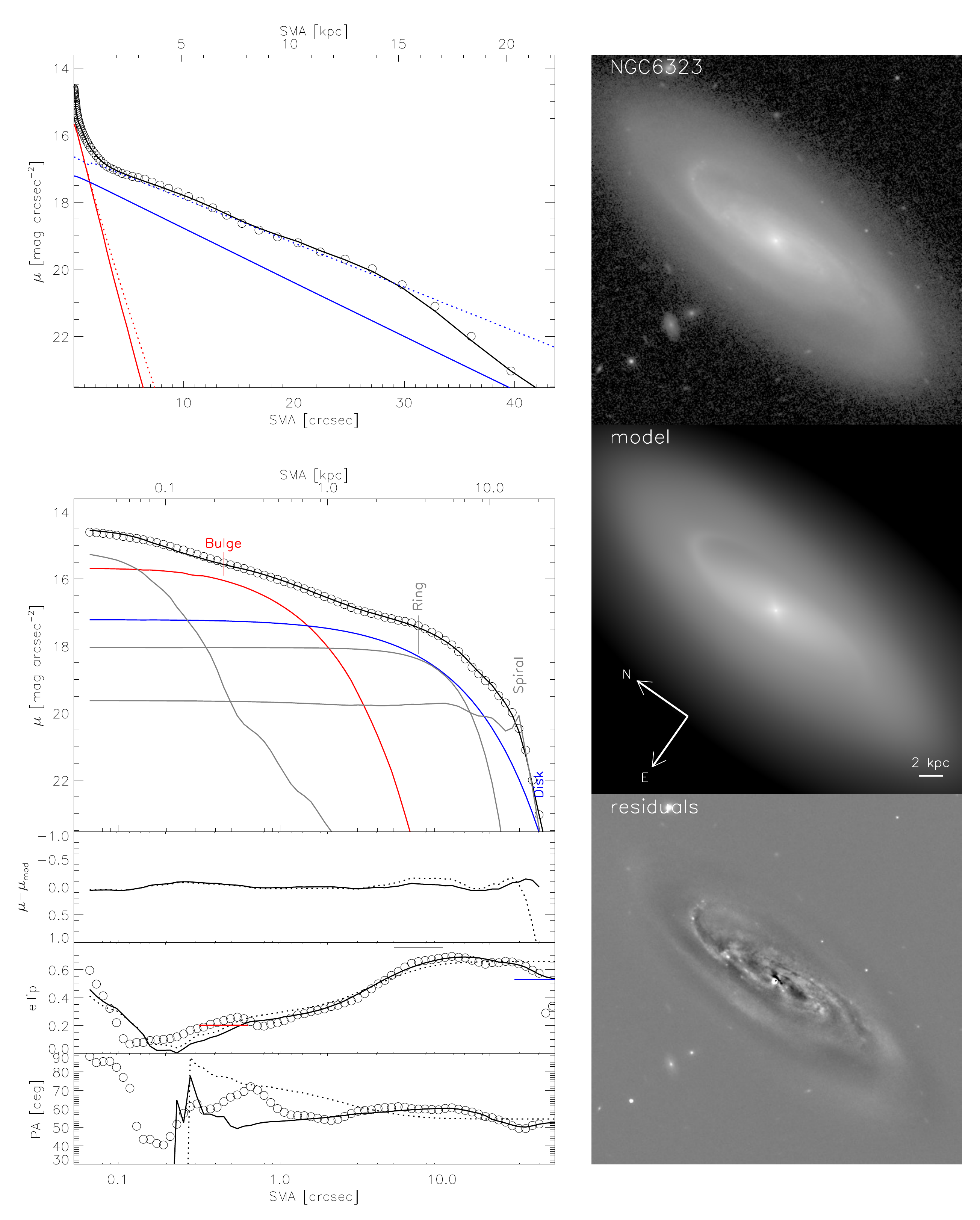}
  \caption{NGC6323 photometric data and model, with layout as in Figure
\ref{fig:IC2560-1}. \textit{Left panels:} semi-major axis (SMA)
profiles of $H$-band surface brightness ($\mu$), data-model residuals
($\mu-\mu_\mathrm{mod}$), ellipticity (ellip), and east-of-north
position angle (PA). Open circles: data, solid lines: full model,
dashed lines: basic (bulge+disk) model. Thick black: total model image
profiles, red: bulge, blue: disk, and thin grey: all other
components. Only select profiles are shown in the $\mu-\text{SMA}$
(top panel) and $\mu-\log\text{SMA}$ (second from top) plots (see also
Figure \ref{fig:NGC6323-2}). Ellipticities of individual components
are indicated by horizontal bars. {\it Right panels:} Images of the
data, model and residuals. The overall structure is simple, dominated
by the disk and a clear profile steepening (bulge) in the
centre. However, spiral arms and varying PA in the inner $\sim\kpc$
are also evident, The spiral arms are not strong or sharply defined in
the $H$-band; we model them nevertheless by an exponential profile,
modified by coordinate rotation and an inner truncation. They appear
to emerge from an elongated structure that nearly forms a ring, which
likely delineates an inner disk (a putative pseudobulge), but possibly
also represents a bar. We model this structure separately by a
\sersic\ component (best-fit axis ratio $q\sim 0.2$ and $n\sim0.4$)
and thus distinguish it from the very round ($q\sim 0.8$) small
($R_e\sim 0.7\kpc$) classical bulge, for which we allowed a 4th-order
isophotal harmonic to account for its boxiness.}
  \label{fig:NGC6323-1}
\end{figure*}

\begin{figure*}
  \centering
  \includegraphics[width=17cm]{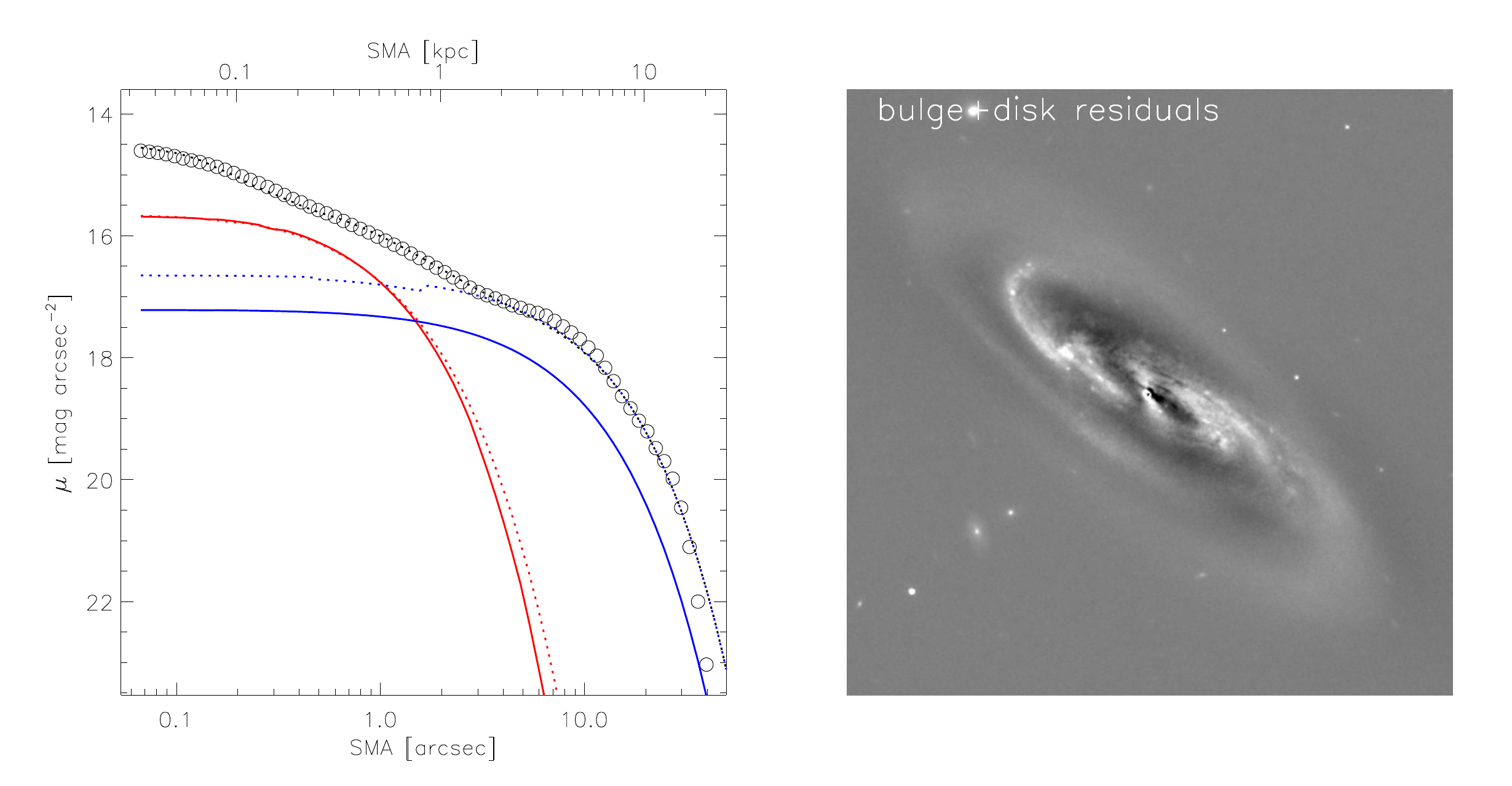}
  \caption{NGC6323 photometric data and model, continued from Figure
\ref{fig:NGC6323-1}. {\it Left panel:} SMA surface brightness ($\mu$)
of the data (open circles), full model (solid lines) and basic
bulge+disk model (dashed), separately for total light (black), bulge
(red) and disk (blue). {\it Right panel:} image of basic model
residuals, which highlight the spiral arms as well as the inner disk
(or bar) from which the spiral arms emerge.}
  \label{fig:NGC6323-2}
\end{figure*}

NGC6323 (Figures \ref{fig:NGC6323-1} and \ref{fig:NGC6323-2}) is a
spiral galaxy at high inclination with an apparent disk axis ratio of
$\sim 0.4$. There is a ring at $\text{SMA}=8"$ ($4~\kpc$), from which
two flocculent and unequal-strength spiral arms emerge and extend to
$\text{SMA} \approx 10~\kpc$ from the center. The main disk profile
exhibits a weak truncation at $\text{SMA}>20~\kpc$ that is
recognizable in the image as the visible boundary of the disk.  An
inner ($4~\kpc$) ring is clearly seen in the surface brightness
profile, delineating the transition from a low-ellipticity, bulge-dominated
inner region to a flatter disk-dominated region.  A nucleus or small
bulge can be distinguished visually within the inner two arcseconds.

The basic bulge+disk+nucleus model fares surprisingly well at
emulating the radial profile. However, significant structure
  remains in the residual image, including spiral arms and a ring from
  which they appear to emerge, as well as a central misaligned
  elongated structure whose ends coincide with the ring. We therefore
  construct a model that includes a spiral component modified by
  coordinate rotation and inner truncation, as well as a very compact,
  low-$n$ \sersic\ profile with high flattening that accounts for
  light between the bulge and the spiral arms and may be interpreted
  as a large-scale bar.  This additional component also effectively
  also removes the ring-like residuals at the onset of the spiral
  structure. Coincidentally, despite the addition of two components,
the best-fit bulge parameters of the reference model are only
marginally different from that of the basic model
($m_\mathrm{bul}=15.5\mg$ instead of $15.4\mg$, and similarly for the
bulge $R_e$ and $n$). The bulge is about $2\mg$ fainter than the disk,
$R_e=1.3''$ ($0.65~\kpc$) in size, and has a near-exponential
profile. Allowing a 4th-order Fourier mode for the bulge isophotes
gives a boxy shape (amplitude $\sim0.1$). The spiral is modified by
Fourier modes (4th order and lower), which enables fitting of the
asymmetry in the spiral arms and a better convergence of the rotation
function.

We explore multiple alternative models and find that, while formally
increasing residuals ($\chi^2$), for some of them the residual images
and radial profile mismatches differ only in details. Removing the
bar/ring component decreases the bulge brightness by $0.2\mg$, and
omitting the spiral arm component instead changes $m_\mathrm{bul}$ by
less then $0.1\mg$. Testing a model where the disk has a \sersic\
profile instead of an exponential gives a $0.2\mg$ brighter bulge, and
a disk \sersic\ index of 0.9. The lowest change in residuals, but
biggest change in $m_\mathrm{bul}$, occurs when we apply a
truncation to the bar/ring instead of the spiral arm component,
obtaining a $1\mg$ brighter and three times larger bulge, with $n=3.0$
instead of $1.1$.Taken together, these alternatives indicate a
systematic $m_\mathrm{bul}$ uncertainty of $0.5\mg$.

\subsection{UGC 3789}

\begin{figure*}
  \centering
  \includegraphics[width=15cm]{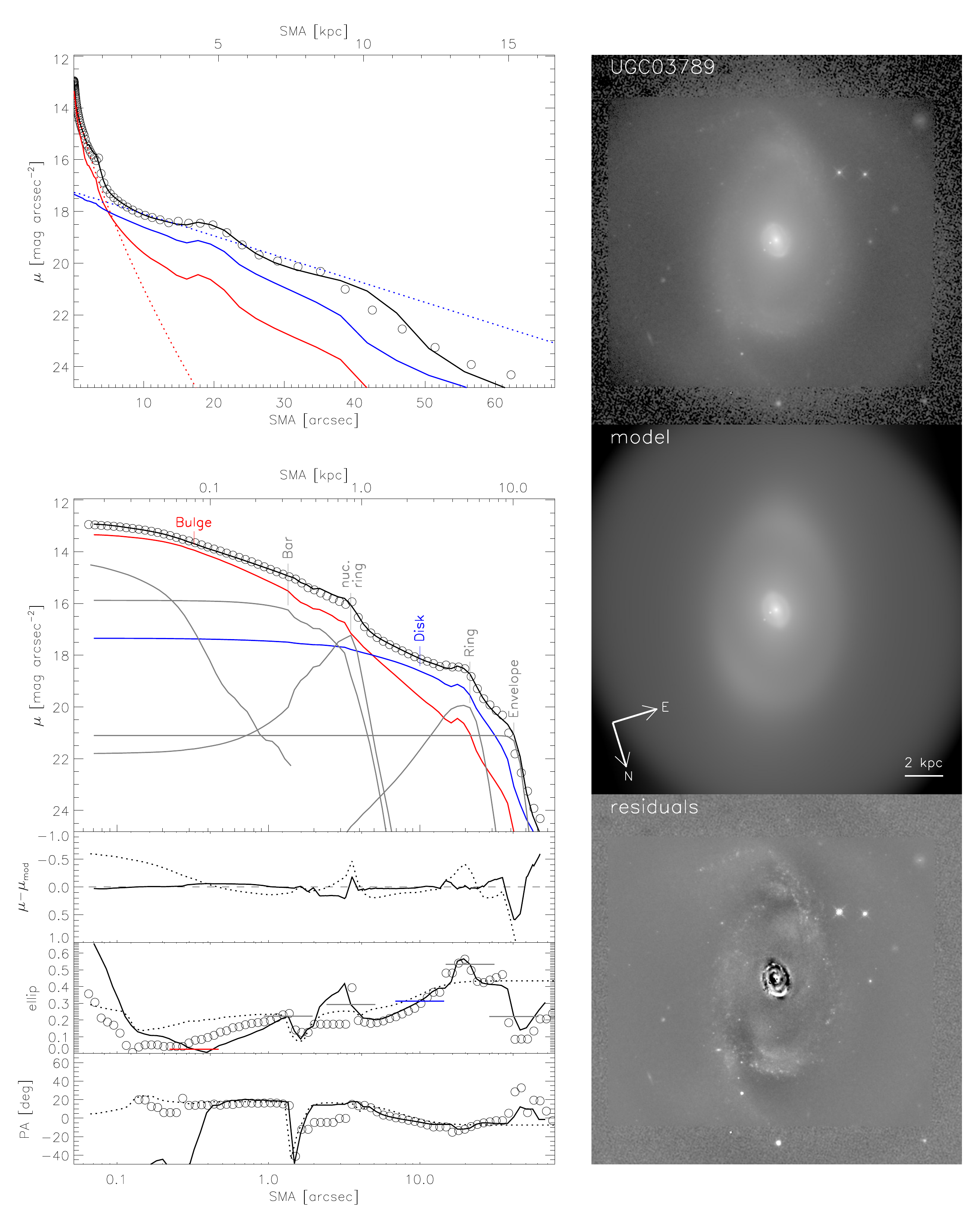}
  \caption{UGC3789 photometric data and model, with layout as in Figure
\ref{fig:IC2560-1}. \textit{Left panels:} semi-major axis (SMA)
profiles of $H$-band surface brightness ($\mu$), data-model residuals
($\mu-\mu_\mathrm{mod}$), ellipticity (ellip), and east-of-north
position angle (PA). Open circles: data, solid lines: full model,
dashed lines: basic (bulge+disk) model. Thick black: total model image
profiles, red: bulge, blue: disk, and thin grey: all other
components. Only select profiles are shown in the $\mu-\text{SMA}$
(top panel) and $\mu-\log\text{SMA}$ (second from top) plots (see also
Figure \ref{fig:UGC3789-2}). Ellipticities of individual components
are indicated by horizontal bars. {\it Right panels:} Images of the
data, model and residuals. The structure of 
UGC3789 is similar to NGC3998,
with a large-scale round outer disk on which faint and weakly defined
spiral arms and star-forming regions are superposed, a bright large
elongated inner disk (or bar) delineated by a bright star forming
ring, and a round nuclear disk/ring with an inset nuclear bar. We
model the bar by a \sersic\ profile (best-fit $n\sim 0.3$), and the
rings with Gaussian profiles with an inner
truncation (dropping the truncation for
the outer ring due to excessive degeneracy). The
best-fit model reproduces the profiles accurately compared to the basic
model, especially in terms of the steep central ($\lesssim 100\pc$)
brightness and variations in the ellipticity.}
  \label{fig:UGC3789-1}
\end{figure*}

\begin{figure*}
  \centering
  \includegraphics[width=17cm]{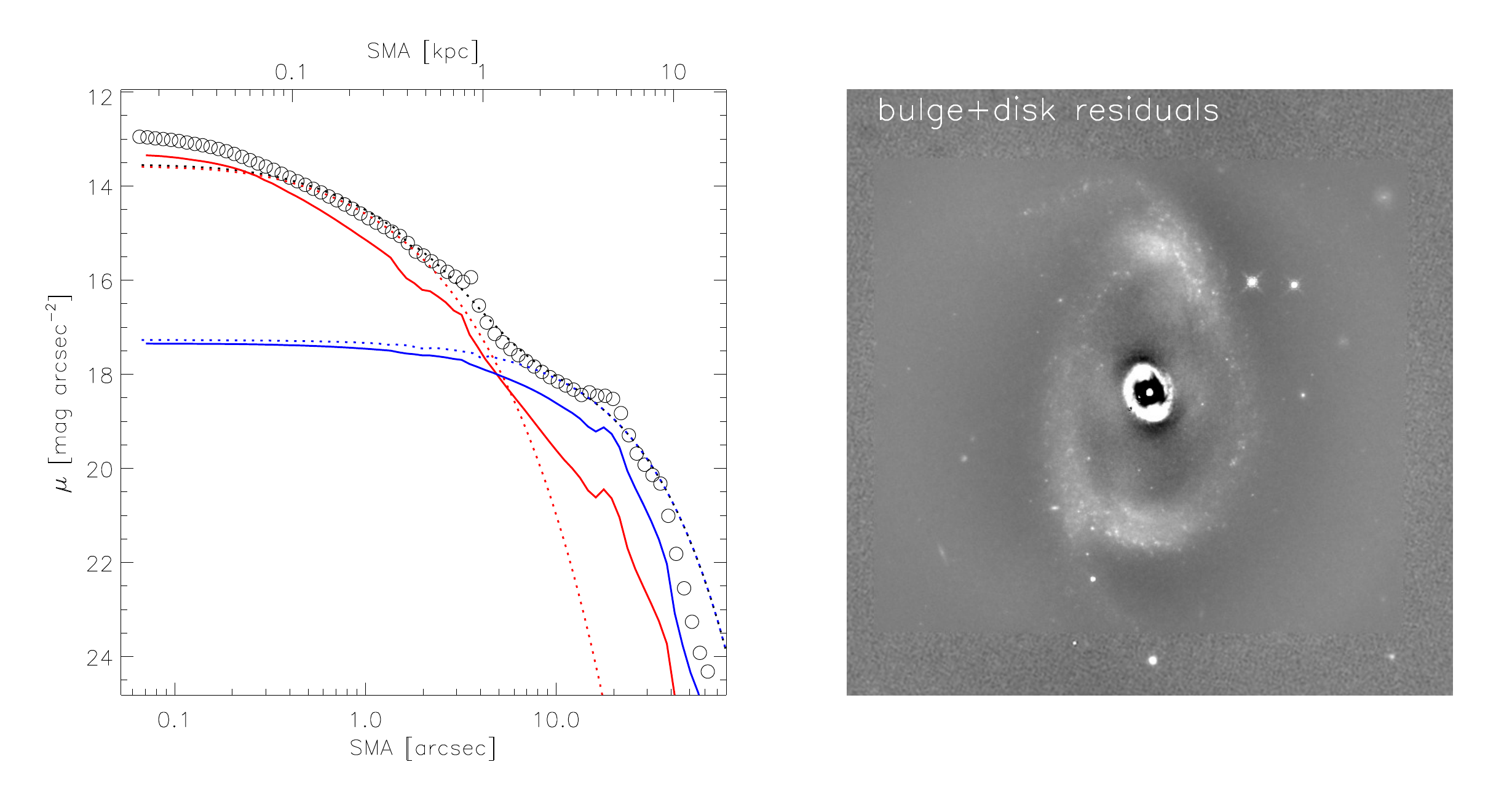}
  \caption{UGC3789 photometric data and model, continued from Figure
\ref{fig:UGC3789-1}. {\it Left panel:} SMA surface brightness ($\mu$)
of the data (open circles), full model (solid lines) and basic
bulge+disk model (dashed), separately for total light (black), bulge
(red) and disk (blue). {\it Right panel:} image of basic model
residuals, which exposes the intermediate elongated ring (extending
about half the way to the edge of the shown image), as well as the
nuclear ring and bar. The basic model bulge is biased by this nuclear
ring and bar to a too compact profile, which we avoid in the full
model by separate nuclear bar and ring components. The basic model
also renders an overprediction for the disk (and hence the total)
luminosity due to the influence of the $\kpc$-scale bar/ring that is
included as a component in the full model as well.}
  \label{fig:UGC3789-2}
\end{figure*}

UGC3789 (Figures \ref{fig:UGC3789-1} and \ref{fig:UGC3789-2}) is a
late-type galaxy that is seen nearly face-on and dominated by two
rings, one close to the center at $\text{SMA}=3.8''$ ($900~\pc$), and
the second one at $\text{SMA}=23''$ ($5.6~\kpc$). Both rings contain
star forming regions and form the boundary of a disk, respectively.
but the larger ring, as in NGC3393, might as well delineate a
large-scale bar. The inner ring is nearly round, while the second ring
shows marked flattening and asymmetry. Two short spiral arms emerge at
a PA$ \approx 170\deg$ but varies by several degrees from the center
outwards. A third, weaker ring is discernible in the $\mu$ profile at
$\sim 40''$ ($10~\kpc$) and marks the edge of the visible large-scale
disk. \\

A basic bulge+disk+AGN model is clearly unsuitable to model this
galaxy, due to the luminous rings and bar. The ``bulge'' component
fitted by the basic model largely fits the inner ring and is therefore
very compact ($n\sim 1$). The ``disk'' in the basic model roughly
accounts for the light of the second ring. 
In our best-fit model, the rings are modeled
by Gaussian profiles with inner truncation, except for the outer ring,
which is an untruncated \sersic\ profile with low index $\sim0.1$ in
the best-fit solution. Finally, the bar component becomes readily fit
by a typical geometry ($q\sim0.3$) and compact profile ($n\sim0.3$).

The resulting reference model is a vast improvement over the basic
model in terms of residuals and interpretation. Coincidentally, the
bulge magnitude is almost identical to that of the basic model, but
the bulge $R_e=3'' / 700~\pc$ and $n=3.3$ are $\sim50\%$ and $300\%$
larger, respectively. Fitting the outer ring proves essential to keep
the size of the main disk from growing extremely large. The truncation
of the two inner rings improve the residuals considerably and allow
bulge and disk to account for the central and inter-ring light. Even
if these models are less precise representations of the data, they are
still superior to the simple model and acceptable alternatives. When
removing the outer ring, or the truncation of the second ring, the
classical bulge magnitude differs by $-0.8\mg$ ($+0.2\mg$ from the
reference model respectively.  Removing the truncation of the inner
ring increases the residuals, but leaves $M_\mathrm{bul}$ nearly
unchanged. Finally, we note that using a \sersic\ profile for the main
disk also leads to a very similar overall model, as the \sersic\ index
of the disk is $0.8 \approx 1$ in the best fit. On average, these
alternatives provide for a systematic uncertainty estimate of the
bulge magnitude of $0.4\mg$.

\section{Ancillary information on \galfit\ image modeling}
\label{subsec:analysis:ancillary}

\subsubsection{Providing an accurate PSF}
\label{subsubsec:analysis:psf}

\begin{figure}
 \centering
 \includegraphics[width=8cm]{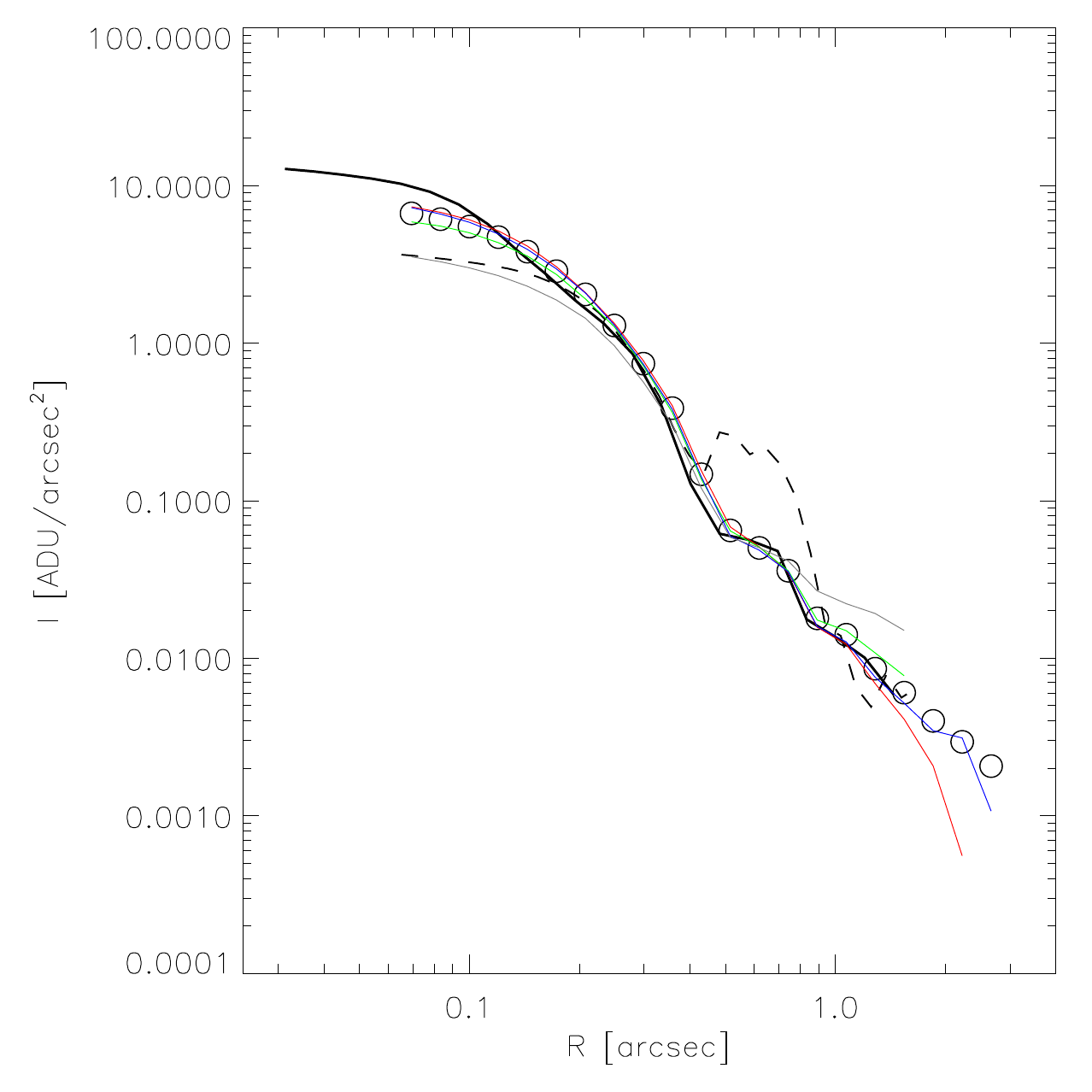}
 \caption{Comparison of our empirical PSF (open circles) with the analytic
TinyTim-based profile (dashed curve). The two are quite different
in the center and around $0.5''$, and the analytic version causes
significant mismatch with observed stars and bright galaxy
centers. The profiles of individual stars used to construct the PSF
image (thin colored curves) agree well with their sum and with one
another, indicating that broadening by centering errors, residual
background, contamination and saturation are minimal. Confirmation of
the accuracy of our empirical PSF also comes from comparing it with an
independent derivation by the CANDELS collaboration (solid thick black
curve), which albeit used upsampling before co-addition (to $0.06''$),
and therefore probes closer to the PSF center}
 \label{fig:psf_comparison}
\end{figure}

In order to account for the effects of the point-spread function
(PSF), \galfit\ convolves each model with a PSF
image. A scaled version of the PSF image is also used as a model of a
point source (AGN in our case). The accuracy of the PSF image effects
the fit results of small-scale components near the galaxy center,
including the bulge, and could in principle be important here. A PSF model can
be provided by detailed modeling of the optical path and detector
properties, as is commonly done for \hst\ images. However, we found
that the PSF that we obtained from TinyTim \citep[][]{TinyTim} is not
adequate to describe the surface brightness distributiosn of stars that 
we observe on our target fields. 
The problem is large enough to leave
characteristic ring-like residuals near the galaxy centers after
modeling, and to notably impact the AGN magnitude and bulge parameters
in some of our targets. We therefore derive the PSF
empirically using cutout images of non-saturated and isolated stars
found on our (co-added) science frames. The co-addition of several of
such star cutouts improves S/N and reduces residual background
uncertainty. It has the drawback of potentially broadening the PSF
profile near the center, due to finite pixel sampling and unavoidable
centering error (we do not resample onto fractional pixels). We tested
the broadening incurred in our co-added image and find it to be
marginal (see Figure \ref{fig:psf_comparison}).

The PSF image must be large compared to the PSF FWHM to include most
of the PSF image flux.  We use a 43x43 pixel common cutout area, which
is $\sim27$ times the FWHM of $0.2''$ that we measure. Thus, our PSF
image includes nearly 100\% of the total PSF flux.

\subsubsection{The sigma image}
\label{subsubsec:analysis:sigimg}

An image of the standard deviation of the flux per pixel (noise, or
''sigma'') is required to compute $\chi^2$. Providing a realistic
estimate of the sigma map is necessary for obtaining the ``true'' best-fit
solution of a given model. 
We obtain the sigma image by first computing it
on the ground-based and \hst\ image stacks separately. This consists
of measuring the background noise across the image as a whole (with
objects masked) for the ground-based images, and adding in quadrature
the Poisson noise from the object flux using the flux
itself and the local effective gain. For the \hst\ stacks, the noise
can be computed since the four
exposures are weighted evenly and all background
levels are known precisely. Afterwards, the two images are scaled and
combined, with the \hst\ data replacing ground-based data wherever it is
available, and the noise maps are scaled accordingly. 

\subsubsection{Object masks}
\label{subsubsec:analysis:masks}

Since we want to model the galaxy light unbiased by fore- and
background objects, masks for the latter are indispensable. We create
masks based on automatic object detection by \sex. We
subsequently add masks by hand for stars that are particularly bright
and thus have extended PSF wings, or those that overlap with the
galaxy so that they are not picked up by the automatic detection. In
one case (UGC3789), we opt not to mask the two stars near the galaxy
center, but include them in our model and thereby avoid masking much
of the area containing important constraints on the central
profile. None of our fields are particularly
crowded. 



\label{lastpage}

\end{document}